\documentclass[11pt,a4paper]{article} 
\pdfoutput=1
\usepackage{jheppub} 
\usepackage{graphicx} 
\usepackage{amsmath} 
\usepackage{amssymb} 
\usepackage{slashed} 
\usepackage{bigstrut} 
\usepackage{mathtools} 
\usepackage{multirow} 
\usepackage[utf8]{inputenc}
\usepackage{todonotes}
\usepackage{xspace}
\usepackage{lscape}

\newcommand{\msbar}{\ensuremath{\overline{\rm MS}}\xspace}

\newcommand{\polem}[1][]{\ensuremath{m_{#1}^{\mathrm{pole}}}\xspace}
\newcommand{\msbarm}[1][]{\ensuremath{m_{#1}(m_{#1})}\xspace}
\newcommand{\msrm}[1][]{\ensuremath{m_{#1}^{\text{MSR}}}\xspace}

\newcommand{\dz}{\ensuremath{D^0}\xspace}
\newcommand{\dch}{\ensuremath{D^{+}}\xspace}
\newcommand{\dstar}{\ensuremath{D^{*+}}\xspace}
\newcommand{\ds}{\ensuremath{D_s^{+}}\xspace}
\newcommand{\lamc}{\ensuremath{\Lambda_{c}}\xspace}

\title{Heavy-flavor hadro-production with heavy-quark masses renormalized in the \msbar, MSR and on-shell schemes}

\author[a]{M.V.~Garzelli,} 
\author[a]{~L.~Kemmler,}
\author[a,b]{~S.~Moch,} 
\author[c]{~O.~Zenaiev}

\affiliation[a]{II.~Institute for Theoretical Physics, Hamburg University \\
Luruper Chaussee 149, D--22761 Hamburg, Germany}
\affiliation[b]{MTA-DE Particle Physics Research Group, PO Box 105, HU--4010 Debrecen, Hungary}
\affiliation[c]{CERN, CH--1211, Geneva 23, Switzerland}

\emailAdd{maria.vittoria.garzelli@desy.de} 
\emailAdd{lkemmler@physnet.uni-hamburg.de} 
\emailAdd{sven-olaf.moch@desy.de} 
\emailAdd{oleksandr.zenaiev@cern.ch}

\abstract{
  We present predictions for heavy-quark production at the Large Hadron
  Collider making use of the \msbar and MSR renormalization schemes for the heavy-quark mass
  as alternatives to the widely used on-shell renormalization scheme. 
  We compute single and double differential distributions including QCD corrections 
  at next-to-leading order and investigate the renormalization and
  factorization scale dependence as well as the perturbative convergence in these mass renormalization schemes.
  The implementation is based on publicly available programs, {\texttt{MCFM}} and {\texttt{xFitter}}, extending their capabilities.
  Our results are applied to extract the top-quark mass using measurements of
  the total and differential $t\bar{t}$ production cross-sections and to investigate
  constraints on parton distribution functions, especially on the gluon distribution at low $x$ values, 
  from available LHC data on heavy-flavor hadro-production.
}  
 
\keywords{QCD,
  radiative corrections,
  heavy quarks, hadron colliders, re\-nor\-mali\-za\-tion, 
  parton distribution functions} 

\preprint{DESY 20-151}
\arxivnumber{arXiv:2009.07763} 
\begin{document} 
\maketitle

\section{Introduction}
\label{intro}

Charm-anticharm and bottom-antibottom pair-production are among the most
frequent inelastic processes occurring in hadronic collisions at the Large Hadron Collider (LHC), 
with cross-sections smaller than for the dijet case but
well above those for the top-antitop case, as follows from the hierarchy of the heavy-quark masses, 
the available phase-space, and the structure of the Standard Model (SM) Lagrangian.  

Heavy quarks are not observable as free asymptotic states. 
Charm- and bottom-quarks hadronize, due to confinement, whereas the top-quark decays before
hadronizing, due to its large decay width. 
Collider experiments implement procedures for reconstructing 
top-quarks from their decay products and are able to detect the products of the 
hadronization / fragmentation of intermediate-mass quarks, i.e. heavy mesons and baryons, 
as well as the jets associated to them, i.e. $b$-jets and $c$-jets.
$B$-hadrons and $D$-hadrons are reconstructed by their decay products, 
with decay vertices displaced with respect to the primary vertex. 
This operation requires good tracking, vertexing and particle identification capabilities. 
The measurements are indeed easier to perform in case of $B$-hadrons than for
$D$-hadrons, because the proper lifetimes of the first ones are longer than those of the latter.

On the theory side, analytical formulae for the hadro-production of massive quark pairs
are known since many years in quantum chromodynamics (QCD) perturbation theory at 
the next-to-leading order (NLO) accuracy, both for the total cross-sections 
as well as for one-particle inclusive differential distributions~\cite{
Nason:1987xz,Beenakker:1988bq,Nason:1989zy,Beenakker:1990maa,Czakon:2008ii}.
More recently, next-to-next-to-leading order (NNLO) QCD predictions have been computed 
for the total cross-sections of heavy-quark pair-production~\cite{Baernreuther:2012ws,Czakon:2012zr,Czakon:2012pz,Czakon:2013goa}. 
On the other hand, differential predictions at NNLO are available for top-quark pair
production~\cite{Czakon:2015owf,Catani:2019hip},
but not yet for the case of charm-quark pair. Very recently, a
  first computation of differential cross-sections for bottom-quark pair
  production including NNLO QCD corrections has
  appeared~\cite{Catani:2020kkl}.

Beyond fixed-order perturbation theory, the resummation of logarithms in the ratio $p_T/m$, relevant when the transverse momentum of the heavy-quark $p_T$ 
is much larger than its mass $m$, up to the next-to-leading-logarithmic accuracy, 
performed through the fragmentation function approach~\cite{Cacciari:1993mq}, 
has been combined with NLO predictions~\cite{Cacciari:1998it,Kniehl:2004fy,Kniehl:2005mk}. 
The resummation of recoil logarithms related to soft gluon radiation from initial state partons, 
as well as the one of threshold logarithms and high-energy logarithms, have
also been worked out (up to various degrees of accuracy) and presented 
in a number of papers (see e.g.~\cite{Banfi:2004xa,Kidonakis:2009ev,Kidonakis:2019yji,Catani:1994sq,Ball:2001pq,Moch:2012mk}). 
The transition from quarks to hadrons is described either by fragmentation
functions (FFs)~\cite{Tanabashi:2018oca, Kneesch:2007ey, Corcella:2017jjf}, or
by matching NLO matrix elements to parton shower and hadronization
approaches~\cite{Frixione:2003ei,Cacciari:2012ny,Garzelli:2015psa,Garzelli:2019kce,Mazzitelli:2020jio}. 

One of the inputs of all aforementioned calculations are the values of the heavy-quark masses.
The SM by itself does not predict the values of the quark masses, 
which are thus fundamental parameters and subject to renormalization.
The ultraviolet divergences appearing in the heavy-quark self-energies, 
to be eliminated by renormalization, require to fix a specific renormalization
scheme for relating the bare masses to the renormalized ones.
The most common choice is the mass in the on-shell scheme, defining the pole mass by the
relation that the inverse heavy-quark propagator with momentum $p$ vanishes
on-shell, i.e. for $p^2 = (m^{\text{pole}})^2$, and it is known at four loops in QCD~\cite{Chetyrkin:1997dh,Vermaseren:1997fq}.
Another option, also known at four loops~\cite{Marquard:2015qpa,Kataev:2015gvt}, is the \msbar prescription.
In complete analogy to the renormalization of the strong coupling constant $\alpha_S$, 
only divergent terms are absorbed such that the quark propagator becomes
finite after wave-function and mass renormalization. 
Finally, the MSR scheme~\cite{Hoang:2008yj} defines a scale-dependent short-distance mass, 
that interpolates smoothly between a valid mass definition at low scales much below the mass and 
the \msbar mass for scales larger than the mass, using 
infrared renormalization group flow.
Thus, predictions for physical observables in perturbation theory carry scheme dependence through the
choice of a particular mass renormalization scheme. 
For a given observable, the behavior of the truncated expansion in perturbative QCD,
including the apparent convergence and the residual scale dependence, can vary 
significantly between different schemes employed.

In this paper we describe the phenomenological effects of the use of the 
\msbar and MSR schemes for the renormalization of the heavy-quark
masses in charm, bottom and top production at hadron colliders.
We investigate the perturbative convergence in these schemes,
by providing comparisons between physical quantities calculated at various
levels of accuracy, and we discuss applications concerning the extraction of
mass values and parton distribution functions (PDFs) from collider data.

  Various phenomenological and experimental investigations on top
  quark hadro-pro\-duc\-tion with top mass renormalized in the \msbar scheme
  already exist in the literature~(see e.g. \cite{Langenfeld:2009wd,
    Alioli:2013mxa, Dowling:2013baa, Moch:2014tta, Fuster:2017rev,
    Aad:2019mkw, Catani:2020tko}). On the other hand, the MSR mass is
  discussed in various theory papers, mainly focusing on its definition and
  properties~(see e.g. \cite{Hoang:2017suc, Corcella:2019tgt,
    Hoang:2020iah}). As for the top quark case, the main novelties of our work
  are the phenomenological predictions for NLO differential
  cross-sections in $pp$ collisions using the MSR mass, which we compare to
  those with the \msbar and pole masses in a consistent framework, the
  implementation of a dynamical mass renormalization scale in the computation
  of transverse momentum distributions of the heavy-quarks, as an alternative
  to the static  $m(m)$ case, and the extraction of a top MSR mass value from the
  analysis of state-of-the-art LHC triple-differential cross-section data,
  previously used for extracting $m(m)$~\cite{Sirunyan:2019zvx}. The 
  extraction is done preserving correlations with the strong coupling constant $\alpha_S$
  and the PDFs, fitted simultaneously to the heavy-quark mass.

  Total cross-sections for charm and bottom production at
    hadron colliders with the mass renormalized in the \msbar scheme are available
    since a while~\cite{Aliev:2010zk}. The \msbar scheme has also been used
    for heavy-quark production in deep-inelastic scattering (DIS), which has allowed
    extractions of the charm mass $m_c(m_c)$ and bottom mass $m_b(m_b)$ (see
    e.g.~\cite{Alekhin:2010sv, Alekhin:2017kpj, Gizhko:2017fiu}). On the other
    hand, differential cross-sections for charm and bottom production at
    hadron colliders with \msbar and MSR masses have never been presented in a
    dedicated phenomenological paper before this one, to the best of our
    knowledge. We show examples of them here for the first time and discuss
    the relative uncertainties due to scale and mass variation, working at NLO
    accuracy. In case of \msbar cross-sections we also show the role of
    variations of the heavy-quark mass renormalization scales, comparing their
    effects to those induced by the variation of other scales appearing in
    fixed-order computations, i.e. the factorization and $\alpha_S$
    renormalization scales.
 
    In recent years heavy-flavor hadro-production data at the LHC
  have been used to constrain gluon and sea quark distributions at small $x$
  in PDF fits at NLO, working in the on-shell mass renormalization
  scheme~\cite{Zenaiev:2015rfa, Gauld:2015yia, Gauld:2016kpd,
    Bertone:2018dse}. In this work we comment on the simultaneous extraction
  of proton PDFs and the charm mass in the \msbar scheme, as an alternative to
  the pole scheme, using LHCb and HERA data and a fitting procedure were
  everything else is the same, except the charm mass renormalization
  scheme. Our findings at NLO support the use of masses renormalized in the
  \msbar scheme for further fits of this kind (see e.g.~\cite{Zenaiev:2019ktw,
    Alekhin:2018pai}). We also show for the first time how LHC charm
  production data extrapolated to the full phase-space can be used to
  constrain NNLO PDF fits, suggesting a reduction of the uncertainties in the
  gluon distributions of various widely used global PDF
  fits~\cite{Harland-Lang:2014zoa,Dulat:2015mca,Hou:2019efy}.

The implementation of the \msbar and MSR schemes for renormalizing the heavy
quark masses, as an alternative to the on-shell scheme, is described in
Sec.~\ref{imple}. The obtained theoretical predictions for differential
distributions including NLO QCD corrections and mass renormalization in these
three schemes, are presented in Sec.~\ref{diff}, together with considerations on the convergence
of the perturbative expansion in the strong coupling constant.   
The results are applied in Sec.~\ref{appli} to investigate the impact of LHC
data on possible extractions of PDFs from collider data and on determinations
of the heavy-quark mass values in different mass renormalization schemes. In
Sec.~\ref{nnlo-charm-fits}, we also use predictions for total cross-sections
of charm hadro-production up to NNLO accuracy in QCD.   
Finally, our conclusions are summarized in Sec.~\ref{conclu}. 

\section{Implementation of heavy-quark mass renormalization schemes}
\label{imple}

In this work light quarks are assumed to be massless. 
For the heavy-quark masses, on the other hand, different renormalization
schemes can be adopted and we briefly recall the relevant relations for the 
above mentioned cases of the \msbar,  MSR and on-shell schemes.
Other choices for the mass renormalization are possible. 
For physical observables inherently connected to the heavy-quark production threshold, for instance,  
the potential subtracted mass~\cite{Beneke:1998rk} was suggested as a possibility to produce an improved perturbative convergence
at energies slightly above the quark-pair production threshold
and the 1S mass has been presented~\cite{Hoang:1999zc} as a way of stabilizing the position of the peak of
the vector-current-induced total cross-section for $t\bar{t}$ production in
$e^+e^-$ collisions, as a function of the center-of-mass energy $\sqrt{s}$, for $\sqrt{s} \sim 2m$. 
In boosted kinematics, limited to the case of top-quarks, the top-quark jet-mass~\cite{Fleming:2007qr}, 
was introduced in the framework of Soft Collinear Effective Theory. 
Further mass renormalization schemes are described, e.g., in Refs.~\cite{Corcella:2019tgt,Hoang:2020iah} and references therein.

The on-shell mass coincides with the pole in the propagator of the renormalized quark field 
and is known up to four loops in QCD~\cite{Chetyrkin:1997dh,Vermaseren:1997fq}.
Thus, it is the same at all scales and infrared finite to all orders in perturbation theory.
This definition of the pole mass \polem, although being gauge invariant, 
has its short-comings~\cite{Bigi:1994em,Beneke:1994sw}.
It does not extend beyond perturbation theory, i.e. to full QCD, 
since it is based on the (unphysical) concept of colored quarks as asymptotic states.
Therefore, \polem must acquire non-perturbative corrections, 
which leads to an intrinsic uncertainty in its definition of the order 
$\mathcal{O}(\Lambda_{\mathrm{QCD}})$ related to the renormalon ambiguity~\cite{Smith:1996xz}.
The latter manifests itself as a linear sensitivity to infrared momenta in Feynman diagrams,
leading to poorly convergent perturbative series for the observables expressed in terms of \polem.

On the other hand, short-distance mass definitions such as the \msbar or the MSR schemes are renormalon-free. 
In general, such short-distance masses $m_{\mathrm{sd}}$ are related to the pole mass through the relation
\begin{equation}
  \label{eq:pole-sd}
  \polem \,=\, m^{\mathrm{sd}}(R,\mu_R) + \delta m^{\mathrm{pole-sd}}(R,\mu_R)\, ,
\end{equation}
where the term $\delta m^{\mathrm{pole-sd}}$ removes the renormalon 
and the dependence of the short-mass definition on long-distance aspects of QCD. 
Here, $\mu_R$ denotes renormalization scale, connected with the
ultraviolet divergences, whereas the scale $R$ is associated with the
infrared renormalization group equation (RGE)~\cite{Hoang:2008yj}.
In many short-distance mass renormalization schemes, $R$ coincides with the renormalized mass itself, 
as for instance in the \msbar scheme, where $R = m(\mu_R)$. 
However, the possibility to consider other choices of $R$ through the associated RGE 
allows to improve the stability of the conversion between short-distance mass schemes 
characterized by different values of $R$.

In the \msbar scheme, the renormalized mass of the heavy quark evolves with
the RGE in the renormalization scale $\mu_R$, governed by the 
mass anomalous dimensions $\gamma (\alpha_S(\mu_R))$,
\begin{equation}
  \label{eq:rge-msbar}
  \mu_R^2\, \frac{d m(\mu_R)}{d \mu_R^2} \,=\, - \gamma(\alpha_S(\mu_R))\, m(\mu_R)
  \, ,
\end{equation}
where the perturbative expansion of 
$\gamma (\alpha_S(\mu_R)) \equiv \sum_{i=0}^\infty\, \gamma_i\, (\alpha_S(\mu_R)/\pi)^{i+1}$ 
is known at four loops~\cite{Marquard:2015qpa}.
Precise determinations of the \msbar masses for charm- and bottom-quarks are
summarized by the Particle Data Group (PDG)~\cite{Tanabashi:2018oca}.
For the \msbar mass of the top-quark, see, e.g., 
Refs.~\cite{Tanabashi:2018oca,Alekhin:2017kpj,Fuster:2017rev,Aad:2019mkw,Sirunyan:2019jyn}. 
The conversion to the on-shell scheme proceeds in the standard manner
\begin{equation}
  \label{eq1}
  \polem \,=\, {m(\mu_R)} \left( 1 + \sum^{\infty}_{i=1}\, c_i\, \left(\frac{\alpha_S}{\pi}\right)^i \right)\, ,
\end{equation}
where the first numerical coefficients $c_i$ read~\cite{Gray:1990yh,Chetyrkin:1999qi,Melnikov:2000qh}, 
\begin{eqnarray}
\label{eq:msbar-to-pole1}
c_1 & = & \frac{4}{3} +  L 
\, , \\
c_2 & = & \frac{307}{32} + 2 \zeta_2 + \frac{2}{3} \zeta_2  \mathrm{ln }2 -\frac{1}{6} \zeta_3 + \frac{509}{72} L + \frac{47}{24}L^2  
\nonumber \\
\label{eq:msbar-to-pole2}
& &- \left(\frac{71}{144} + \frac{1}{3} \zeta_2 + \frac{13}{36}L + \frac{1}{12}L^2\right)n_{lf}
+ \frac{4}{3} \sum_{1 \le i \le n_{lf}}\, \Delta \left(\frac{m_i}{m(\mu_R)}\right)
\, .
\end{eqnarray}
Here, $\zeta$ denotes the Riemann zeta-function, $L \equiv \mathrm{ln}(\mu_R^2/m(\mu_R)^2)$ and 
the function $\Delta$ accounts for all quarks with masses $m_i$ smaller than the heavy-quark one. 
As the light quarks are taken massless here, i.e., $m_i=0$, the $\Delta$ term vanishes. 
The strong coupling $\alpha_S$ is evaluated at the scale $\mu_R$ 
and renormalized in the \msbar scheme with the number of
active flavors set to $n_{f}$ = $n_{lf} + 1$ at and above the heavy-quark threshold scale, 
which is assumed to be equal to its running mass. 
The number of light flavors is $n_{lf}=3, 4, 5$ for charm, bottom and top production, respectively.

For the particular choice $m(m) \equiv m(\mu_R=m(\mu_R))$, 
i.e. the \msbar mass renormalized at the specific scale $\mu_R = m(\mu_R)$, 
the logarithmic terms $L$ cancel and 
eq.~(\ref{eq1}) evaluates numerically (up to terms $\mathcal{O}(\alpha_S^4)$) 
as~\cite{Chetyrkin:1999ys}
\begin{eqnarray} 
  \label{fromruntopol}
  \polem & = & m(m) \left[1 + 1.333 \left(\frac{\alpha_S}{\pi}\right) 
    + \left(13.44 - 1.041\,n_{lf}\right) \left(\frac{\alpha_S}{\pi}\right)^2 
  \right. 
  \nonumber \\  & & 
  \left. 
    + \left(190.595 - 27.0\,n_{lf} + 0.653\,n_{lf}^2\right) \left(\frac{\alpha_S}{\pi}\right)^3 
    + \mathcal{O}(\alpha_S^4)\right]
  \, .
\end{eqnarray}
The infrared renormalon ambiguity in the conversion in eqs.~(\ref{eq1}), (\ref{fromruntopol})
manifests itself in practice as factorially growing terms in the perturbative
expansion, that spoil convergence.  
The sizable coefficients in eq.~(\ref{fromruntopol}) indicate the poor convergence 
of \polem for the case of charm and bottom, when $\alpha_S$ at low scales is large.
For top-quarks, the convergence is better due to the smaller value of $\alpha_S$ at larger scales. 
Including the four-loop QCD results~\cite{Marquard:2015qpa,Kataev:2015gvt}, 
the residual uncertainty in \polem for top-quarks, including renormalon contributions, 
is estimated of the order of a few hundred MeV \cite{Beneke:2016cbu}, 
i.e., of the order of $\mathcal{O}(\Lambda_{\mathrm{QCD}})$.
All available relations for scheme changes from $m(\mu_R)$ to \polem and vice versa have been
summarized in the programs {\texttt{CRunDec}}~\cite{Schmidt:2012az} 
and {\texttt{RunDec}}~\cite{Herren:2017osy}.

\begin{figure}
    \centering
    \includegraphics[width=0.48\textwidth]{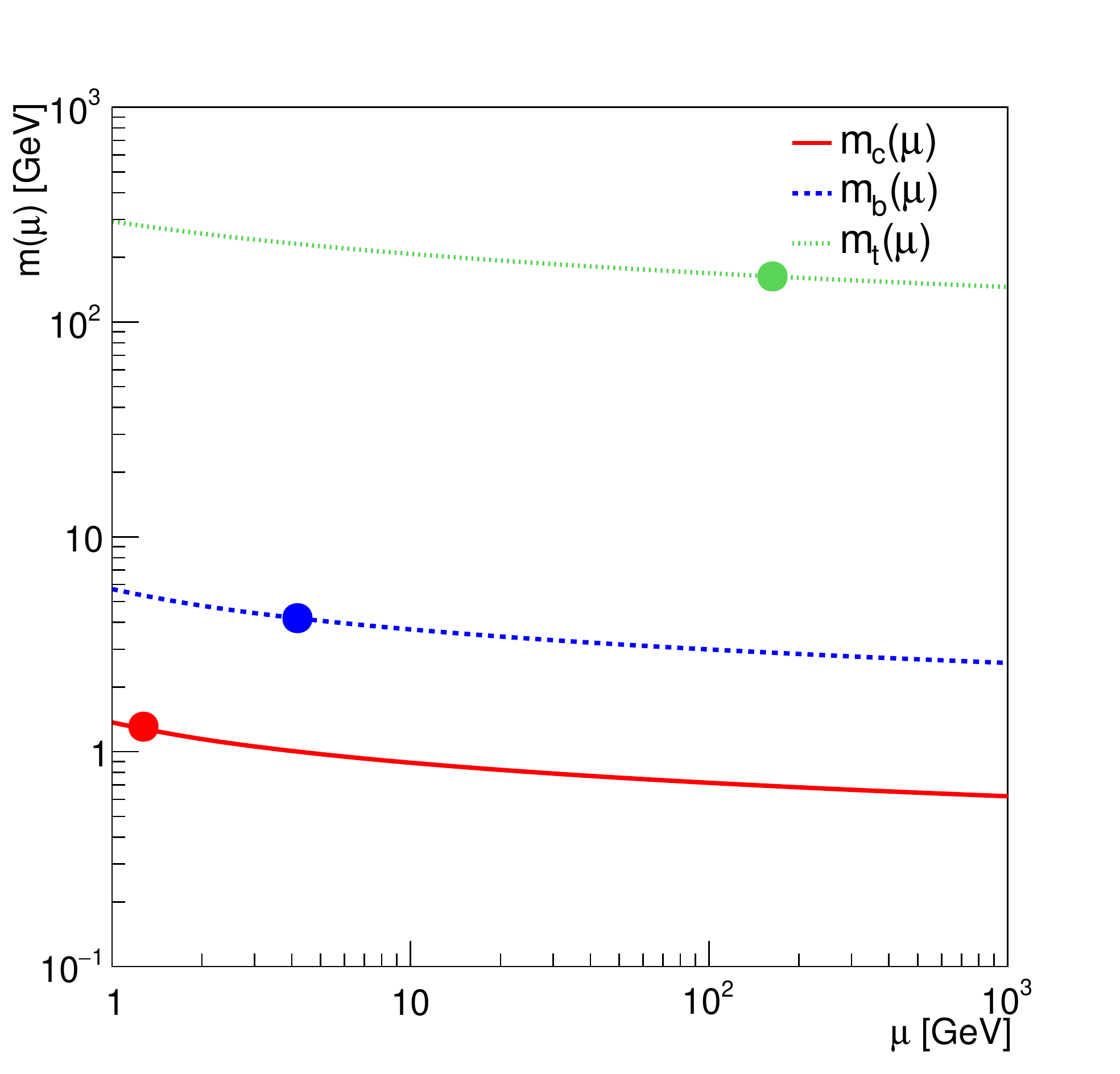}
    \includegraphics[width=0.48\textwidth]{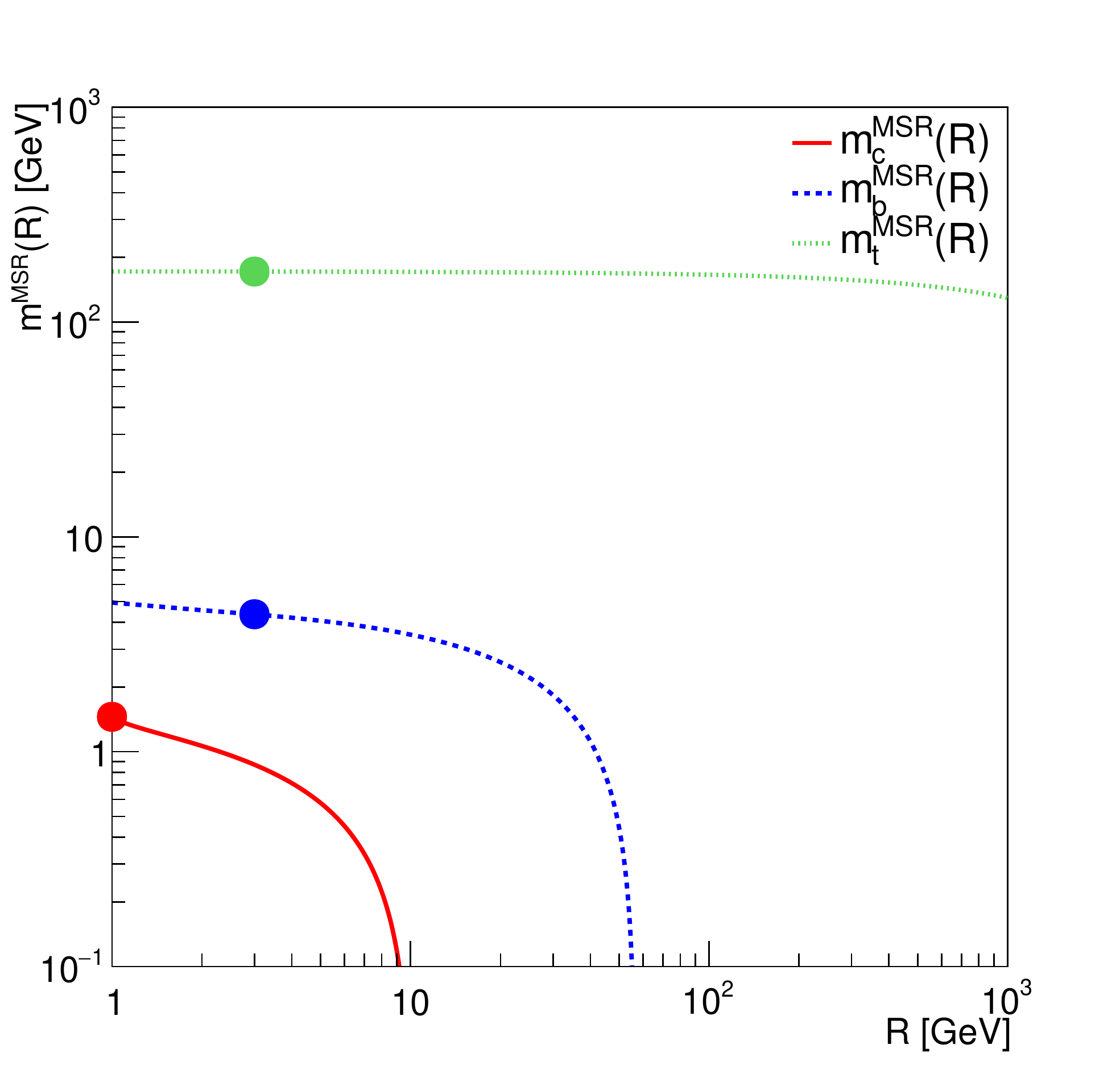}
    \caption{The one-loop evolution of the \msbar charm-, bottom- and top-masses
      for varying renormalization scales $\mu$ (left). 
      The one-loop evolution of the MSR charm-, bottom- and top-masses for varying scales $R$ (right). 
      The values of $\mu$ and $R$ used in the subsequent calculations and figures of this work are marked with
      filled dots. The value of $\alpha_S(M_Z)$ is fixed to 0.118 and $\alpha_S$ is evolved
      at four loops, as in Table~\ref{tab:masses}. 
      \vspace*{-2.mm}
    }
    \label{fig:mcmbmtrun}
\end{figure}

\begin{table}
  \begin{center}
    \begin{tabular}{|ccc|c|ccc|ccc|}
        \hline
        $m^{\text{MSR}}(1)$ & $m^{\text{MSR}}(3)$ & $m^{\text{MSR}}(9)$ & $m(m)$ & $m_{1\text{lp}}^{\text{pl}}$ & $m_{2\text{lp}}^{\text{pl}}$ & $m_{3\text{lp}}^{\text{pl}}$ & $m_{1\text{lp}}^{\text{pl}}$ & $m_{2\text{lp}}^{\text{pl}}$ & $m_{3\text{lp}}^{\text{pl}}$ \\
        \multicolumn{3}{|c|}{
          from $m(m)$}
        & & \multicolumn{3}{|c|}{
          from $m(m)$}
        &  \multicolumn{3}{|c|}{
          from MSR(3)
        }
        \\
        \hline
        \multicolumn{10}{|c|}{top-quark}\\
        \hline
        171.8 &        171.5 &        170.9 &        162.0 &        169.5 &        171.1 &        171.6 &        171.8 &        172.0 &        172.1 \\ 
        172.9 &        172.5 &        171.9 &        163.0 &        170.5 &        172.1 &        172.6 &        172.9 &        173.0 &        173.1 \\ 
        173.9 &        173.6 &        173.0 &        164.0 &        171.5 &        173.2 &        173.6 &        173.9 &        174.1 &        174.2 \\ 
        \hline
        \multicolumn{10}{|c|}{bottom-quark}\\
        \hline
        4.69 &         4.30 &         3.67 &         4.15 &         4.53 &         4.74 &         4.90 &         4.61 &         4.80 &         4.97 \\ 
        4.72 &         4.33 &         3.70 &         4.18 &         4.57 &         4.77 &         4.94 &         4.64 &         4.84 &         5.01 \\ 
        4.75 &         4.36 &         3.74 &         4.21 &         4.60 &         4.81 &         4.97 &         4.68 &         4.87 &         5.04 \\ 
        \hline
        \multicolumn{10}{|c|}{charm-quark}\\
        \hline
        1.33 &         0.94 &         0.31 &         1.25 &         1.46 &         1.68 &         1.98 &         1.25 &         1.44 &         1.61 \\ 
        1.37 &         0.97 &         0.35 &         1.28 &         1.50 &         1.70 &         2.00 &         1.29 &         1.48 &         1.65 \\ 
        1.40 &         1.01 &         0.38 &         1.31 &         1.53 &         1.73 &         2.02 &         1.33 &         1.52 &         1.69 \\ 
        \hline
    \end{tabular}
    \caption{
      Numerical values for heavy-quark MSR, \msbar and pole masses. 
      Columns 1--3 and 4 show the MSR masses at different $R$ scales (1, 3 and 9~GeV)
      and the \msbar mass from which they are obtained~\cite{Tanabashi:2018oca,Sirunyan:2019jyn}
      using eq.~(\ref{eq:massrunning}) with the anomalous dimensions at three-loop for the $R$-evolution of
      the MSR mass from the scale $R_0 = m(m)$ to $R$. 
      Columns 5--7 show the one-, two- and three-loop pole masses obtained from the
      conversion of the \msbar mass in eq.~(\ref{eq1}). 
      Columns 8--10 show the one-, two- and three-loop 
      pole masses obtained from the conversion of the MSR mass at $R=3$~GeV 
      using eq.~(\ref{eq:massrelation2}).
      All values are given in GeV.
      In the conversion formulas between the expression of masses in different
      renormalization schemes, we use the coupling constant $\alpha_S$ of the
      effective theory including 5 active flavors in case of top and 3 active flavors in case of charm
      and bottom, obtained through decoupling from the theory including one
      additional quark, supposed to be massive. 
      We fix $\alpha_s(M_Z)^{n_f=5} = 0.118$ ($\alpha_s(M_Z)^{n_f=3} = 0.106$)
      and we evolve $\alpha_S$ at four loops in all cases.  
    }
    \label{tab:masses}
    \end{center}
\end{table}

While $\alpha_S$ in eqs.~(\ref{eq1}), (\ref{fromruntopol}) is
renormalized in the \msbar scheme, the matrix elements, 
as well as the PDFs and the associated $\alpha_S$ evolution 
used in the fixed-order massive calculations presented in this paper
are all defined with a fixed number of light flavors $n_{lf} = 3$ for charm and bottom production~\footnote{
  The use of $n_{lf}$ = 3 even for bottom production is justified for 
  bottom-quark production at very low $p_T$ (see e.g. the available
  measurements of $B$-meson production by the LHCb
  collaboration~\cite{Aaij:2013noa,Aaij:2016avz,Aaij:2017qml}).} 
and $n_{lf} = 5$ for top production, even at scales well above the heavy-quark mass value.
For $\alpha_S$ renormalization in the decoupling
  scheme~\cite{Collins:1978wz}, subtractions in graphs with light-quark loops
  are done at zero mass, as in the \msbar scheme, whereas those in graphs with
  heavy-quark loops are done at zero momentum. The heavy quarks therefore do
  not contribute as active flavors to the running of $\alpha_S$ in the
  effective theory. Factorization is performed in the same scheme, which
  implies the use of non-vanishing PDFs only for light partons, and a
  calculation of partonic cross-sections done consistently. The latter is
  achieved by including contributions of virtual amplitudes
  corresponding to Feynman diagrams with massive heavy-quark fermionic
  loops. For the case of bottom production in the 3-flavor scheme, we verify
  that including or not the relevant charm-loop diagrams in our computations,
  assuming $m_c = 0$, produces differences well within 1\% on the total
  cross-section of $b\bar{b}$ hadro-production. We argue that including a
  charm mass different from zero, as appropriate for a fully consistent calculation in
  the 3-flavor scheme, makes also a small effect, well below the
  uncertainties on cross-sections due to scale variations. 
  This effect compensates only partially the differences due to
  the modified $\alpha_s$ value, that, at renormalization scales of the order
  of the bottom quark mass, is a few percent lower in the scheme with $n_f=3$
  active flavors, with respect to the $n_f=4$ case. Possible advantages of
  computing bottom production in the 3-flavor scheme have also been claimed
  in the framework of DIS calculations~\cite{Ablinger:2017tan,
    Alekhin:2020edf}, where the effects of including a finite charm-mass
  in the fermionic loop corrections turn also out to be small. We
  observe that our predictions for bottom production typically differ by some
  percent from those of fully consistent calculations with matrix-elements,
  PDFs and $\alpha_S$ in the 4-flavor scheme, depending on the input.

Using the standard decoupling relations, it is possible to
 relate the PDFs, $\alpha_S$ and the partonic cross-sections in the \msbar and decoupling schemes,
once the matching scale is fixed.
In this way, also eqs.~(\ref{eq1}), (\ref{fromruntopol}) can be re-expressed  
in terms of $\alpha_S$ with the heavy degrees of freedom decoupled.
If the decoupling is performed at a scale equal to the \msbar mass \msbarm, 
the coefficients $c_1$ and $c_2$ in eqs.~(\ref{eq:msbar-to-pole1}), (\ref{eq:msbar-to-pole2}) 
remain identical due to the fact that the leading order coefficient 
in the decoupling relation for $\alpha_S^{(n_{lf}+1)}$ to $\alpha_S^{(n_{lf})}$ vanishes. 

In practice, although the perturbative expansion in eqs.~(\ref{eq1}),~(\ref{fromruntopol}) 
is known up to four loops~\cite{Marquard:2015qpa}, 
we truncate it in this work to two loops (order $\alpha_S^2$) for computing the NNLO cross-sections and to
one loop (order $\alpha_S$) for computing NLO cross-sections, respectively, 
unless stated otherwise.
In addition, the evolution of $\alpha_S$ as a function of $\mu_R$ and the
corresponding $\alpha_S$ values entering in eqs.~(\ref{eq1}),~(\ref{fromruntopol})
and other parts of the fixed-order computation are evaluated retaining 
three loops for producing NNLO cross-sections and two loops for producing the NLO ones, 
respectively, unless stated otherwise. 

The MSR mass is a specific realization of the short-distance mass introduced in eq.~(\ref{eq:pole-sd}).
It is obtained, e.g., by considering the difference between \polem and \msbarm, 
see eq.~(\ref{eq1}), and substituting $m(\mu_R)$ with $R$ in the terms
proportional to $\alpha_S$ to determine the difference between \polem and $\msrm(R)$ as
%
\begin{equation}
  \label{eq:massrelation2}
  \polem \,=\, \msrm(R) + R\,\sum^{\infty}_{i=1}\, a_i\, \left(\frac{\alpha_s(R)}{\pi}\right)^i
  \, ,
\end{equation}
where the numerical coefficients $a_i$ are given in Ref.~\cite{Hoang:2008yj}.
The evolution of the MSR mass with the $R$ scale follows the RGE
\begin{equation}
  \label{eq:rge2}
  R\, \frac{d \msrm(R)}{d R} \,=\, - R\, \gamma^{\mathrm{MSR}}(\alpha_S(R)) 
  \, ,
\end{equation}
which is linear in the scale $R$ and 
where $\gamma^{\mathrm{MSR}}(\alpha_S(R)) \equiv \sum_{i=0}^\infty\, \gamma_i^{\mathrm{MSR}}\, (\alpha_S(R)/\pi)^{i+1}$ 
denotes the $R$ scheme anomalous dimension.
In practice, the MSR mass interpolates between the pole and the \msbar mass \msbarm. 
This occurs through the dependence on the scale $R$, because by construction 
$\msrm(R) \rightarrow \polem$ for $R \rightarrow 0$ and $\msrm(R) \rightarrow \msbarm$ for $R \rightarrow \msbarm$.

In the following we use 
what has been called {\it practical} MSR mass in Ref.~\cite{Hoang:2017suc}, in contrast to the {\it natural} MSR mass. 
For our purposes the numerical differences between those definitions are mostly negligible.

The evolution of the \msbar heavy-quark masses with renormalization scale is
shown in the left panel of Fig.~\ref{fig:mcmbmtrun}. 
It is calculated at one loop using the {\texttt{CRunDec}} program~\cite{Schmidt:2012az,Herren:2017osy} 
($n_{lf} = 3$ for charm and bottom, $n_{lf} = 5$ for top).
The evolution of the MSR heavy-quark masses with the $R$ scale at one loop is
shown in the right panel of Fig.~\ref{fig:mcmbmtrun}. 
It is obtained by solving the RGE in eq.~(\ref{eq:rge2}): 
\begin{equation}
  \label{eq:massrunning}
  {\msrm}(R) \,=\, {\msrm}(R_0) - \int^{\ln R}_{\ln R_0}\, R\, \gamma^{\mathrm{MSR}}(\alpha_s(R))\, {\rm d} \ln R
  \, ,
\end{equation}
where expressions for the first few coefficients of the anomalous dimension $\gamma^{\mathrm{MSR}}$
can be found in Ref.~\cite{Hoang:2008yj,Hoang:2017suc}. 
Here, eq.~(\ref{eq:massrunning}) is expanded up to the lowest non-vanishing order of $\alpha_s$.
As is visible in Fig.~\ref{fig:mcmbmtrun}, 
the \msbar mass values are decreasing with increasing values of the renormalization scale $\mu_R$. 
The MSR mass values are decreasing at increasing $R$ values, 
as follows from the form of the RGE for the $R$ evolution 
and the fact that the first coefficient $\gamma_0^{\mathrm{MSR}}$ in the perturbative 
expansion of the anomalous dimension $\gamma^{\mathrm{MSR}}$ is positive, cf. eq.~(\ref{eq:rge2}).

In Table~\ref{tab:masses} we compare the \msbar masses at the reference scale
$\mu_{\mathrm{ref}} \equiv m (\mu_{\mathrm{ref}})$, 
i.e. \msbarm, for charm-, bottom- and top-quarks~\footnote{
  For the top-quark masses, such comparisons have already been presented in Ref.~\cite{Moch:2014tta}.} 
with the pole masses \polem, obtained from the previous ones by retaining
different numbers of terms in the conversion formula eq.~(\ref{eq1}), 
and the MSR masses $\msrm(R)$ at various numerical values of the $R$ scale 
obtained by using eq.~(\ref{eq:massrunning}) for the evolution.
For top-quarks, the MSR mass value at $R=3$~GeV is numerically close to the
values obtained in the on-shell scheme at two- or three loops.
For bottom- or charm-quarks on the other hand, the conversion of $m(m)$ or $\msrm(R)$ 
to the on-shell scheme demonstrates the poor convergence of the perturbative expansion already
discussed above, cf. eq.~(\ref{fromruntopol}).

\section{Predictions for differential cross-sections}
\label{diff}

Predictions for cross-sections of heavy-quark production with different mass 
renormalization schemes can be obtained from those in the widely used on-shell
scheme by substituting eqs.~(\ref{eq1}) and (\ref{eq:massrelation2})
in the cross-sections and performing a subsequent perturbative expansion in $\alpha_s$, 
see Refs.~\cite{Langenfeld:2009wd,Dowling:2013baa}.
In particular, the cross-sections converted to the \msbar 
or MSR mass schemes are determined to NLO accuracy as follows 
\begin{eqnarray}
\label{eq:conversion}
\sigma^{\overline{\mathrm{MS}}}(m(\mu_R)) &=& 
\sigma^{\text{pole}}(\polem)\bigg|_{\polem=m(\mu_R)} + (m(\mu_R) - \polem)\left(\frac{{\rm d}\sigma^{0}}{{\rm d}m}\right)\bigg|_{m=m(\mu_R)}
\, ,
\\[1ex] \nonumber
\sigma^{\text{MSR}}(\msrm(R)) &=& 
\sigma^{\text{pole}}(\polem)\bigg|_{\polem=\msrm(R)} + (\msrm(R) - \polem)\left(\frac{{\rm d}\sigma^{0}}{{\rm d}m}\right)\bigg|_{m=\msrm(R)}
\, .
\end{eqnarray}
Here $\sigma^{0}$ is the Born contribution to the cross-section (proportional to $\mathcal{O}(\alpha_s^2)$), 
and the differences $m(\mu_R) - \polem$ and $\msrm(R)-\polem$
are calculated up to the lowest non-vanishing  order in $\alpha_s$, such that
all terms of order $\mathcal{O}(\alpha_s^4)$ are dropped in eq.~(\ref{eq:conversion}), as
appropriate for a NLO calculation at order $\mathcal{O}(\alpha_s^3)$. 
Formulae for scheme conversions up to NNLO haven been given 
in Refs.~\cite{Langenfeld:2009wd,Dowling:2013baa}.

\begin{table}
\renewcommand*{\arraystretch}{1.5}
\tabcolsep 0.025\textwidth
\centering
    \begin{tabular}{|c|ccc|}
        \hline
       Cross-section & pole mass scheme & \msbar mass scheme & MSR mass scheme \\
        \hline
        ${{\rm d}\sigma}/{{\rm d}p_T}$ & M, X & M, X & X \\
        ${{\rm d}\sigma}/{{\rm d}y}$ & M, X & M, X & X \\
        ${{\rm d}\sigma}/{{\rm d}M}$ & M, X & M & -- \\
        ${{\rm d}^2\sigma}/{{\rm d}p_T{\rm d}y}$ & M, X & X & X \\
        \hline
    \end{tabular}
    \caption{Summary of the capabilities of the {\texttt{MCFM}} (M) and 
      {\texttt{xFitter}} (X) frameworks to compute differential cross-sections for
      heavy-quark hadro-production in different mass schemes.
    } 
    \label{tab:xsec}
\end{table}

We are considering theory predictions for stable heavy quarks 
(in case of bottom and charm augmented by FFs to describe the final state $B$- and $D$-hadrons, as for the applications in Sec.~\ref{appli}). 
The additional impact of parton showers and the dependence of the quark mass parameter on their cutoff~\cite{Hoang:2018zrp} 
as well as the study of renormalon effects in obser\-vables with cuts leading to corrections of order $\Lambda_{\mathrm{QCD}}$ 
in the extracted quark masses~\cite{FerrarioRavasio:2018ubr} are subject of ongoing theory research. 

We have computed double-differential cross-sections as functions 
of the transverse momentum $p_T$ and rapidity $y$ of the heavy quark $Q$, 
and single-differential cross-sections as a function of the invariant mass
$M_{Q{\bar Q}}$ of the heavy-quark pair in the on-shell, \msbar and MSR mass renormalization schemes
at NLO using both frameworks, 
{\texttt{MCFM}}~\cite{Campbell:1999ah,Campbell:2011bn} 
with modifications~\cite{Dowling:2013baa, Kemmler:2019the} 
and {\texttt{xFitter}}~\cite{Alekhin:2014irh}. 
In both cases, the original NLO calculations are done in 
the pole mass scheme~\cite{Nason:1989zy,Beenakker:1990maa,Mangano:1991jk}. 
The modified {\texttt{MCFM}} program~\cite{Dowling:2013baa} is capable of
converting the NLO calculations using a pole mass into those with the heavy-quark
mass renormalized in the \msbar mass scheme,  in case of single-differential 
distributions in $p_T$ of the heavy quark, $y$ of the heavy quark and invariant mass $M_{Q{\bar Q}}$ of the heavy-quark pair.
On the other hand, the developed {\texttt{xFitter}} framework implements the
calculation of one-particle inclusive cross-sections (i.e., with the other
particle integrated over), i.e., it is capable to compute the
double-differential cross-sections as a function of $p_T$ and $y$ of the heavy
quark, but not as a function of the invariant mass $M_{Q{\bar Q}}$ of the
heavy-quark pair. It converts the pole-mass NLO cross-sections into the \msbar
and MSR mass schemes for fully differential distributions. 
The derivative of the Born contribution appearing in eq.~(\ref{eq:conversion})
is calculated semi-analytically in {\texttt{MCFM}} 
(see~\cite{Dowling:2013baa}), whereas it is computed numerically in
{\texttt{xFitter}}, which allows for cross-checks of both methods. 
The differential cross-sections in different schemes which can be computed by
{\texttt{MCFM}} and {\texttt{xFitter}} are summarized in Table~\ref{tab:xsec}. 
For all cross-sections calculated with both programs, {\texttt{MCFM}} and {\texttt{xFitter}}, 
(i.e. all those in the pole mass scheme and the $p_T$ and $y$ differential
distributions in the \msbar mass scheme), agreement within one percent
accuracy is found~\footnote{
  The {\texttt{MCFM}} and {\texttt{xFitter}}
  differential cross-sections for the production of all heavy-quarks
  (the latter are based on the program {\texttt{HVQMNR}}~\cite{Mangano:1991jk}) 
  calculated using $\mu_R=\mu_F$ agree within about $1\%$. 
  In view of the large scale uncertainties at NLO this level of agreement 
  is satisfactory.
}. 
{\texttt{xFitter}} is also interfaced to other codes, like
e.g. {\texttt{aMC@NLO}}, and thus it can be used for computing cross-sections
with heavy-quark masses renormalized in the \msbar scheme (instead of the pole
scheme implemented in the standalone standard version of these codes) for a
wider range of processes (e.g. $t\bar{t}+j$ hadro-production, see
Sec.~\ref{alphas-mt-fit} for the application of this interface to a
phenomenological study).

\begin{figure}[!h]
  \centering
  \includegraphics[width=0.46\textwidth]{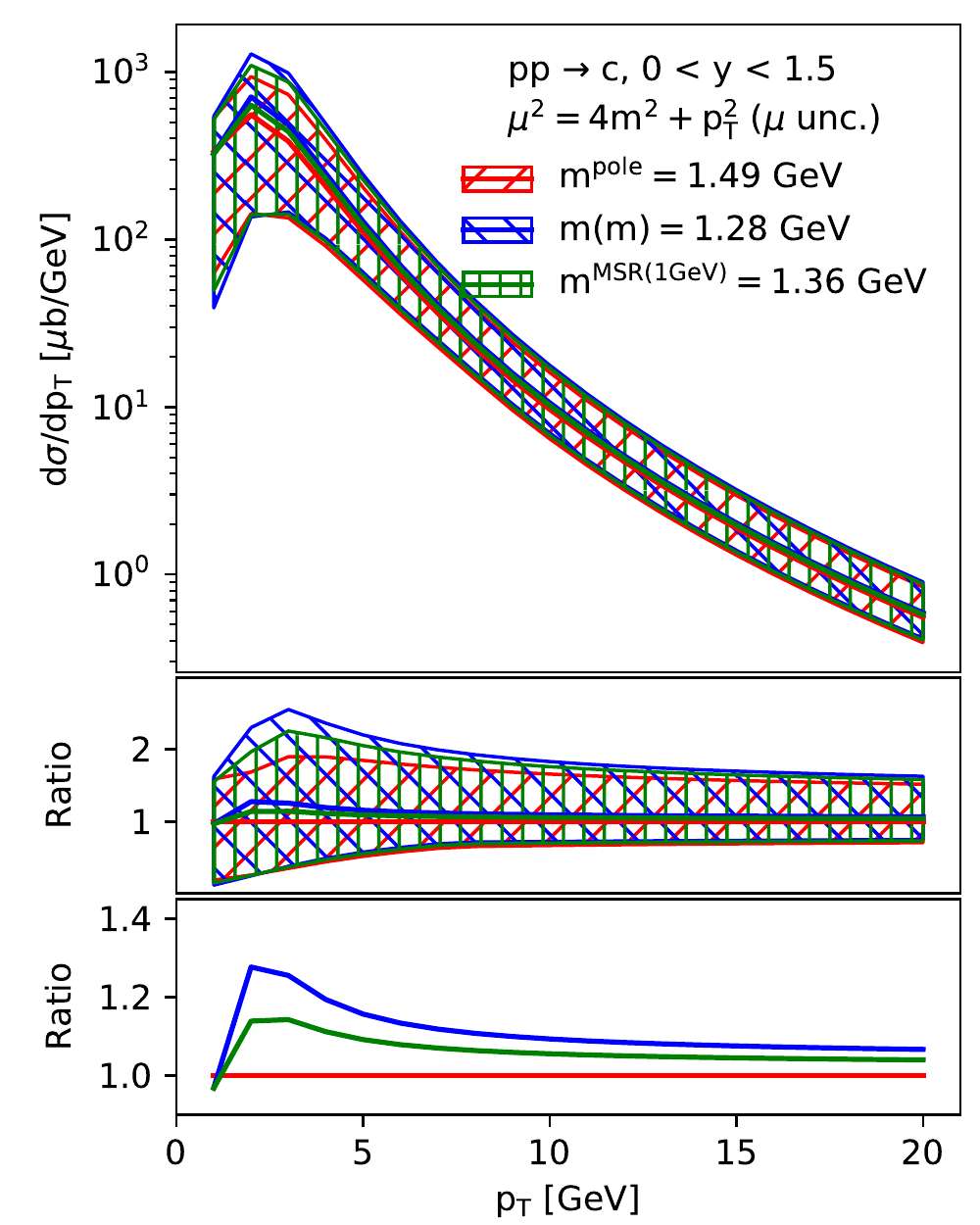}
  \includegraphics[width=0.46\textwidth]{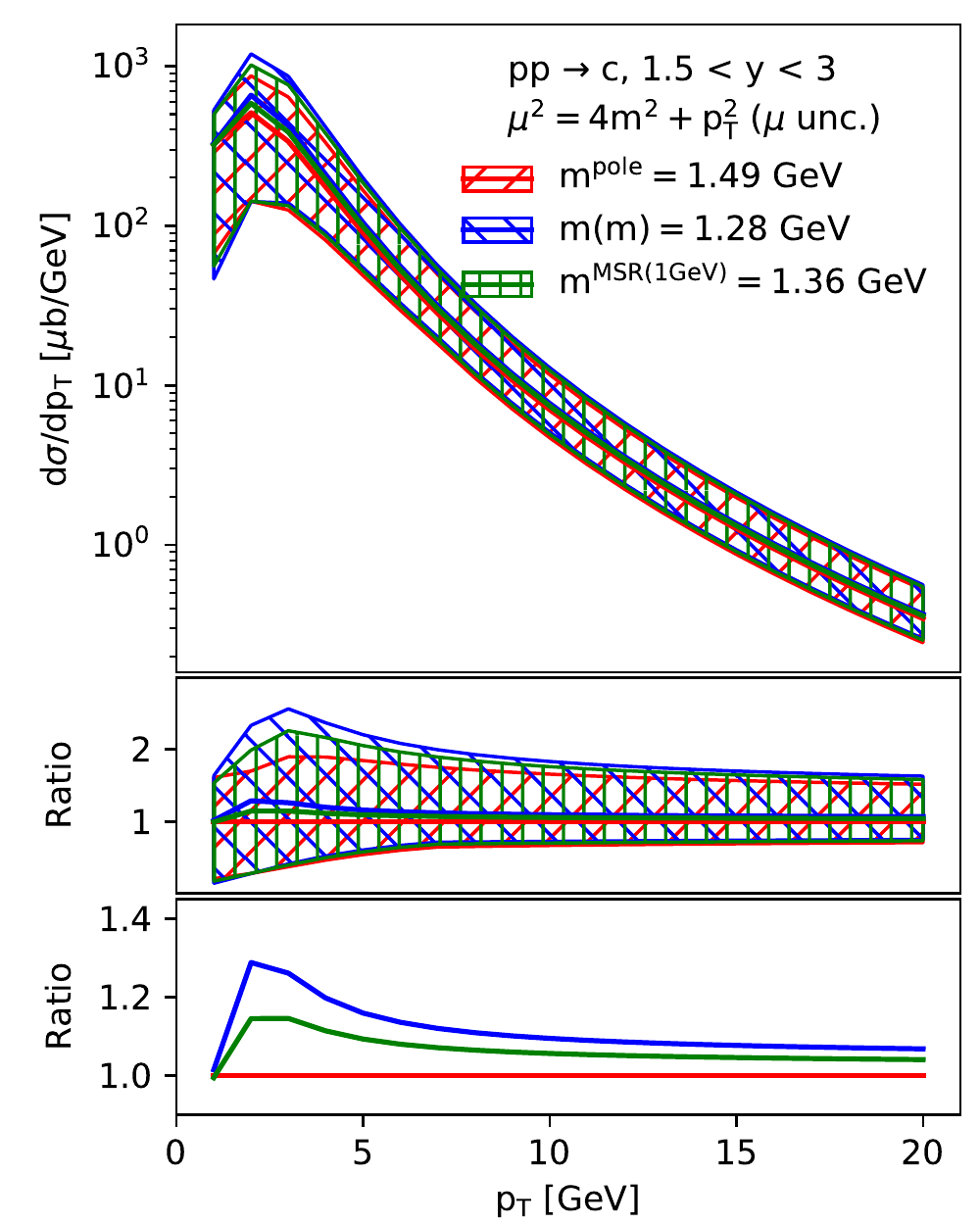}\\[5ex]
  \includegraphics[width=0.46\textwidth]{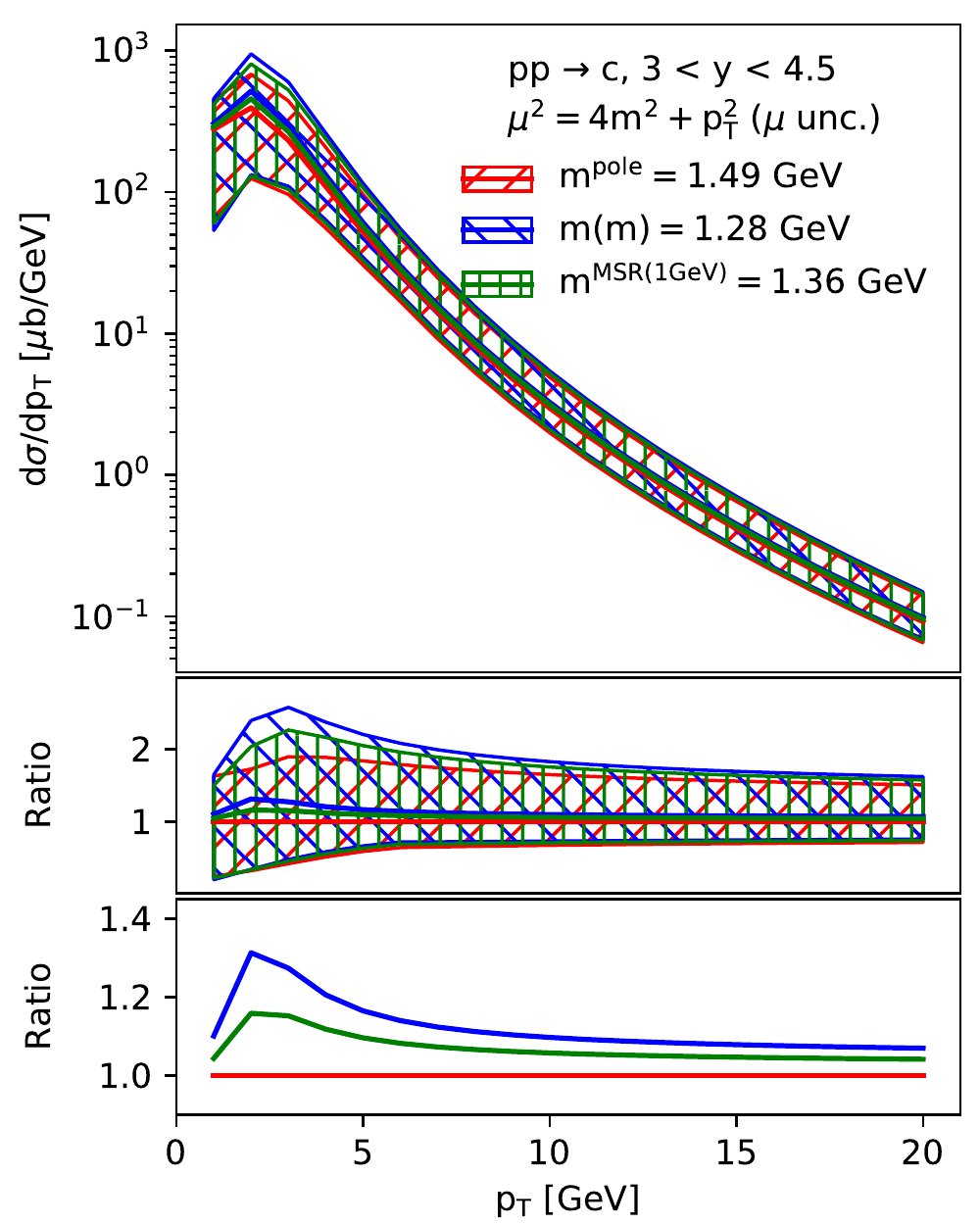}
  \includegraphics[width=0.46\textwidth]{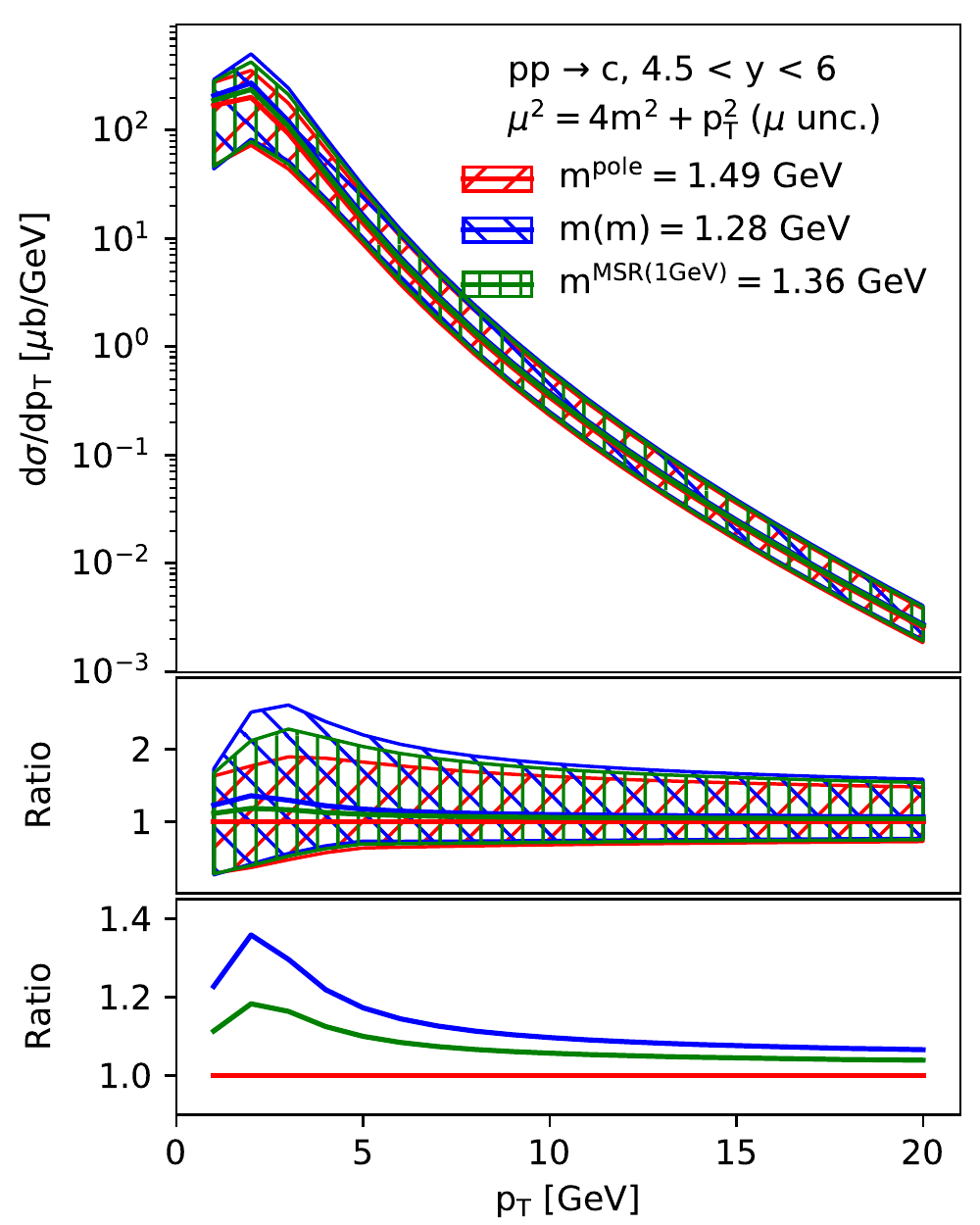}
  \caption{The NLO differential cross-sections for charm production at the LHC
    ($\sqrt{s} = 7$~TeV) 
    with their scale uncertainties as a function of $p_T$ in different intervals
    of $y$ of the charm-quark with the mass renormalized in the pole, \msbar and
    MSR schemes. 
    The lower parts of each panel display the theoretical predictions
    normalized to the central values obtained in the pole mass scheme,
    including scale uncertainties (upper ratio plot), or just
      the ratio of central predictions (lower ratio plot) in order to magnify
      shape differences.
  }
  \label{fig:c-pty-mu}
\end{figure}

\begin{figure}
  \centering
  \includegraphics[width=0.49\textwidth]{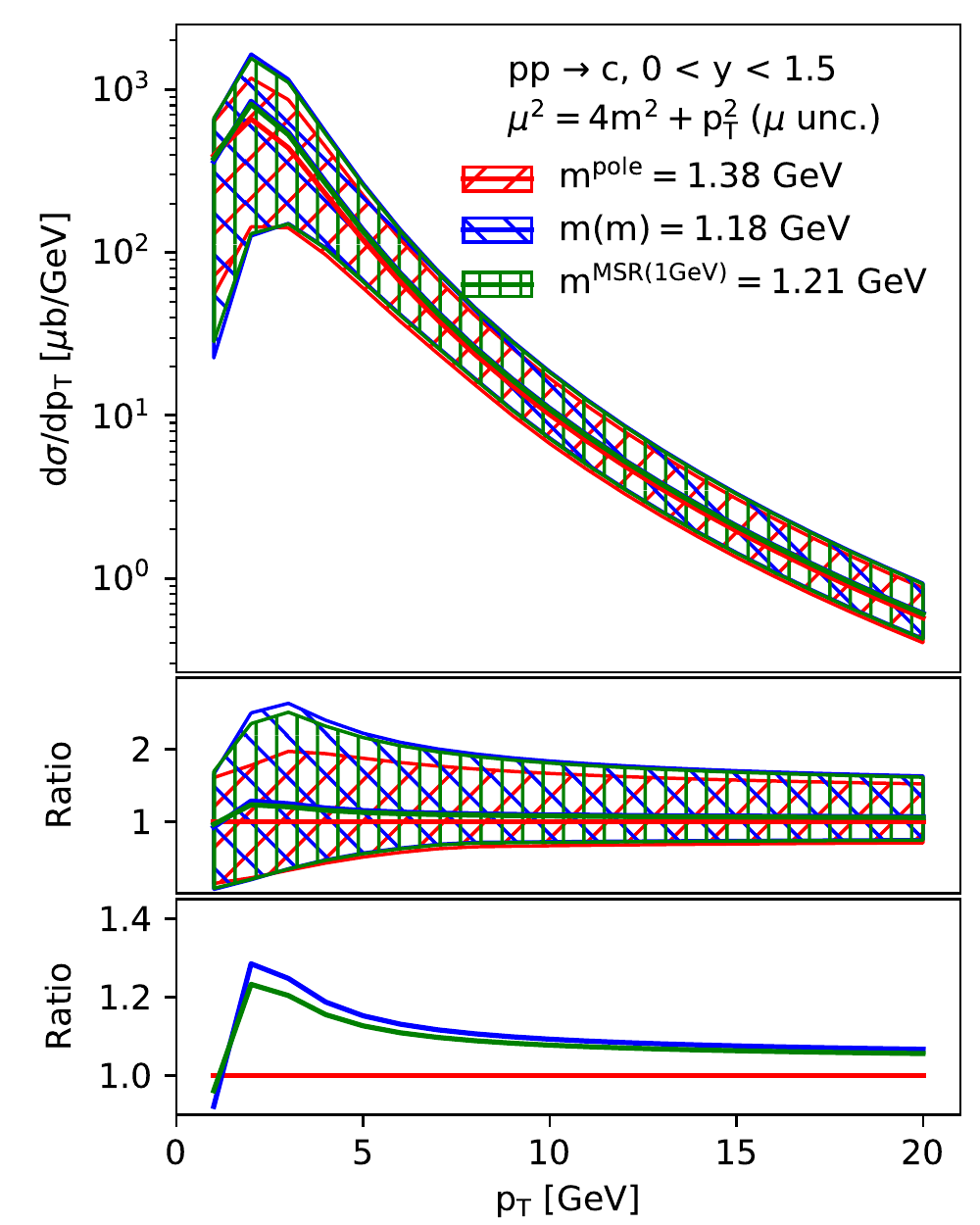}
  \includegraphics[width=0.49\textwidth]{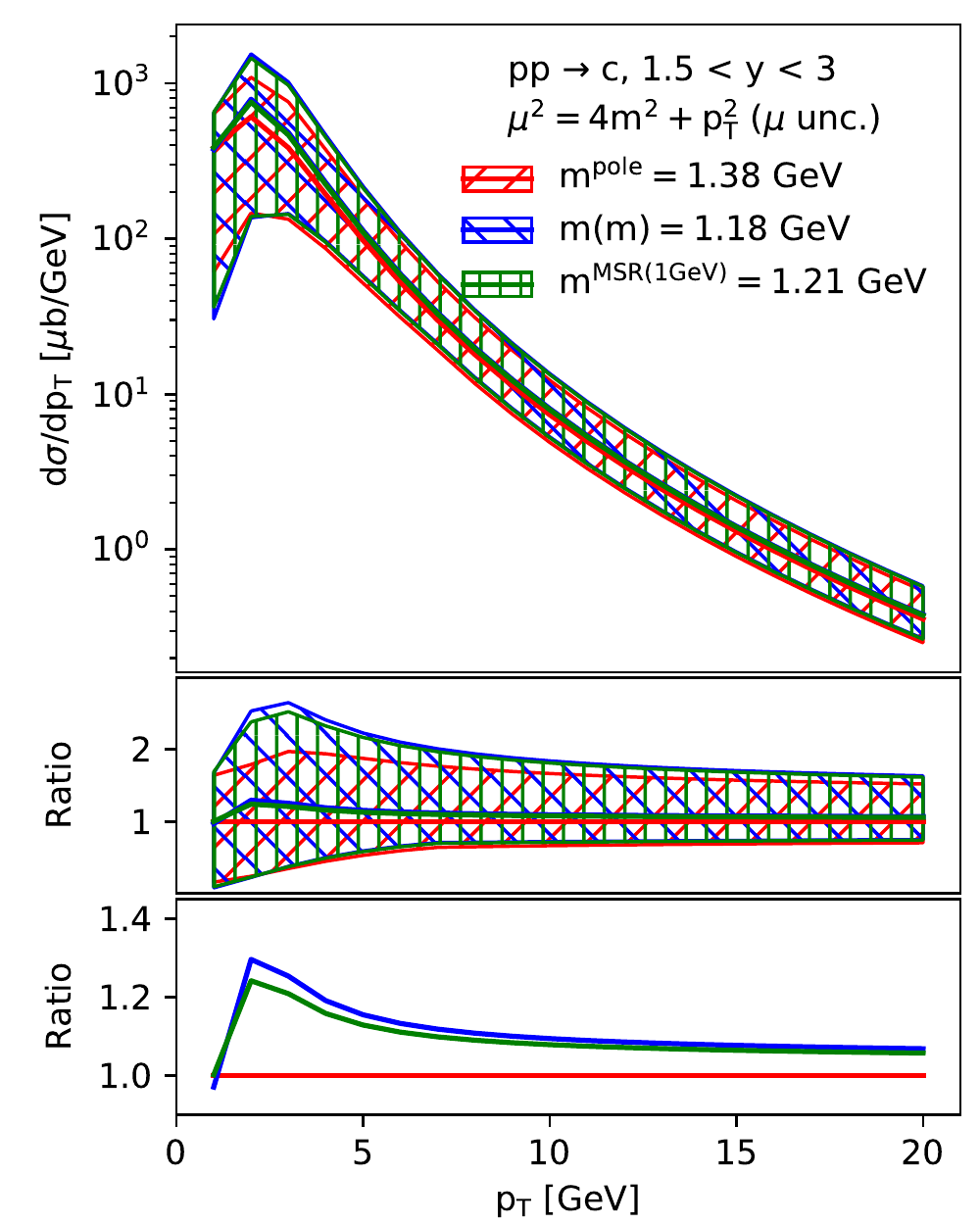}\\[8ex]
  \includegraphics[width=0.49\textwidth]{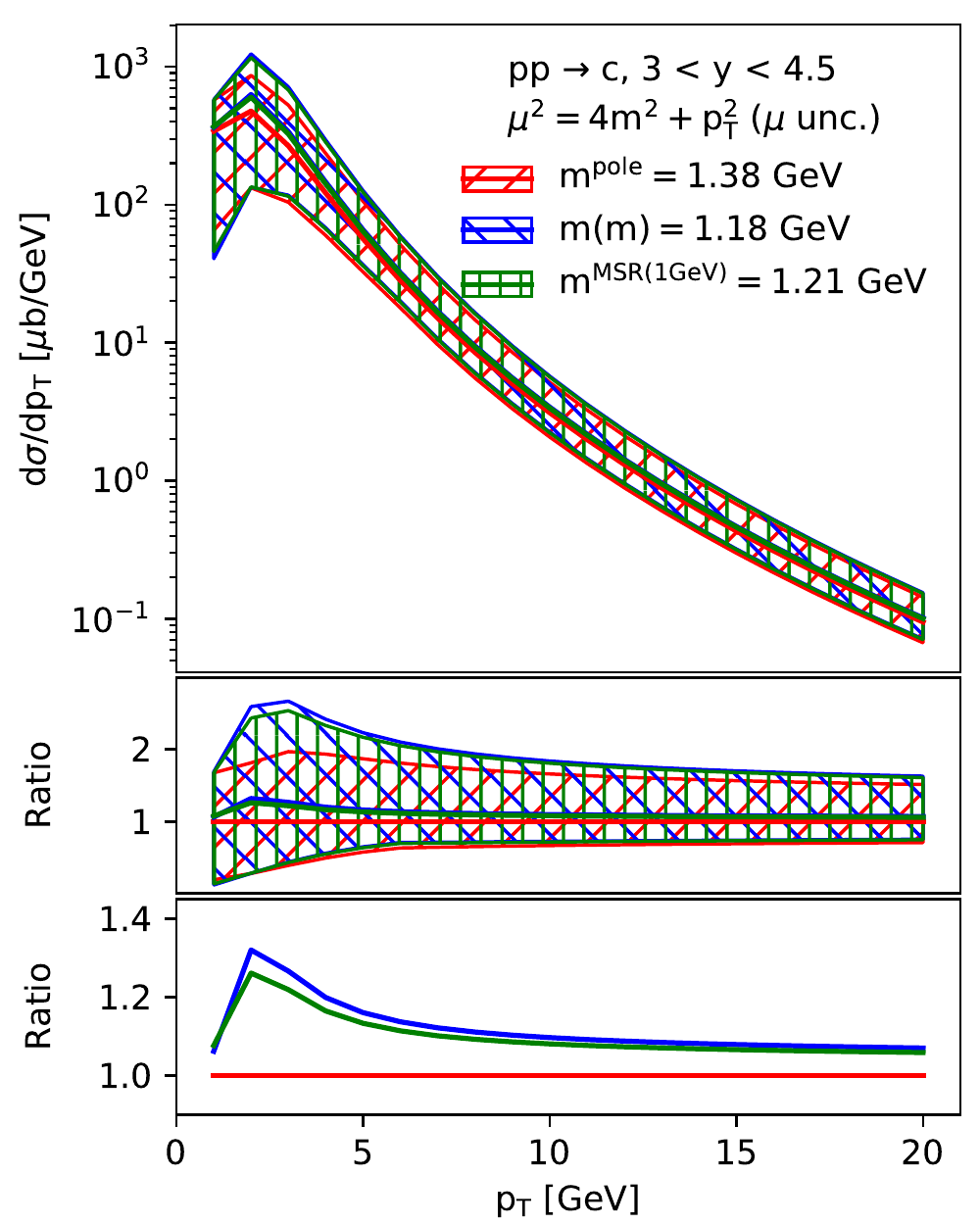}
  \includegraphics[width=0.49\textwidth]{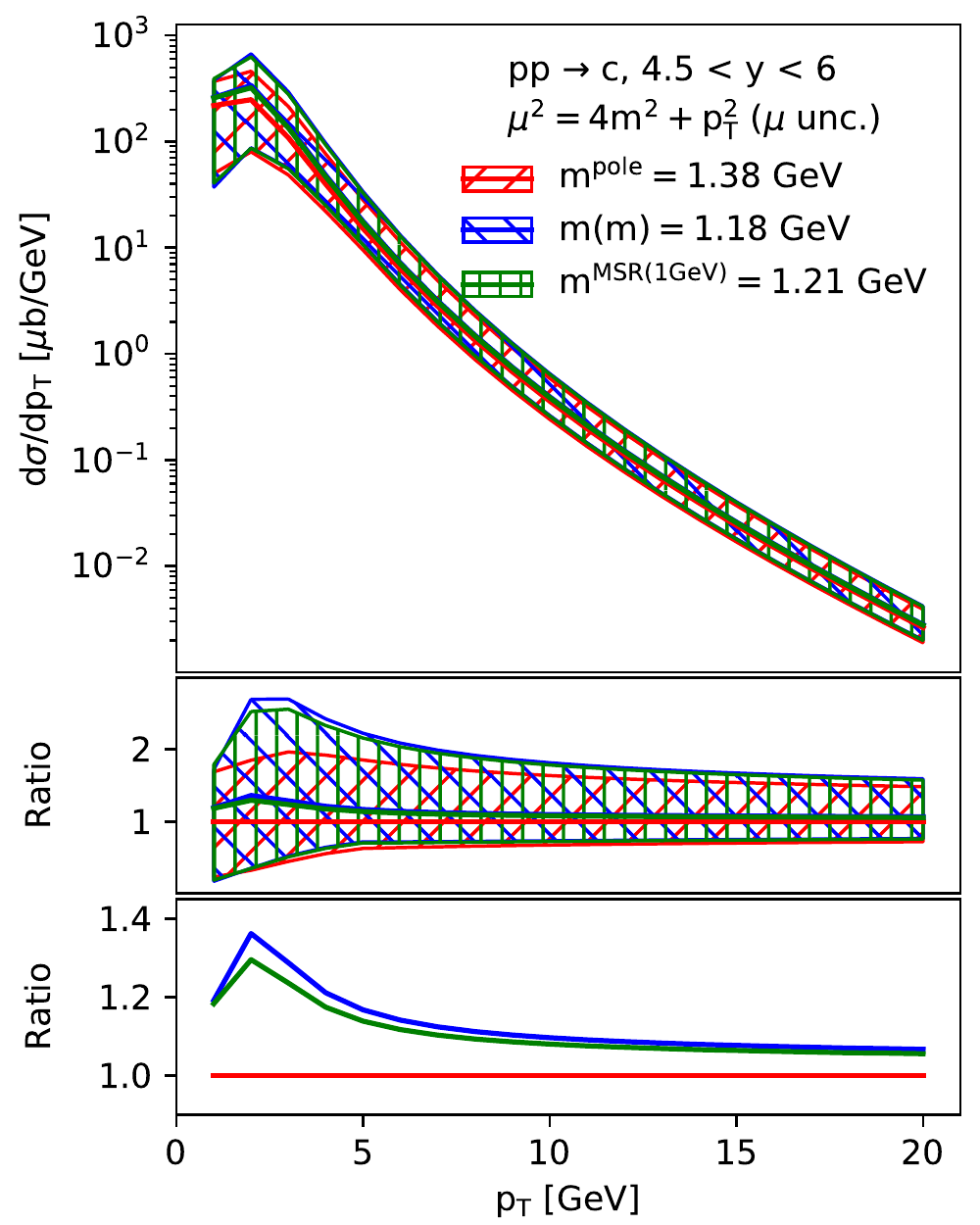}
  \caption{Same as Fig.~\ref{fig:c-pty-mu}, but for the charm-mass value
    $\msbarm[c]$~=~1.18~GeV (converted to $\msrm[c]$(1~GeV)~=~1.21~GeV
    and $\polem[c]$~=~1.38~GeV), as extracted in the ABMP16 NLO fit. 
  }
  \label{fig:c-pty-mu-abm}
\end{figure}

\begin{figure}
  \centering
  \includegraphics[width=0.49\textwidth]{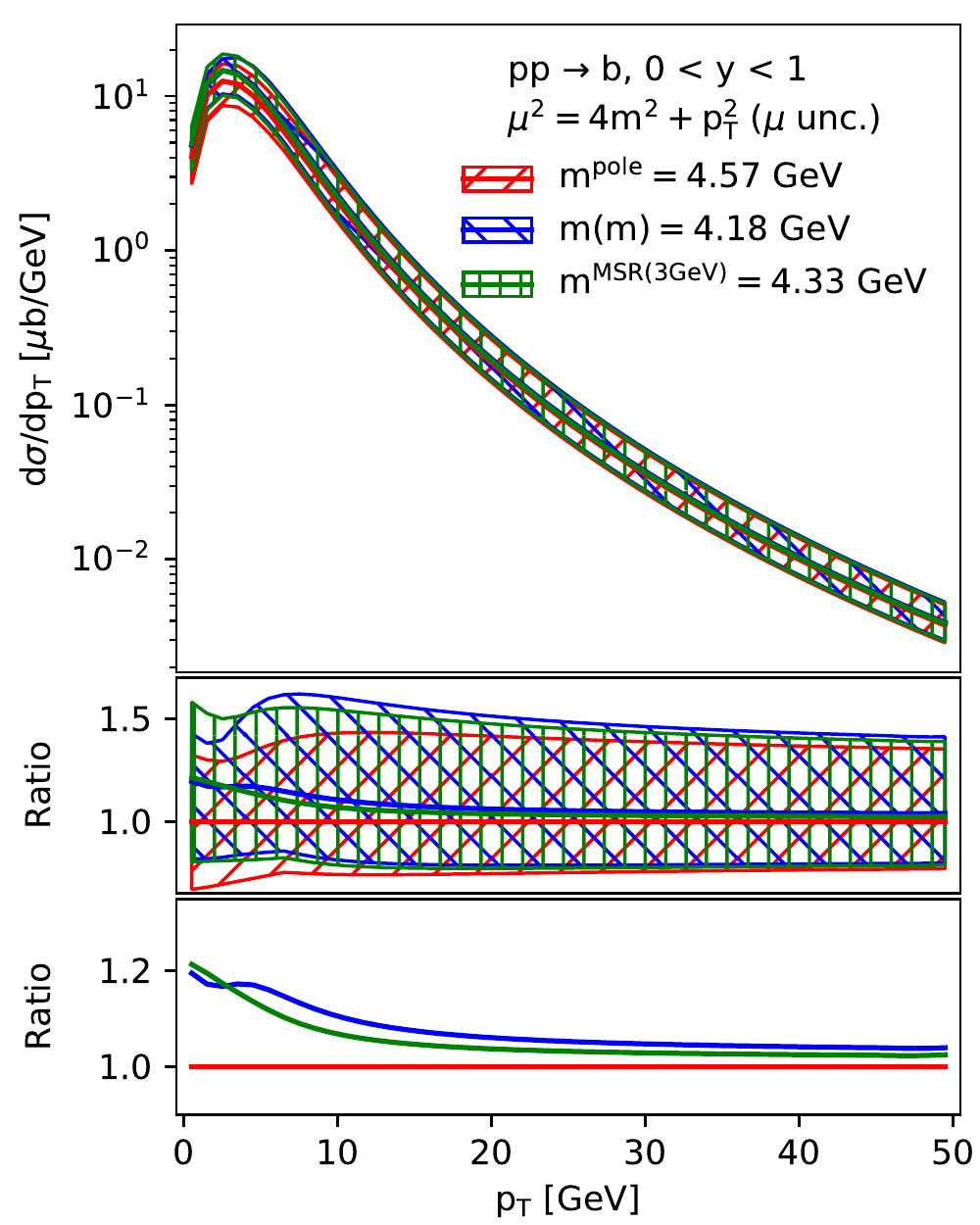}
  \includegraphics[width=0.49\textwidth]{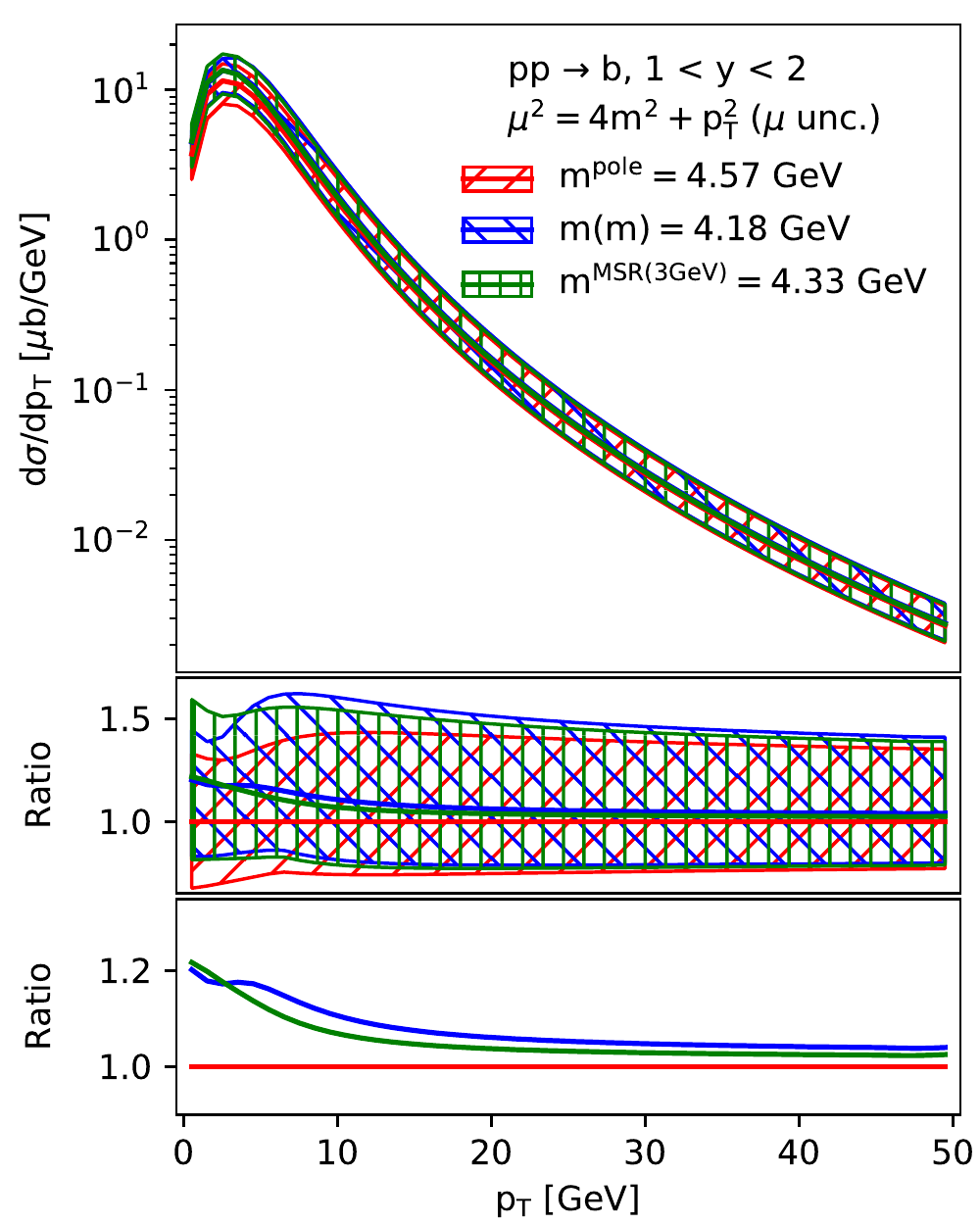}\\[10ex]
  \includegraphics[width=0.49\textwidth]{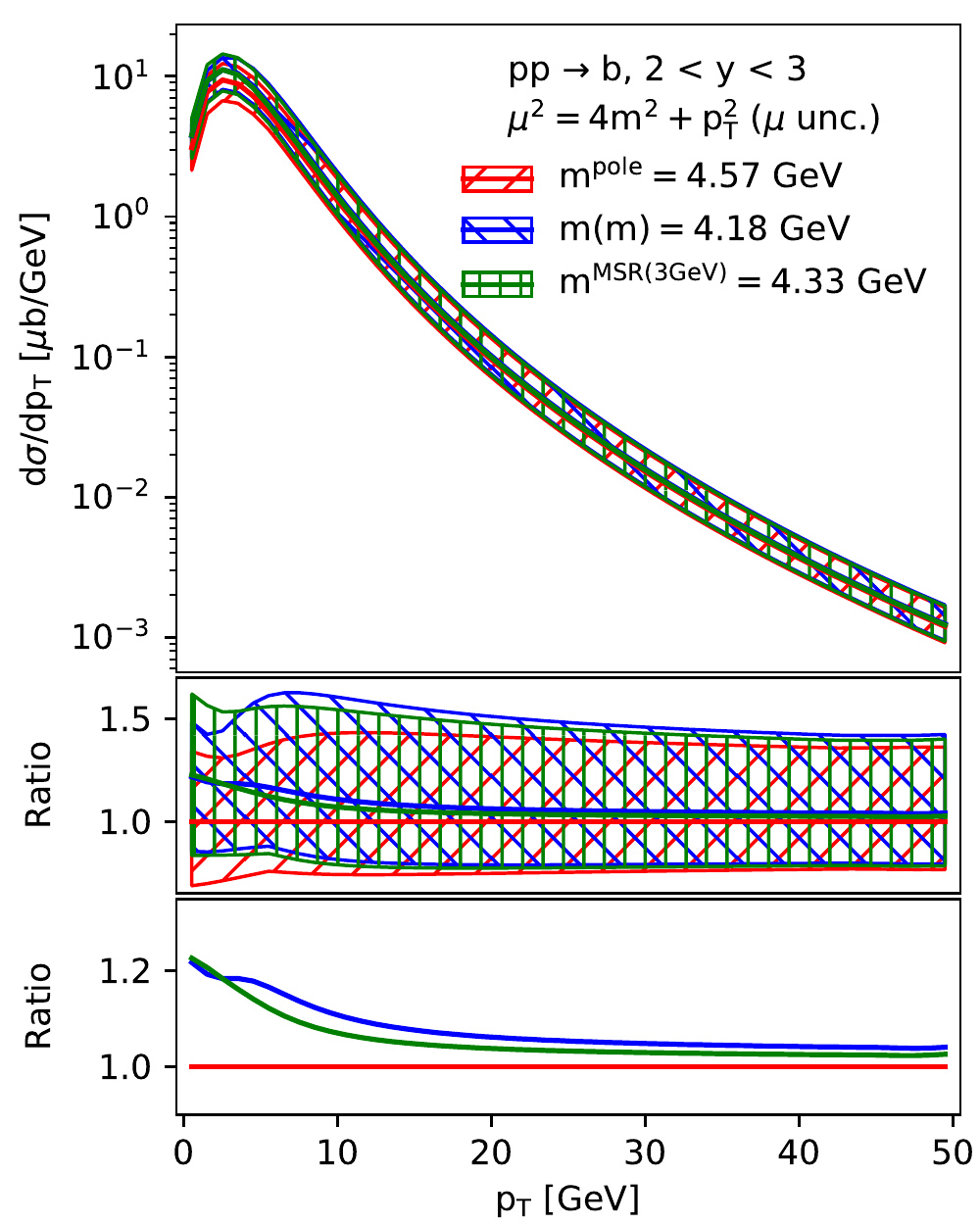}
  \includegraphics[width=0.49\textwidth]{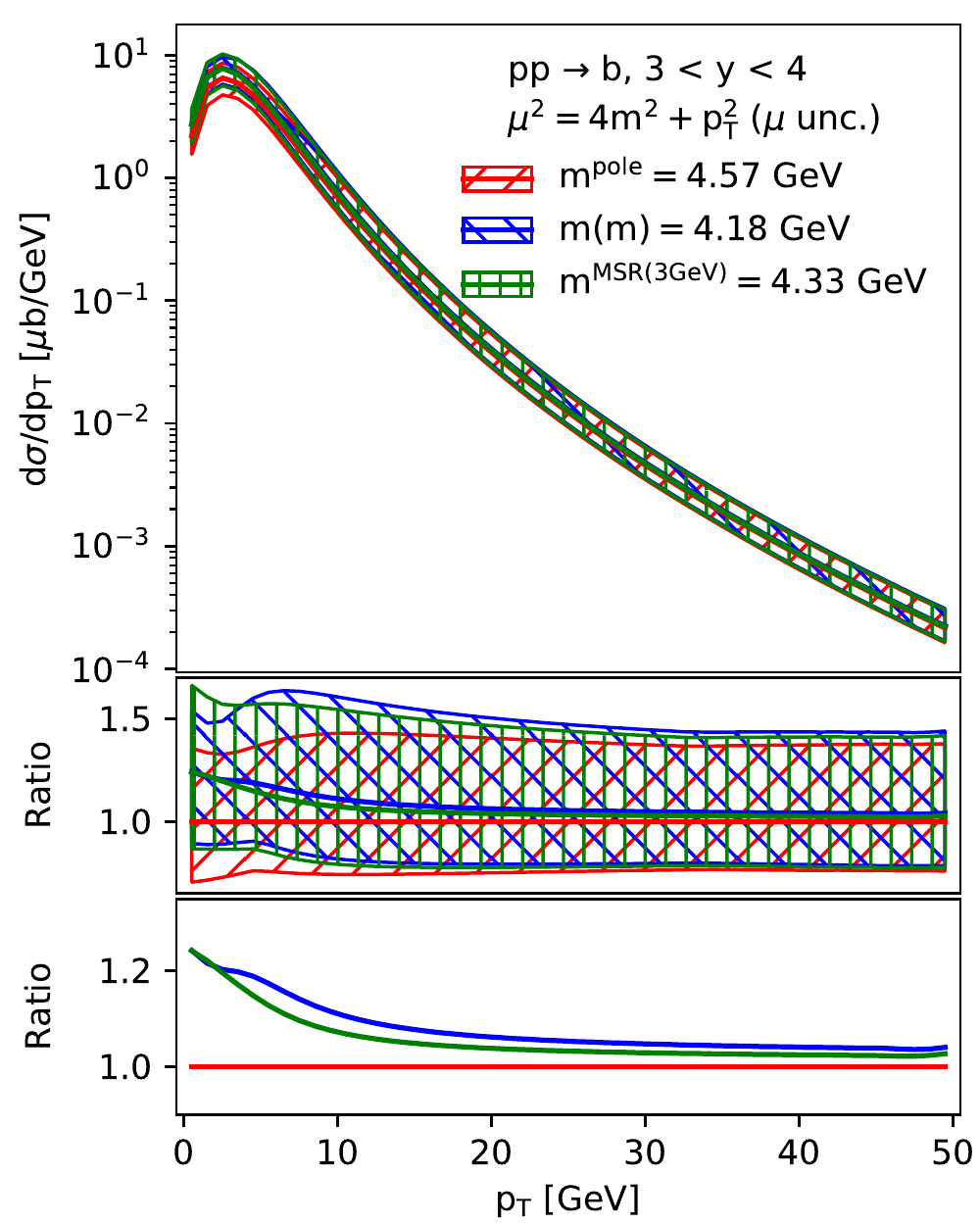}
  \caption{Same as Fig.~\ref{fig:c-pty-mu} for bottom production.
  } 
  \label{fig:b-pty-mu}
\end{figure}

\begin{figure}
  \centering
  \includegraphics[width=0.49\textwidth]{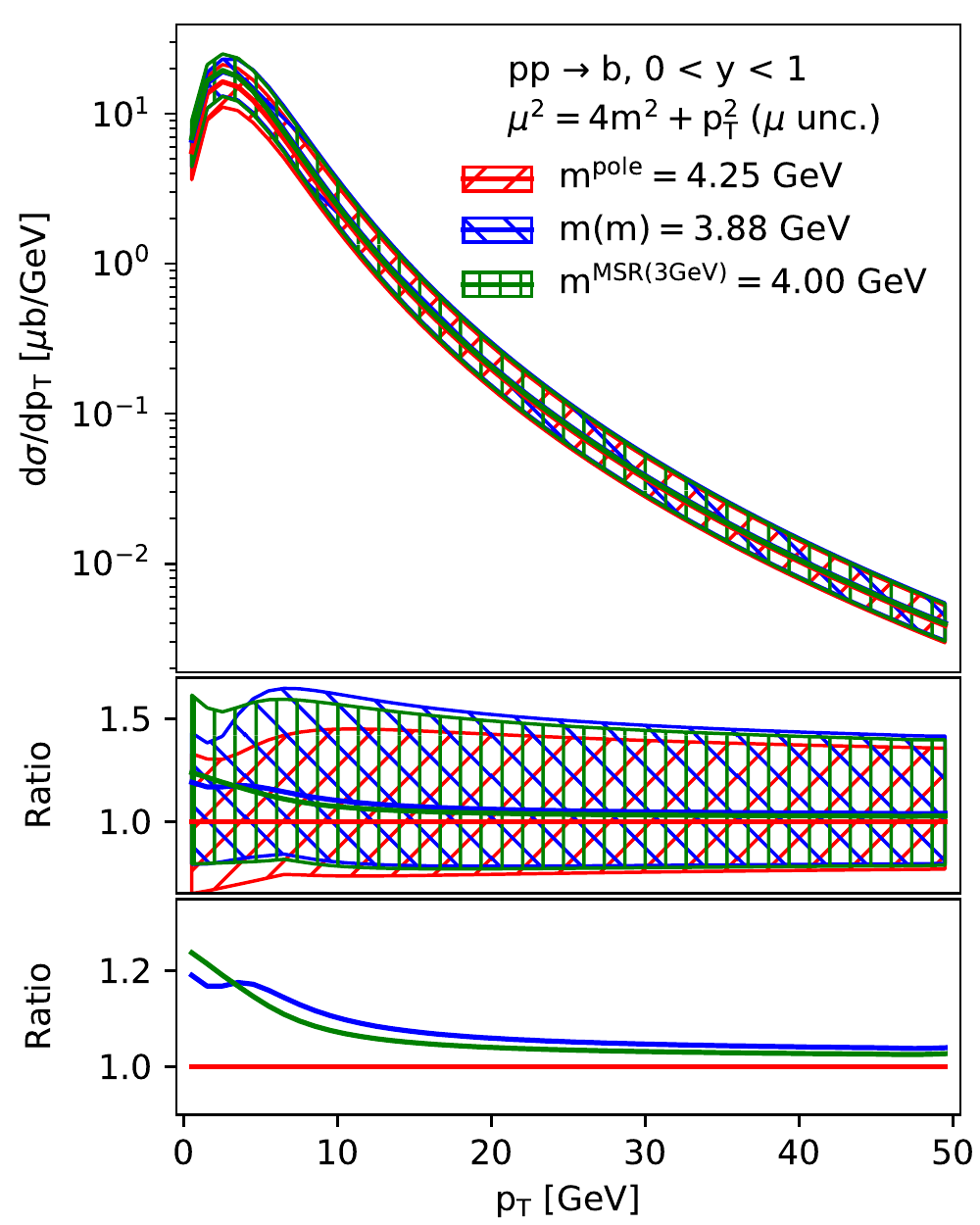}
  \includegraphics[width=0.49\textwidth]{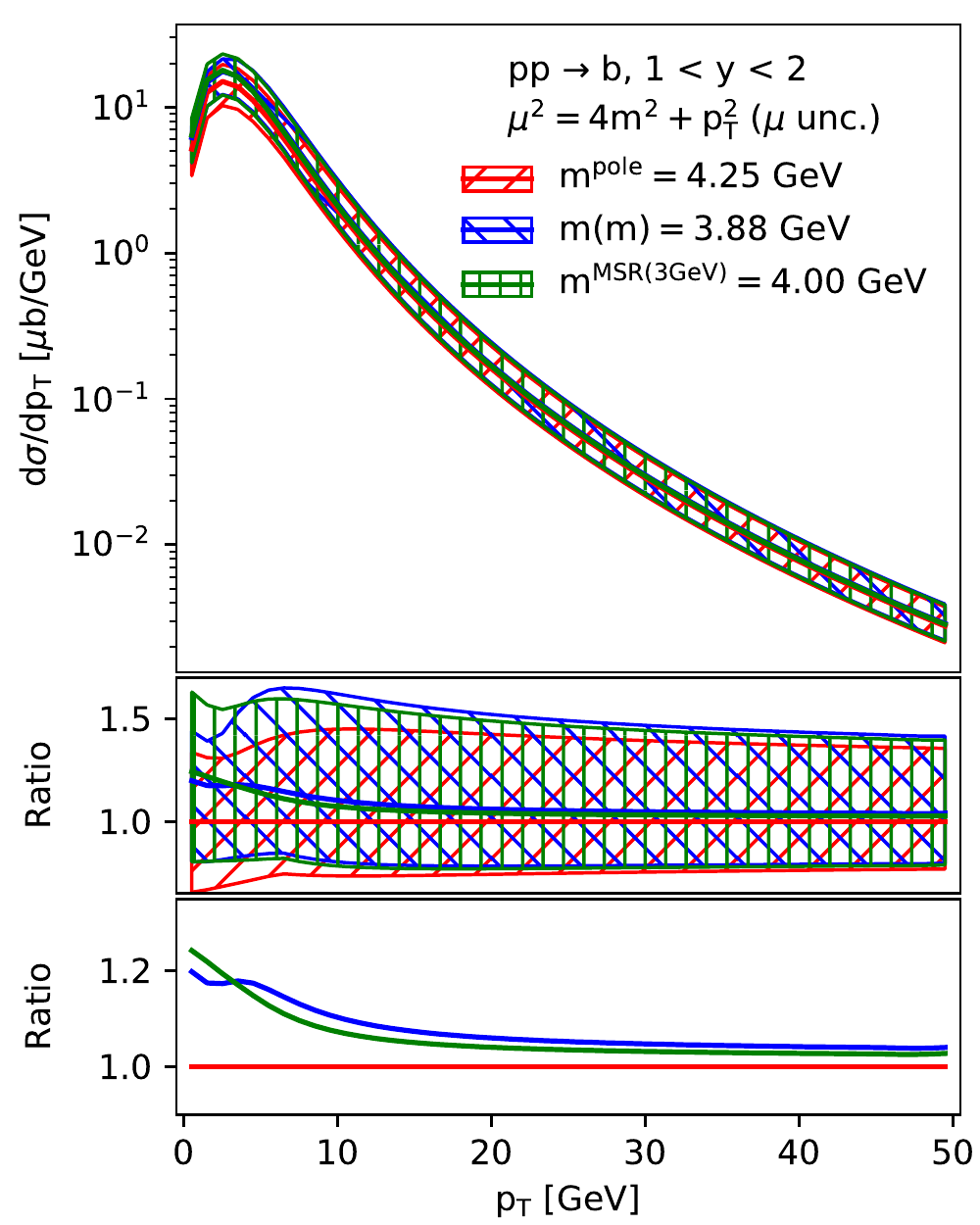}\\[8ex]
  \includegraphics[width=0.49\textwidth]{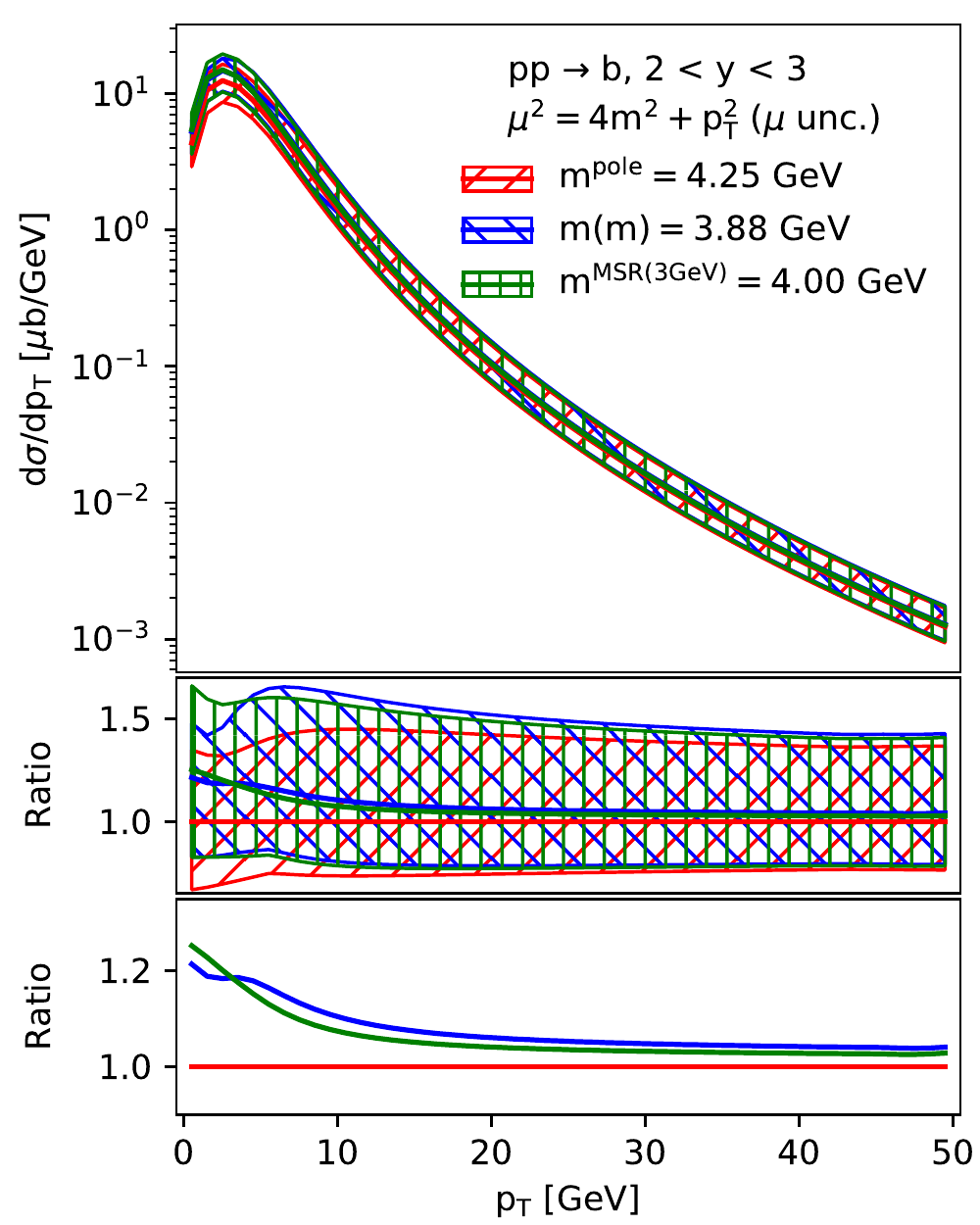}
  \includegraphics[width=0.49\textwidth]{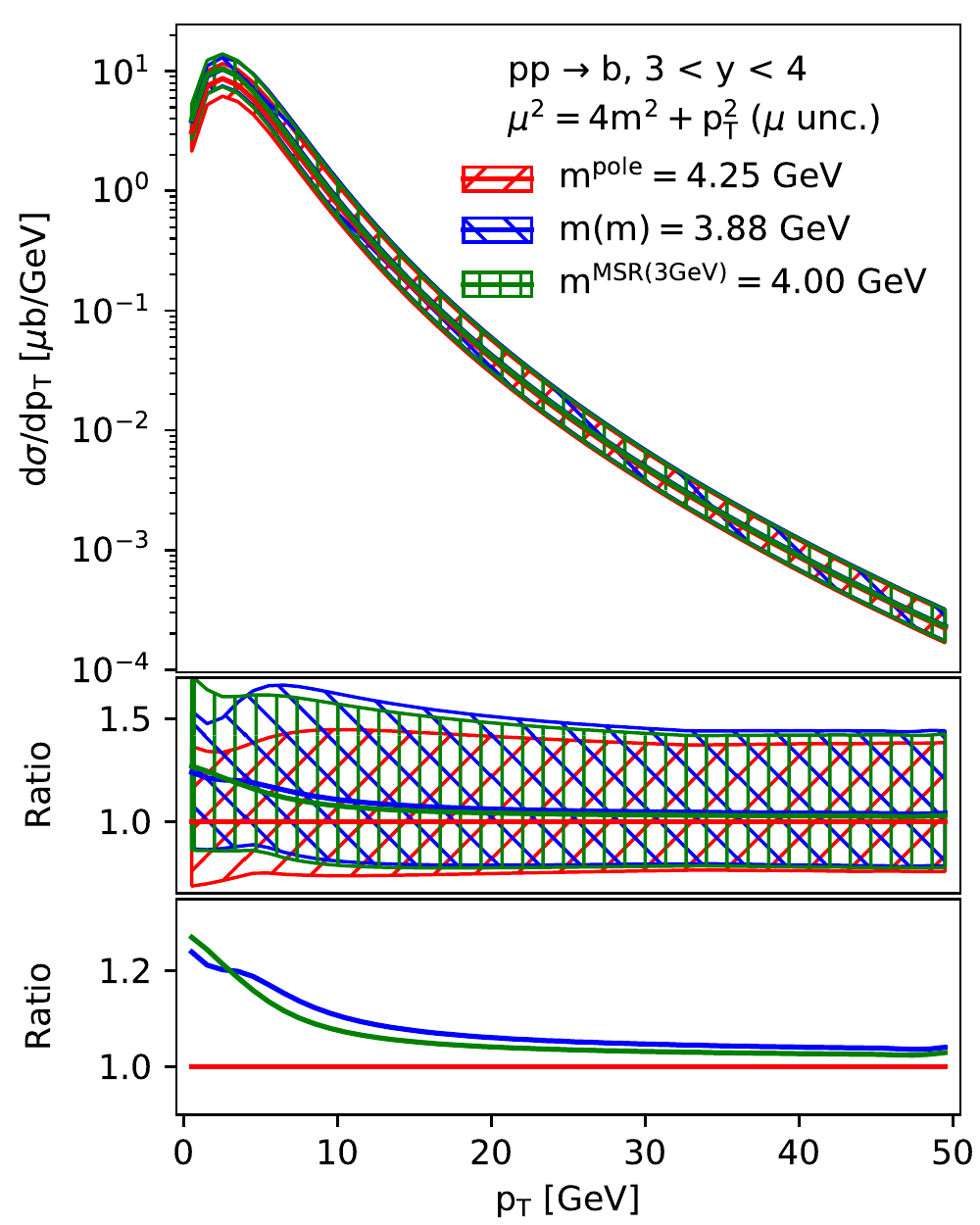}
  \caption{Same as Fig.~\ref{fig:b-pty-mu}, but for the bottom-mass value
    $\msbarm[b]$~=~3.88~GeV (converted to $\msrm[b]$(3~GeV)~=~4.00~GeV
    and $\polem[b]$~=~4.25~GeV), as extracted in the ABMP16 NLO fit. 
  }
  \label{fig:b-pty-mu-abm}
\end{figure}

\begin{figure}
  \centering
  \includegraphics[width=0.49\textwidth]{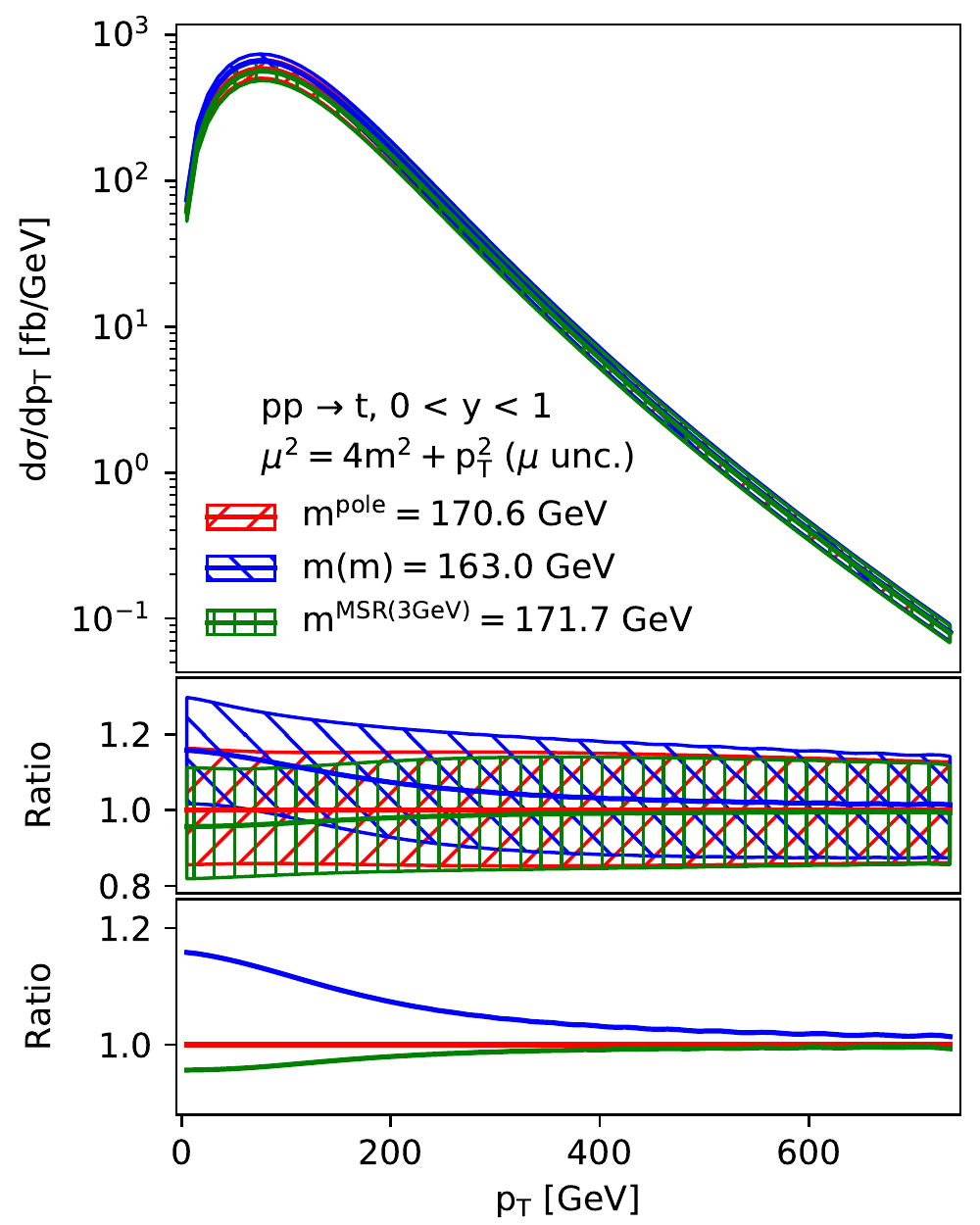}
  \includegraphics[width=0.49\textwidth]{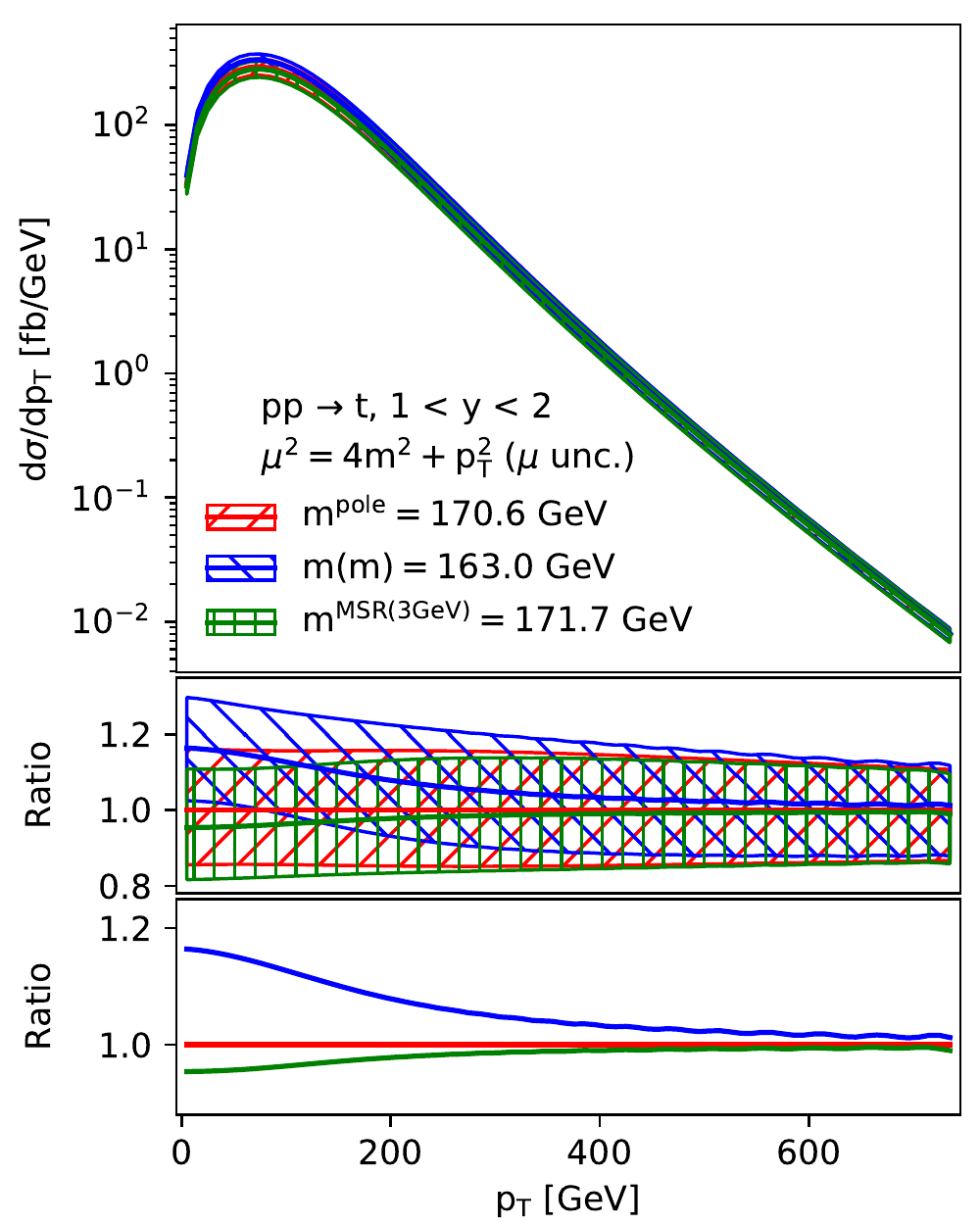}\\[10ex]
  \includegraphics[width=0.49\textwidth]{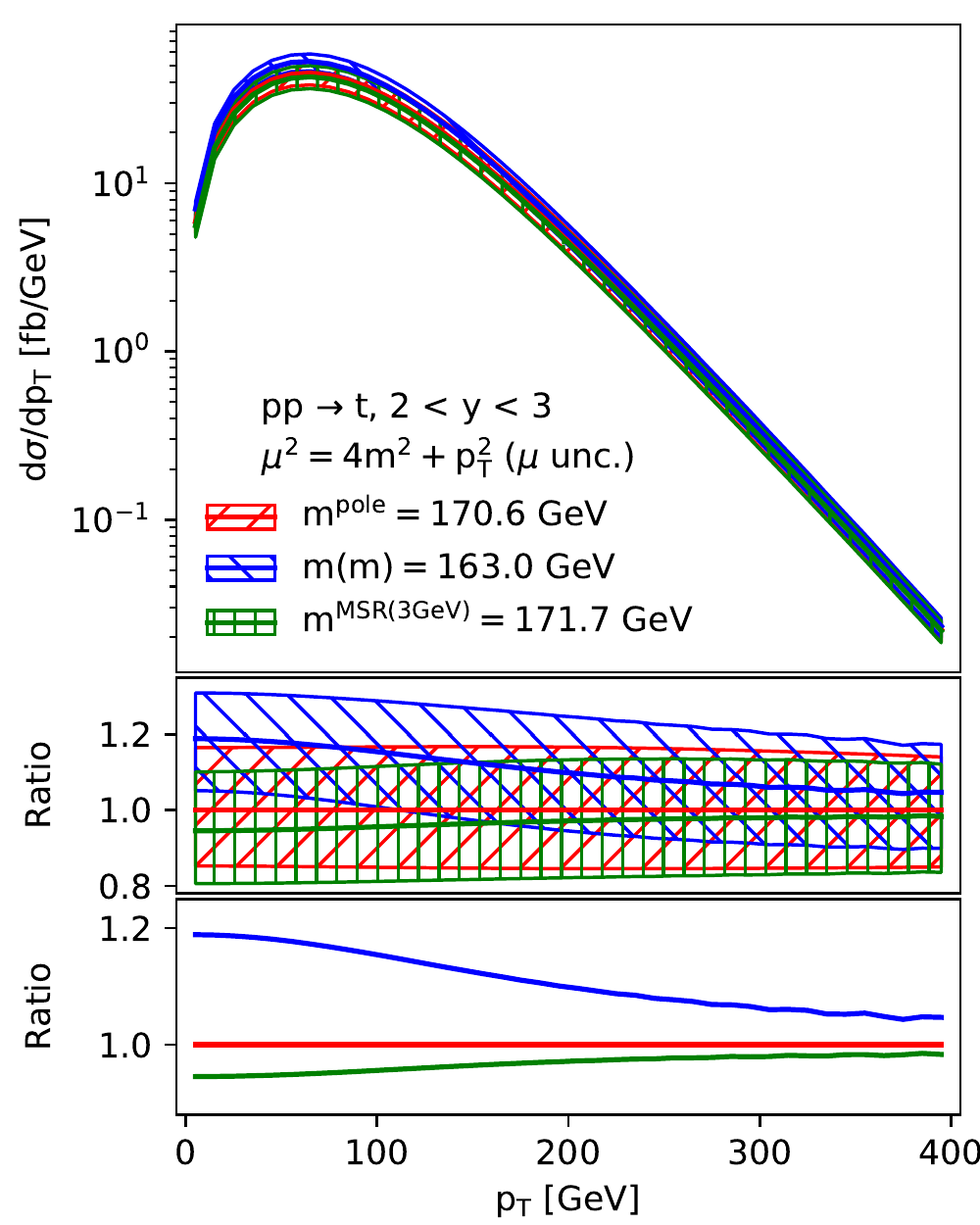}
  \caption{Same as Fig.~\ref{fig:c-pty-mu} for top production.
  }
  \label{fig:t-pty-mu}
\end{figure}

\begin{figure}
  \centering
  \includegraphics[width=0.49\textwidth]{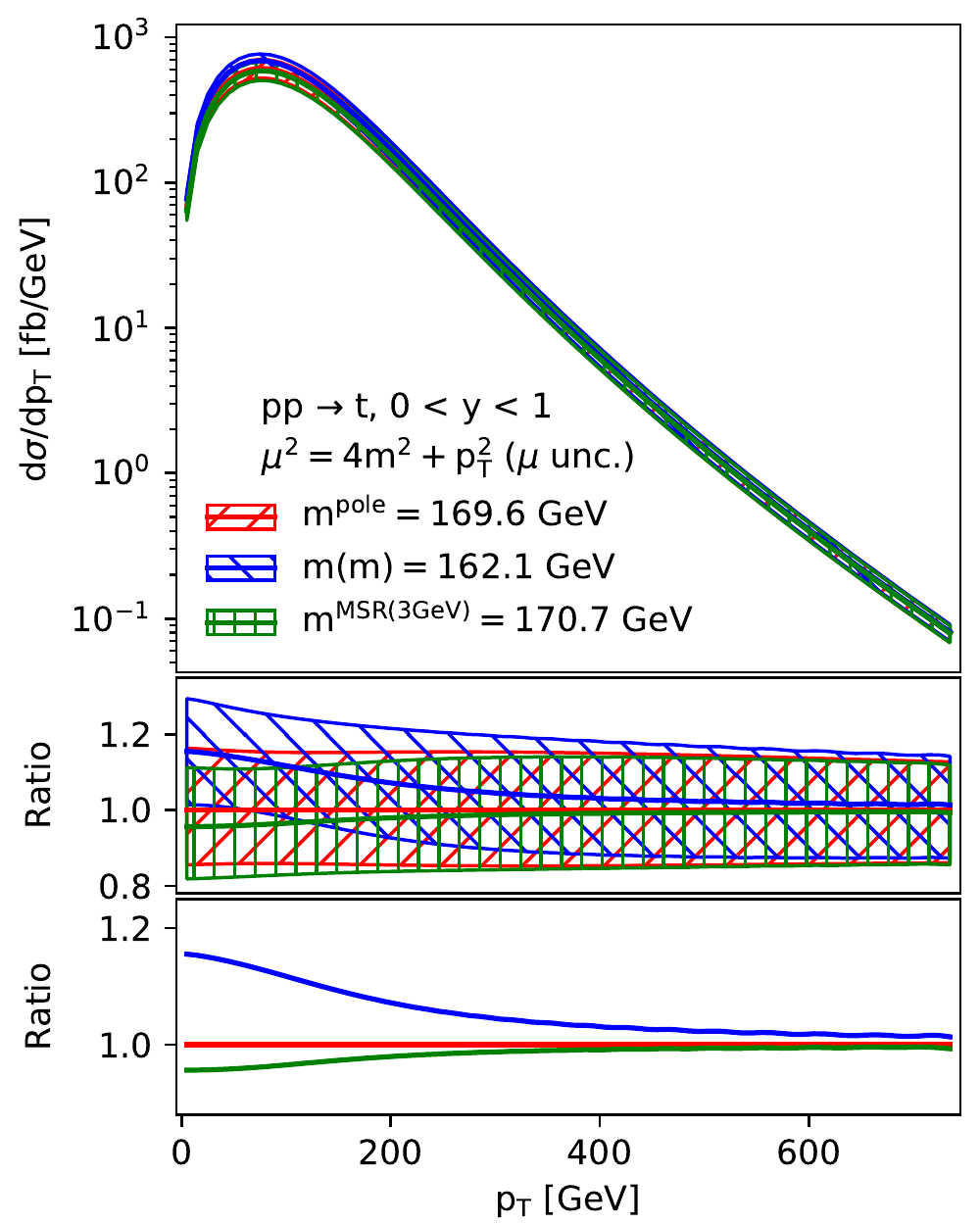}
  \includegraphics[width=0.49\textwidth]{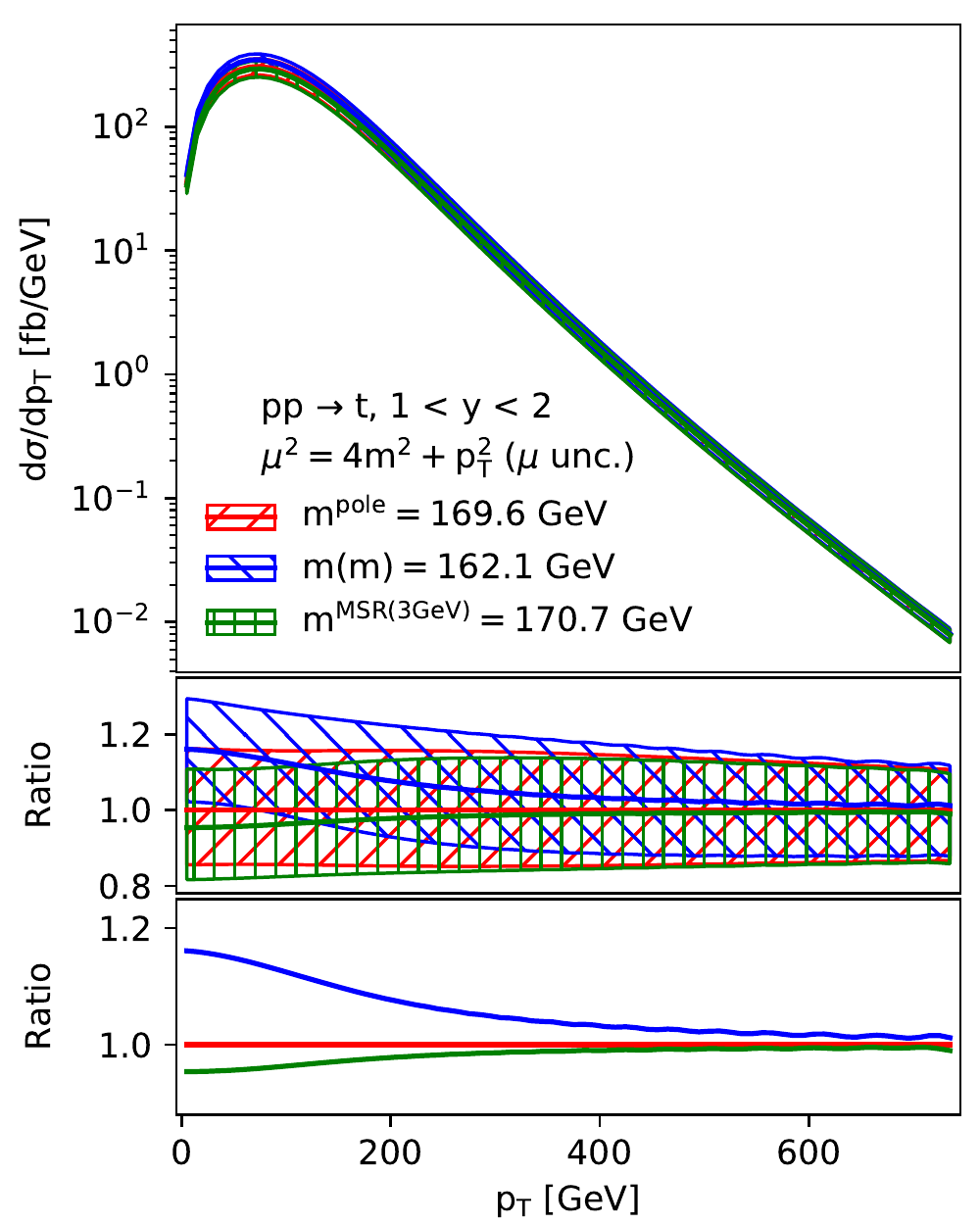}\\[8ex]
  \includegraphics[width=0.49\textwidth]{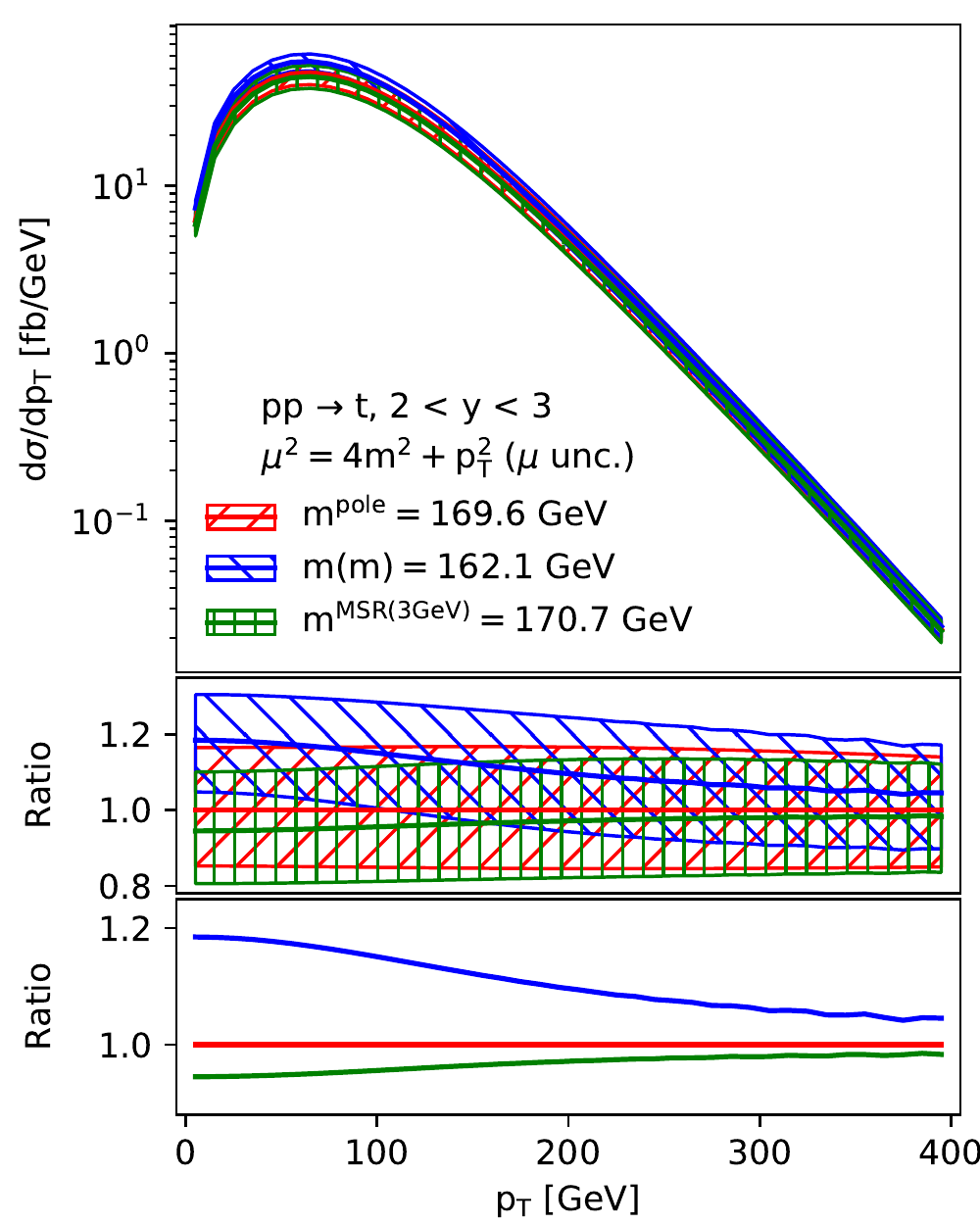}
  \caption{Same as Fig.~\ref{fig:t-pty-mu}, but for the top-mass value
    $\msbarm[t]$~=~162.1~GeV (converted to $\msrm[t]$(3~GeV)~=~170.7~GeV and
    $\polem[t]$~=~169.6~GeV), as extracted in the ABMP16 NLO fit. 
  }
  \label{fig:t-pty-mu-abm}
\end{figure}


In the calculations of the differential distributions presented in the
following, we use the PDG values~\cite{Tanabashi:2018oca}  
for the \msbar charm- and bottom-quark masses, 
$\msbarm[c]= 1.28$~GeV and $\msbarm[b] = 4.18$~GeV, 
and the \msbar top-quark mass value $\msbarm[t] = 163$~GeV~\cite{Aad:2019mkw}.
Alternatively, we use the \msbar charm-, bottom- and top-quark mass central values
extracted from the ABMP16 NLO simultaneous fit of PDFs, $\alpha_S(M_Z)$
and \msbar heavy-quark masses~\cite{Alekhin:2018pai}:
$\msbarm[c]= 1.18$~GeV, $\msbarm[b] = 3.88$~GeV and $\msbarm[t] = 162.1$~GeV. 
Although these values are smaller than those quoted by the PDG, they allow for fully self-consistent
computations when used in association with the ABMP16 $\alpha_S$ values and
PDFs~\footnote{
The values of \msbarm[c] and \msbarm[b] extracted in the ABMP16 fit 
are determined from HERA data on open heavy-flavor production in DIS, see ~\cite{Alekhin:2017kpj}. 
The low value of \msbarm[b] with its larger uncertainty in particular 
is a consequence of using those data. See also Sec.~\ref{nlo-charm-fits}
and eq.~(\ref{eq:dztocc1}).
}.
The pole and MSR mass values are obtained from the previous ones, using a procedure
analogous to that adopted for building Table~\ref{tab:masses}, except that the
$\alpha_S(M_Z)$ values are now those extracted in the ABMP16 NLO fit
($\alpha_s(M_Z)^{n_f=5} = 0.1191$, $\alpha_s(M_Z)^{n_f=3} = 0.1066$),
instead of those used in Table~\ref{tab:masses}, and $\alpha_S$ is evolved at two loops as in the fit. 
Specifically, for the MSR masses \msrm[b] and \msrm[t] for bottom- and top-quarks the scale $R = 3$~GeV is chosen, 
whereas for charm-quarks $R = 1$~GeV, 
in order to avoid using the too small value of $\msrm[c]$ at $R = 3$~GeV (see Fig.~\ref{fig:mcmbmtrun}, right panel).  
For the pole masses, the values from the \msbar mass conversion at one loop are chosen. 
The factorization and renormalization scales $\mu_R$ and $\mu_F$ are set to $\mu_0 = \sqrt{4m_Q^2+p_T^2}$ and 
the proton is described by the PDF set ABMP16 at NLO.
To estimate the theoretical uncertainties, the pair of factorization and
renormalization scales, $(\mu_R, \mu_F)$, are varied by a factor of two up and
down around the nominal value $\mu_0$, both simultaneously and independently, 
and excluding the combinations (0.5, 2)$\mu_0$ and (2, 0.5)$\mu_0$, 
following the conventional 7-point scale variation.
All calculations are provided for $pp$ collisions at the LHC at a center-of-mass
energy of $\sqrt{s} = 7$~TeV.

In Fig.~\ref{fig:c-pty-mu} the NLO differential cross-sections are shown
together with their scale uncertainties as a function of $p_T$ in different
intervals of the rapidity $y$ of the charm-quark, and with the charm-quark
mass renormalized in the pole, \msbar and MSR mass schemes. These
cross-sections are computed using {\texttt{xFitter}}.  
The changes of the cross-sections are in the range of a few percent to $\sim 40\%$, 
when using the \msbar or MSR mass scheme instead of the pole mass scheme, and
they are more evident in the bulk of the phase space.
However, this is a small effect compared to the size of scale uncertainties at NLO.
The latter amount to a factor of $\sim$ $2$ in the bulk of the phase space, 
decreasing slightly towards large $p_T$ values. 
It turns out that the scale uncertainties are very similar in all mass schemes 
for variations around the chosen nominal scale $\mu_R = \mu_F = \sqrt{4m_c^2+p_T^2}$.
Modifying the value of the charm-quark \msbar mass, which is set to the PDG
value in Fig.~\ref{fig:c-pty-mu}, to the value extracted in the ABMP16 NLO
fit, produce results qualitatively similar, shown in Fig.~\ref{fig:c-pty-mu-abm}. 
Differences between predictions in different mass renormalization schemes in
Fig.~\ref{fig:c-pty-mu-abm} are smaller than in Fig.~\ref{fig:c-pty-mu}, due
to the fact that the ABMP16 \msbar masses are smaller than the PDG ones. 

In Fig.~\ref{fig:b-pty-mu} the same comparison of NLO differential 
cross-sections in the various mass renormalization schemes is presented for bottom-quark
production. 
In this case, the impact of converting the pole mass calculations
into the \msbar or MSR schemes vary from
a few percents to $25\%$, which is still small
compared to the NLO scale uncertainties of the order of $50\%$. 
With the choice for the nominal scale $\mu_R = \mu_F = \sqrt{4m_b^2+p_T^2}$, 
the scale uncertainties are similar in the pole and MSR mass
schemes, whereas they are more asymmetric and slightly smaller at low $p_T$ in the \msbar mass scheme.
Again, modifying the value of the bottom-quark \msbar mass, which is set to
the PDG value in Fig.~\ref{fig:b-pty-mu}, to the value extracted in the ABMP16
NLO fit, leads to results qualitatively similar, shown in
Fig.~\ref{fig:b-pty-mu-abm}, with slightly smaller differences (up to $\sim
20\%$ among predictions in different mass renormalization schemes.

Finally, Fig.~\ref{fig:t-pty-mu} displays the same comparison for top-quark production. 
In this case, the impact of converting the pole mass calculations 
into the \msbar mass scheme is about $20\%$ at low $p_T$, 
which is no longer small compared to the NLO scale uncertainties. 
It decreases towards higher $p_T$ values. 
When converting the cross-sections from the pole to the MSR mass scheme, 
the impact is below $10\%$ and is within the NLO scale uncertainties for
variations around the nominal scale $\mu_R = \mu_F = \sqrt{4m_t^2+p_T^2}$. 
The scale uncertainties in the \msbar mass scheme are slightly
smaller than in the pole mass scheme, as was already reported
previously~\cite{Dowling:2013baa}, while the scale uncertainties in the MSR
and pole mass schemes are very similar. 
Again, modifying the value of the top-quark \msbar mass, which is set to the
PDG value in Fig.~\ref{fig:t-pty-mu}, to the value extracted in the ABMP16
fit, leads to predictions qualitatively similar, shown in
Fig.~\ref{fig:t-pty-mu-abm}. 

In general, the differences between predictions in different mass renormalization schemes
slightly increase with the rapidity, as can be seen in all 
Figs.~\ref{fig:c-pty-mu}--\ref{fig:t-pty-mu-abm}.


\begin{figure}
  \centering
  \includegraphics[width=0.49\textwidth]{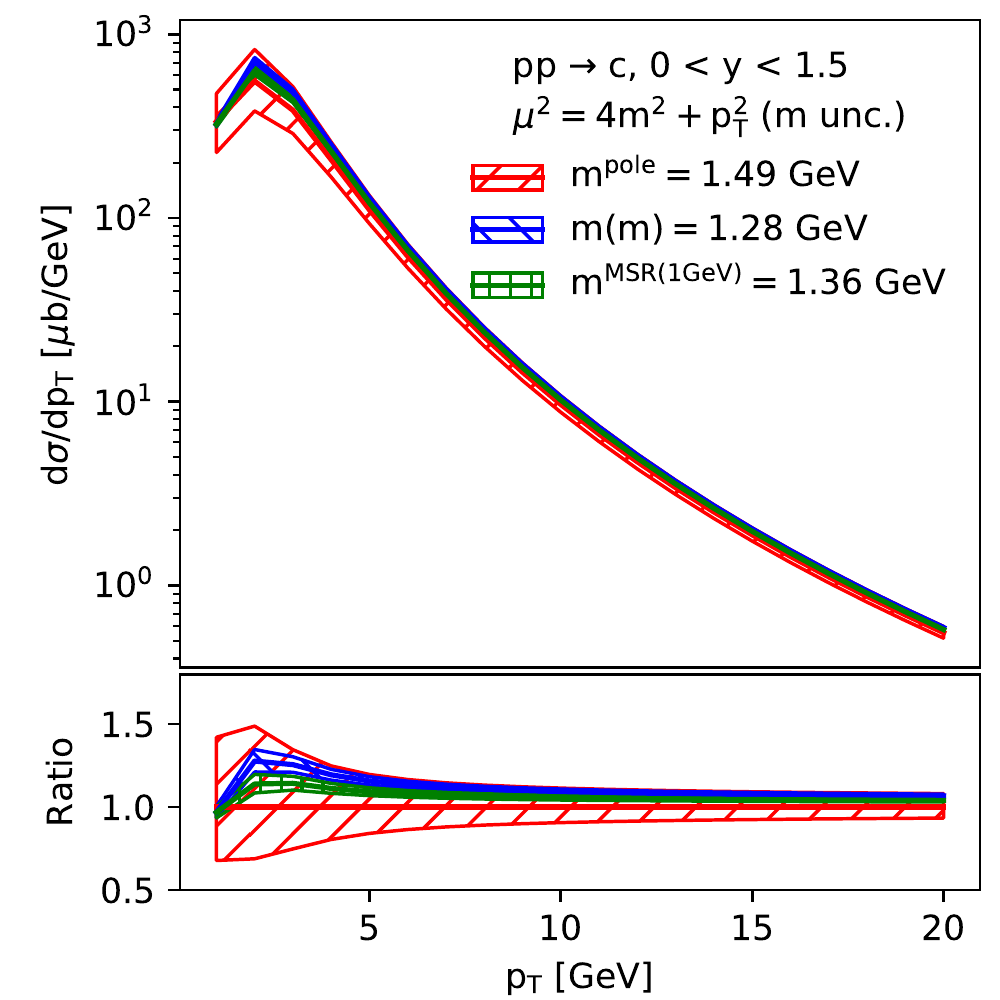}
  \includegraphics[width=0.49\textwidth]{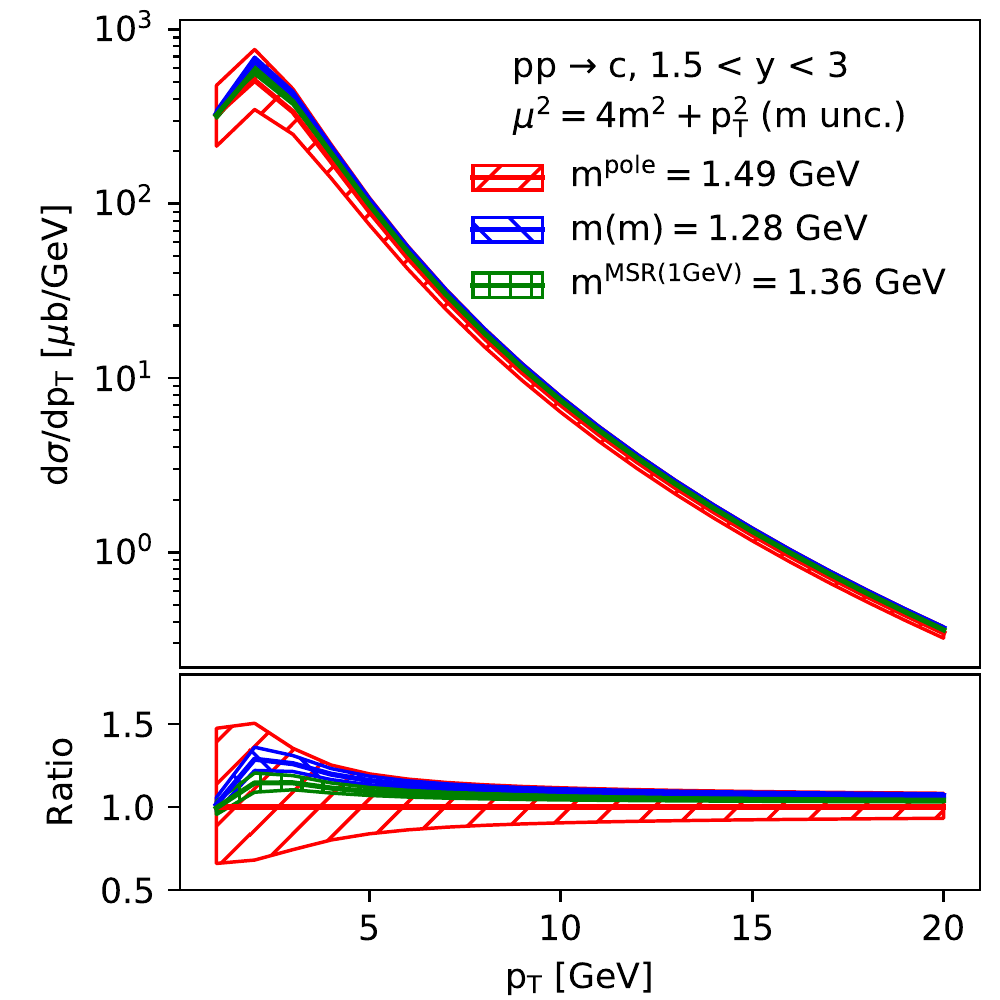}\\[3ex]
  \includegraphics[width=0.49\textwidth]{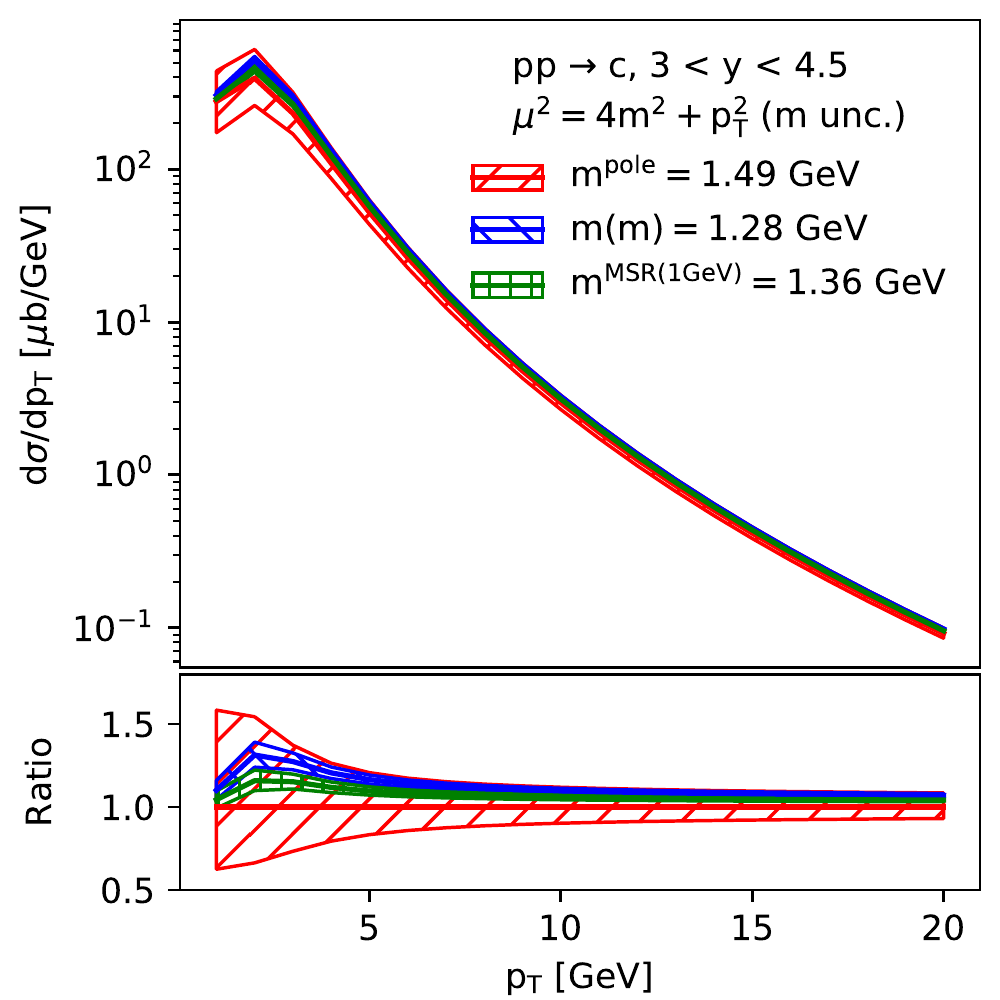}
  \includegraphics[width=0.49\textwidth]{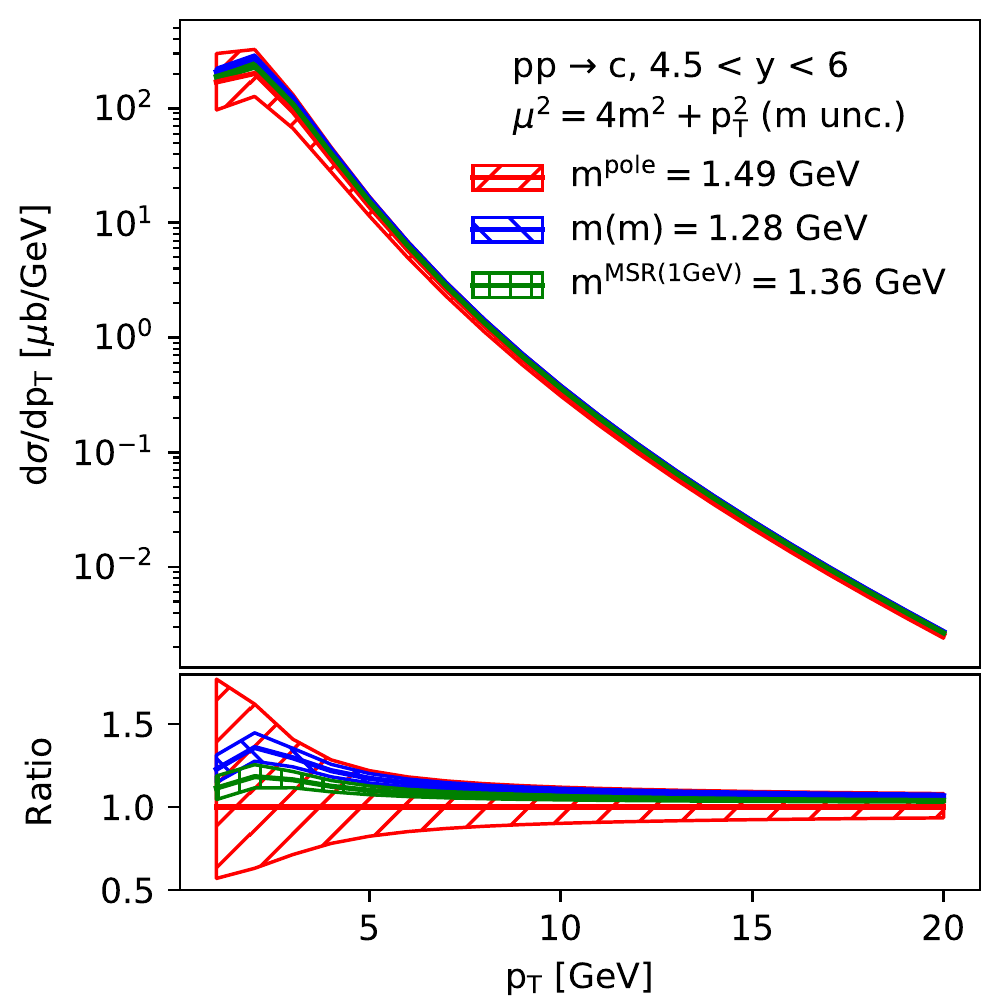}
  \caption{The NLO differential cross-sections for charm production 
    at the LHC ($\sqrt{s} = 7$~TeV) as a function of $p_T$ 
    in intervals of $y$ of the charm-quark in the pole and \msbar mass schemes.
    The bands denote variations of the mass values in the different schemes, 
    $\polem[c] = 1.49 \pm 0.25$~GeV, $\msbarm[c] = 1.28 \pm 0.03$~GeV and 
    $\msrm[c] = 1.36 \pm 0.03$~GeV.
    The lower panels display the theoretical
    predictions normalized to the central values obtained in the pole mass
    scheme.
  } 
  \label{fig:c-pty-mass}
\end{figure}

\begin{figure}
  \centering
  \includegraphics[width=0.49\textwidth]{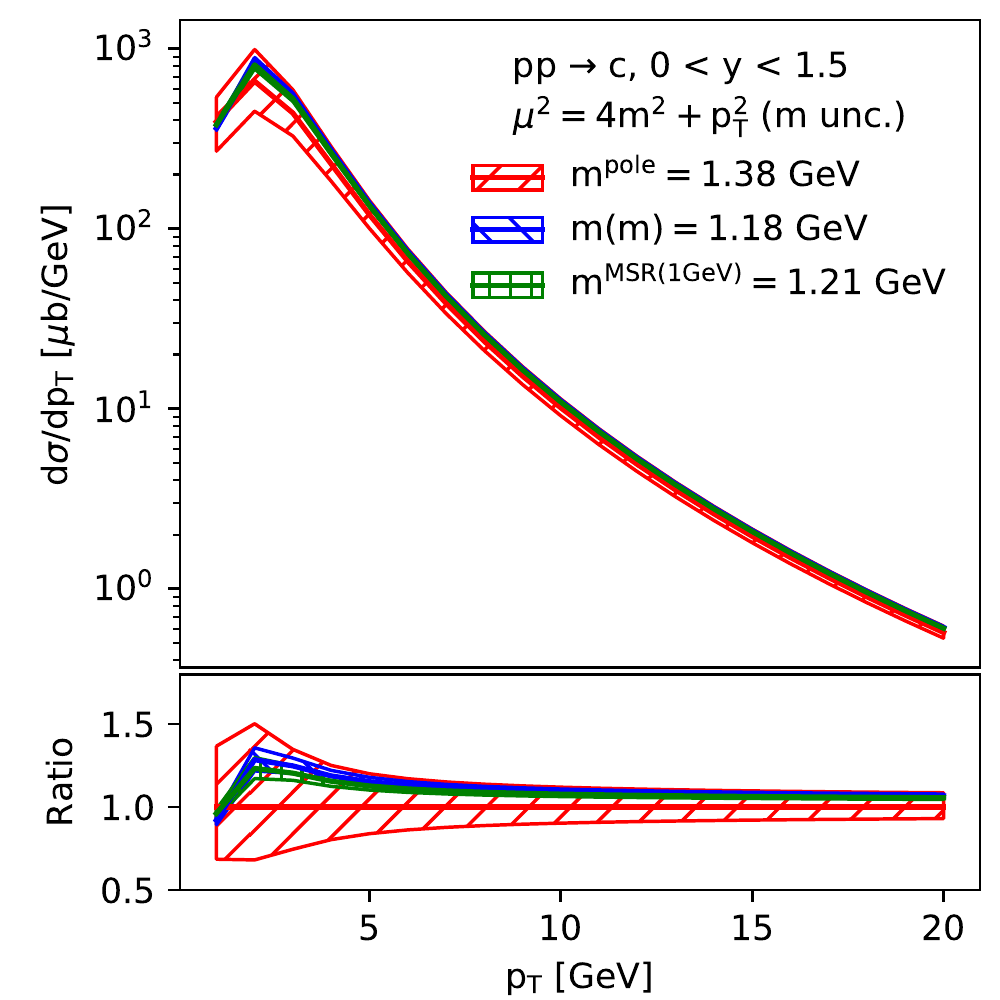}
  \includegraphics[width=0.49\textwidth]{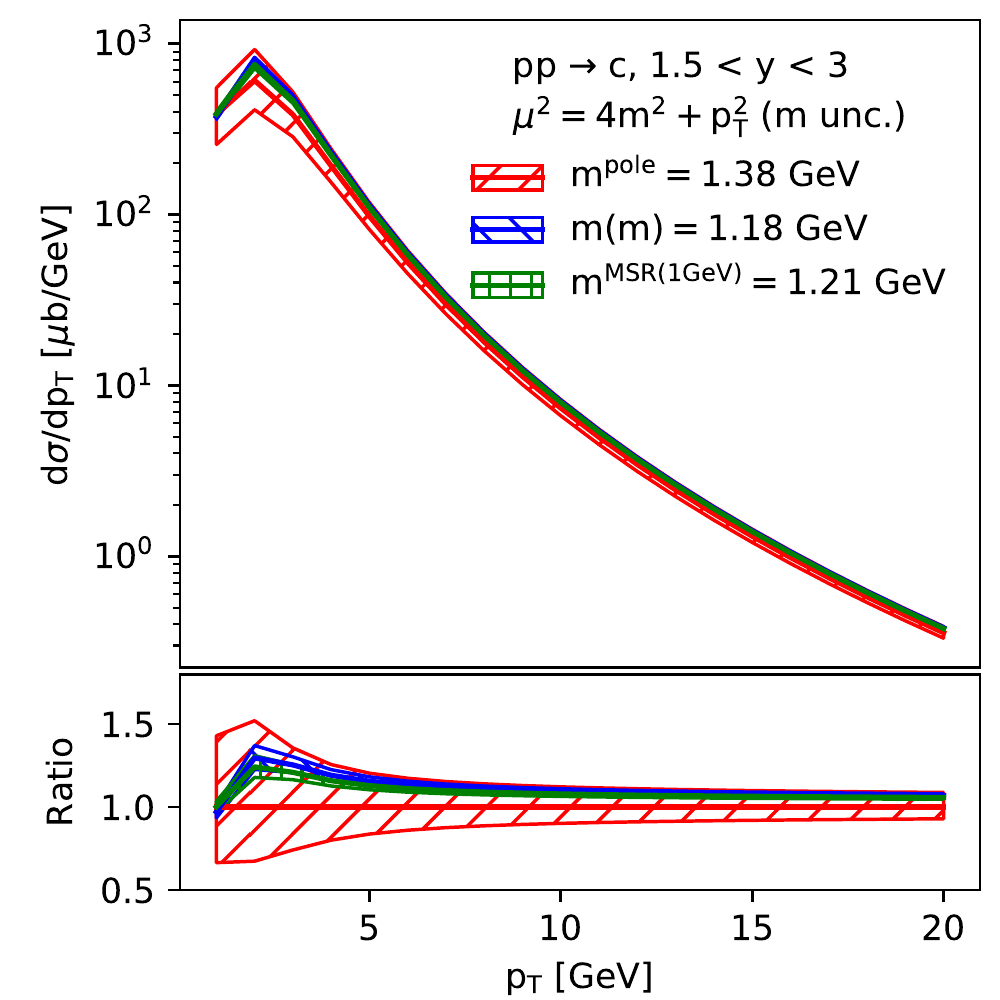}\\[3ex]
  \includegraphics[width=0.49\textwidth]{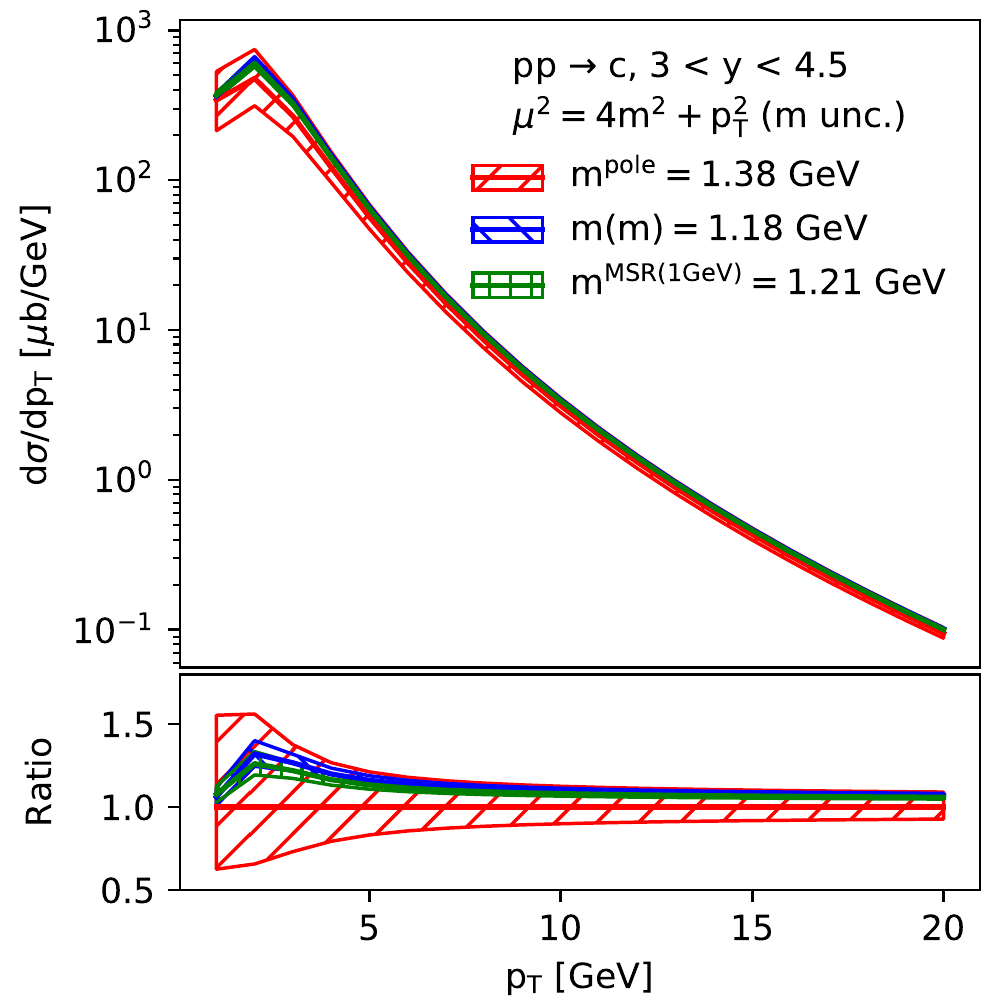}
  \includegraphics[width=0.49\textwidth]{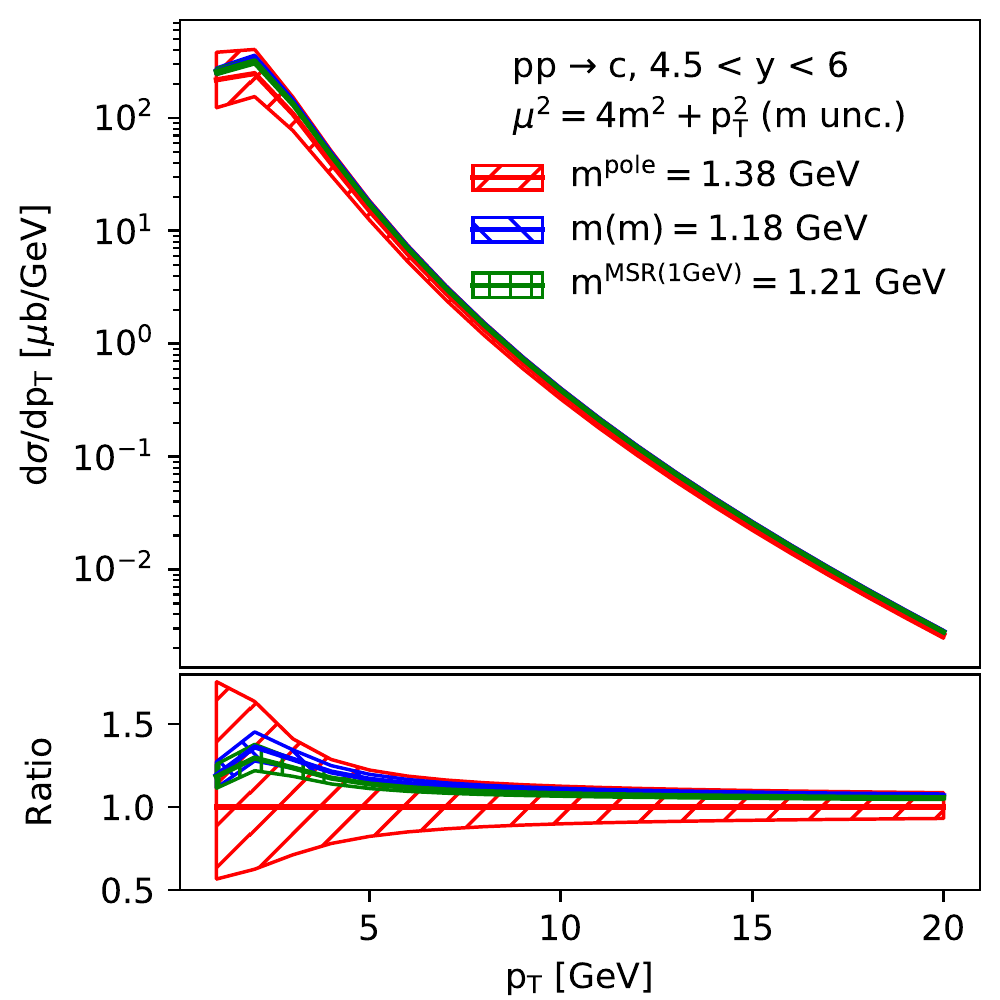}
  \caption{Same as Fig.~\ref{fig:c-pty-mass}, but for the charm-mass value
    $\msbarm[c]$~=~1.18~$\pm$~0.03~GeV (converted to
    $\msrm[c]$(1~GeV)~=~1.21~$\pm$~0.03~GeV and
    $\polem[c]$~=~1.38~$\pm$~0.25~GeV), as extracted in the ABMP16 NLO fit. 
  }
  \label{fig:c-pty-mass-abm}
\end{figure}

\begin{figure}
  \centering
  \includegraphics[width=0.49\textwidth]{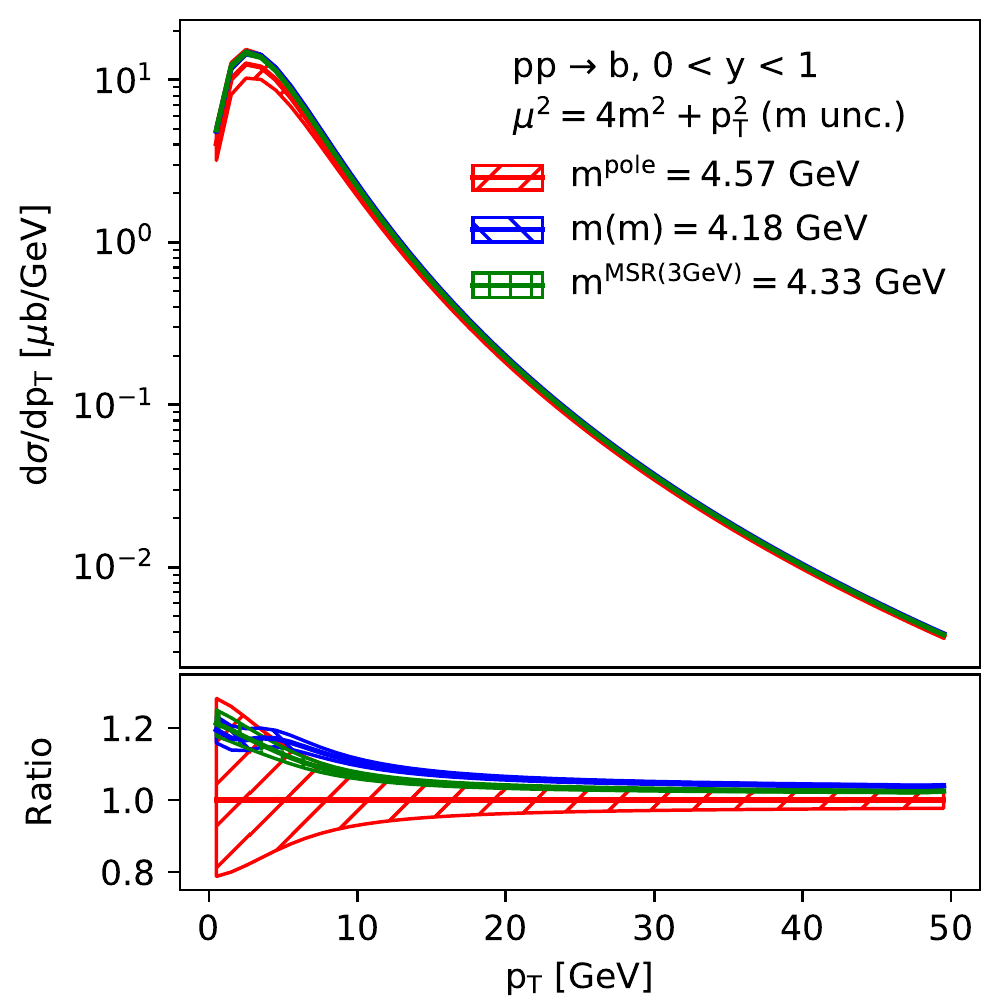}
  \includegraphics[width=0.49\textwidth]{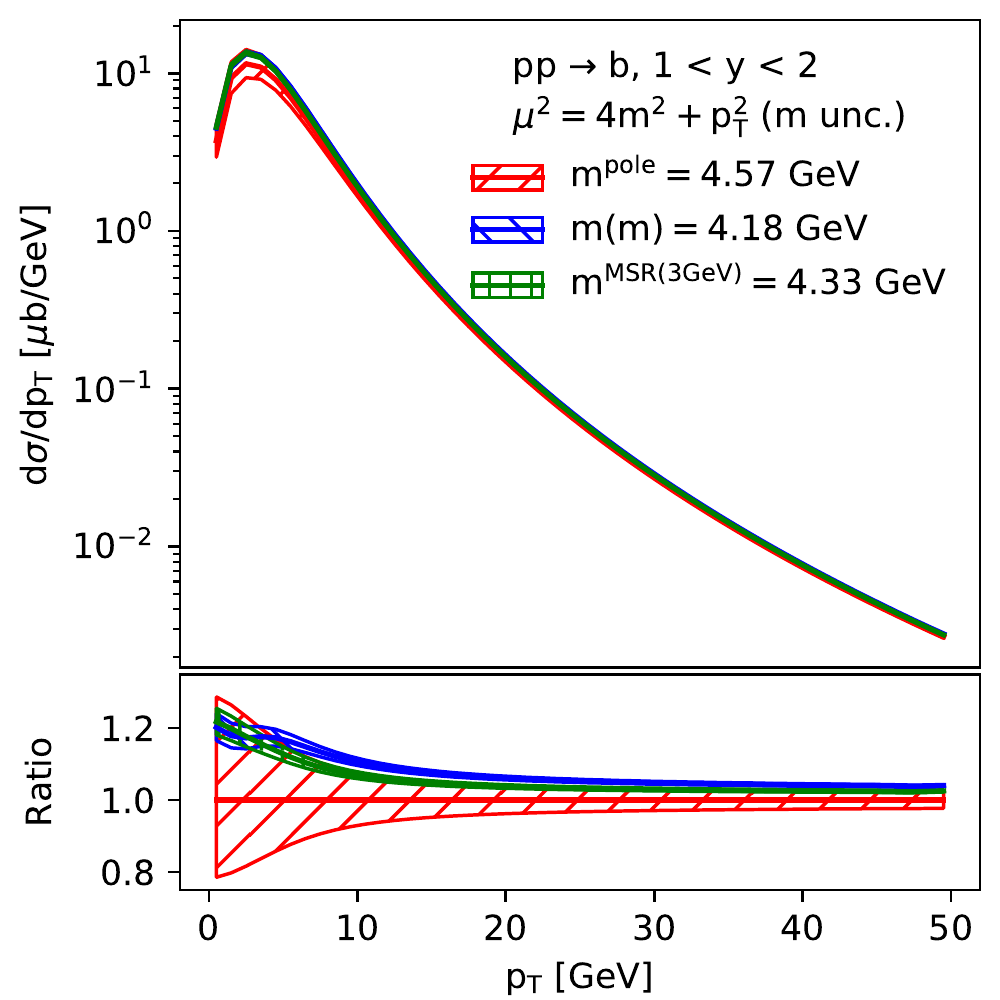}\\[3ex]
  \includegraphics[width=0.49\textwidth]{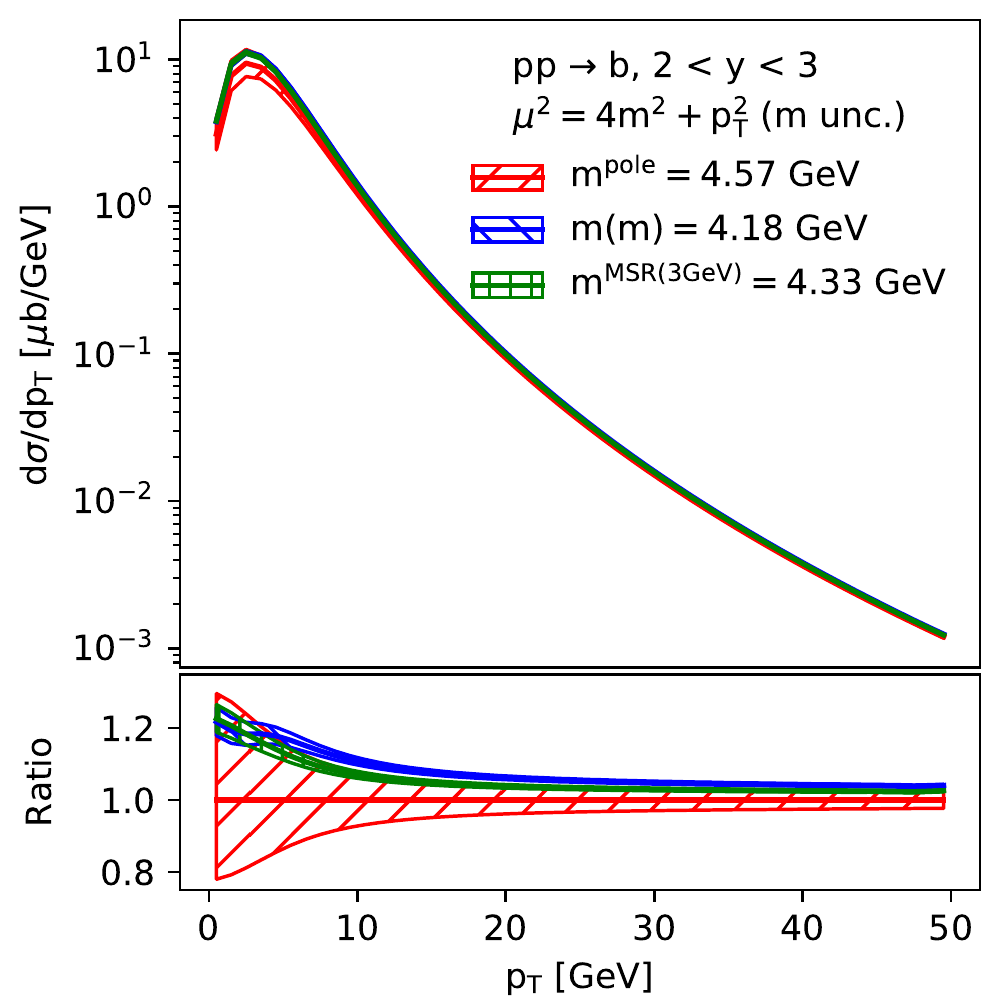}
  \includegraphics[width=0.49\textwidth]{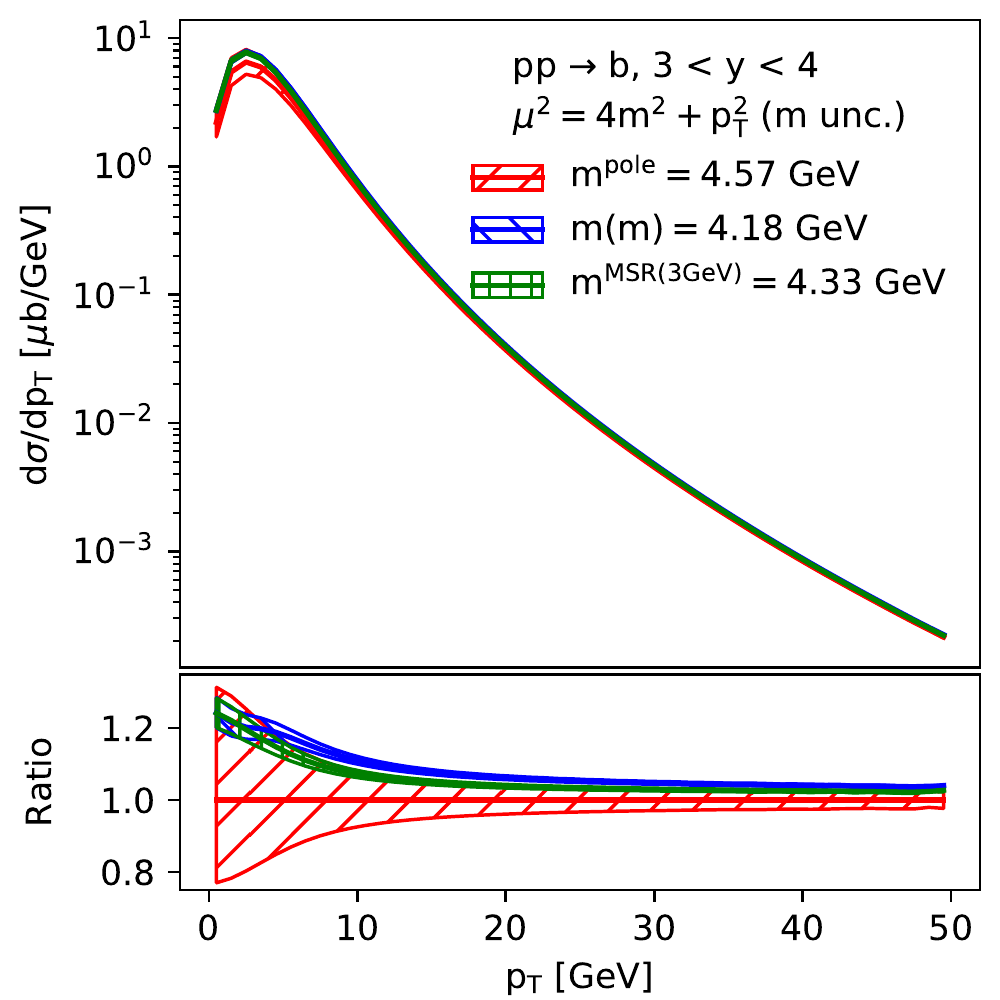}
  \caption{Same as Fig.~\ref{fig:c-pty-mass} for bottom production with
    variations of the mass values in the different schemes as  
    $\polem[b] = 4.57 \pm 0.25$~GeV, 
    $\msbarm[b] = 4.18 \pm 0.03$~GeV
    and $\msrm[b] = 4.33 \pm 0.03$~GeV.
  } 
  \label{fig:b-pty-mass}
\end{figure}

\begin{figure}
  \centering
  \includegraphics[width=0.49\textwidth]{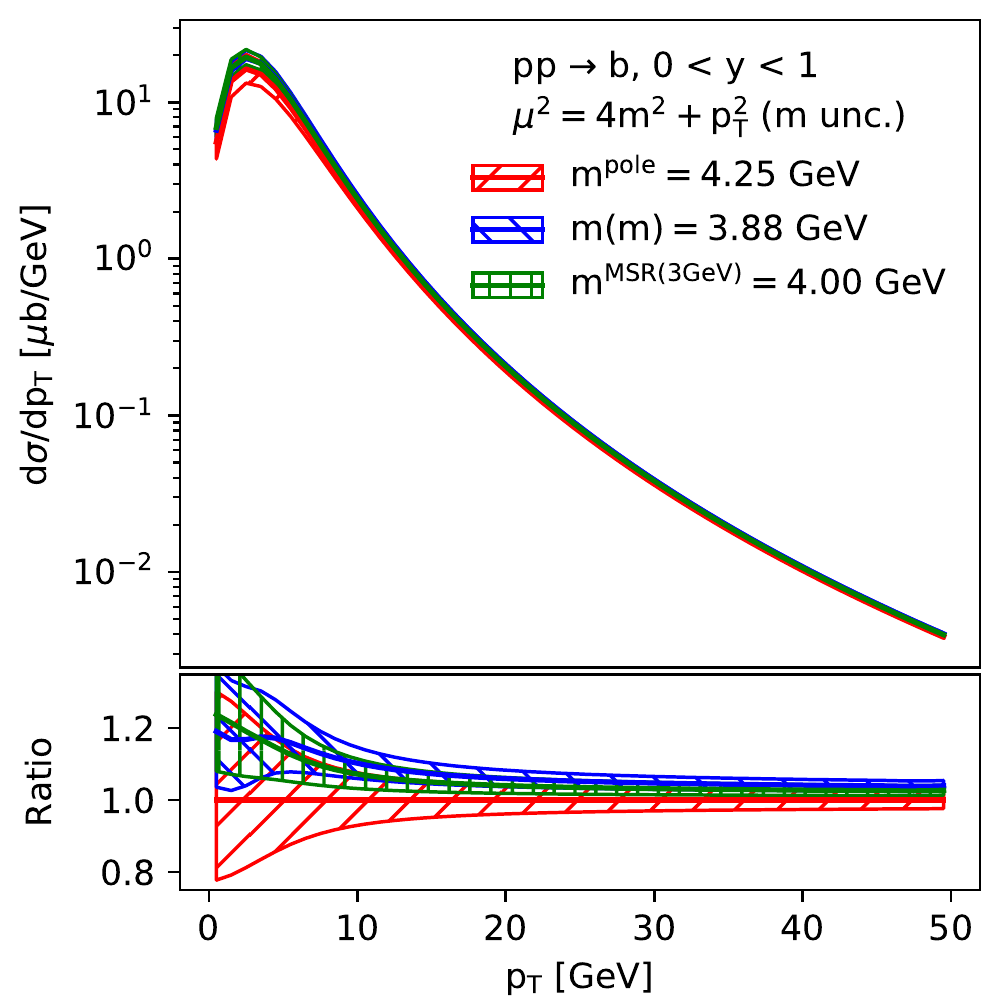}
  \includegraphics[width=0.49\textwidth]{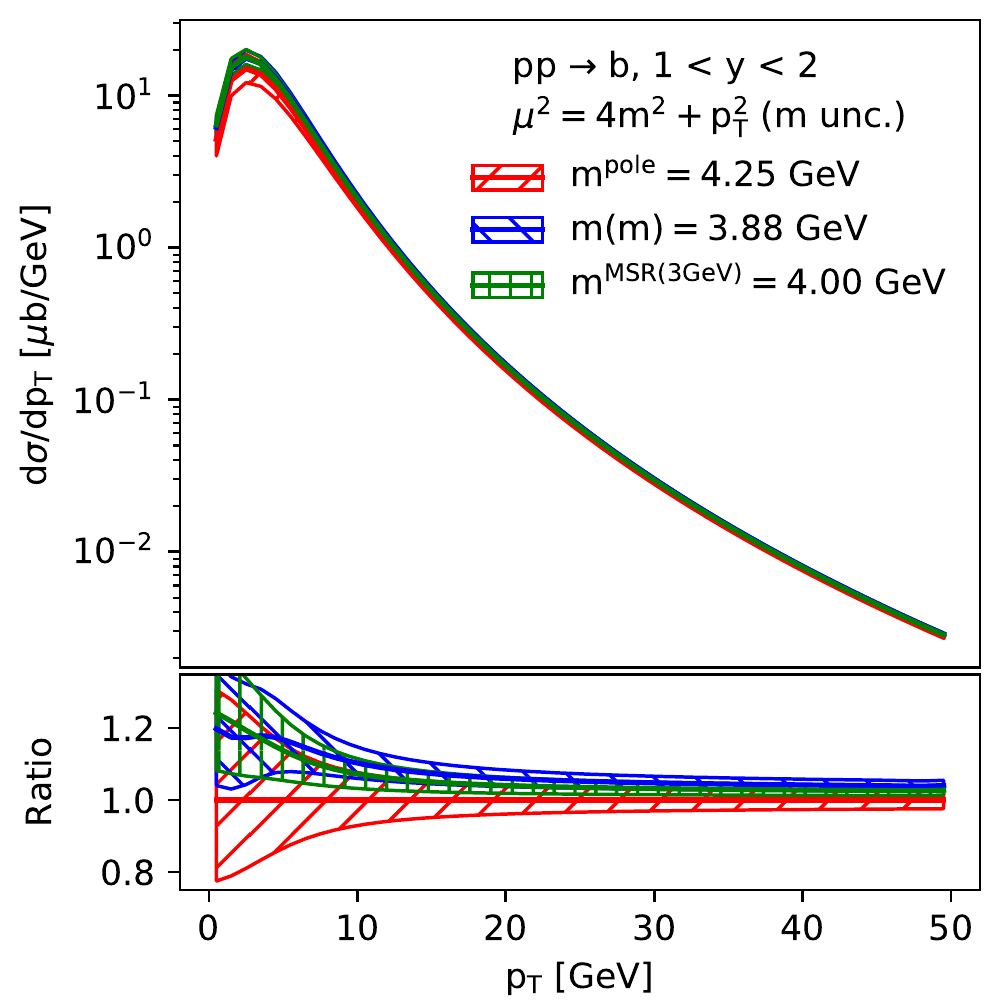}\\[3ex]
  \includegraphics[width=0.49\textwidth]{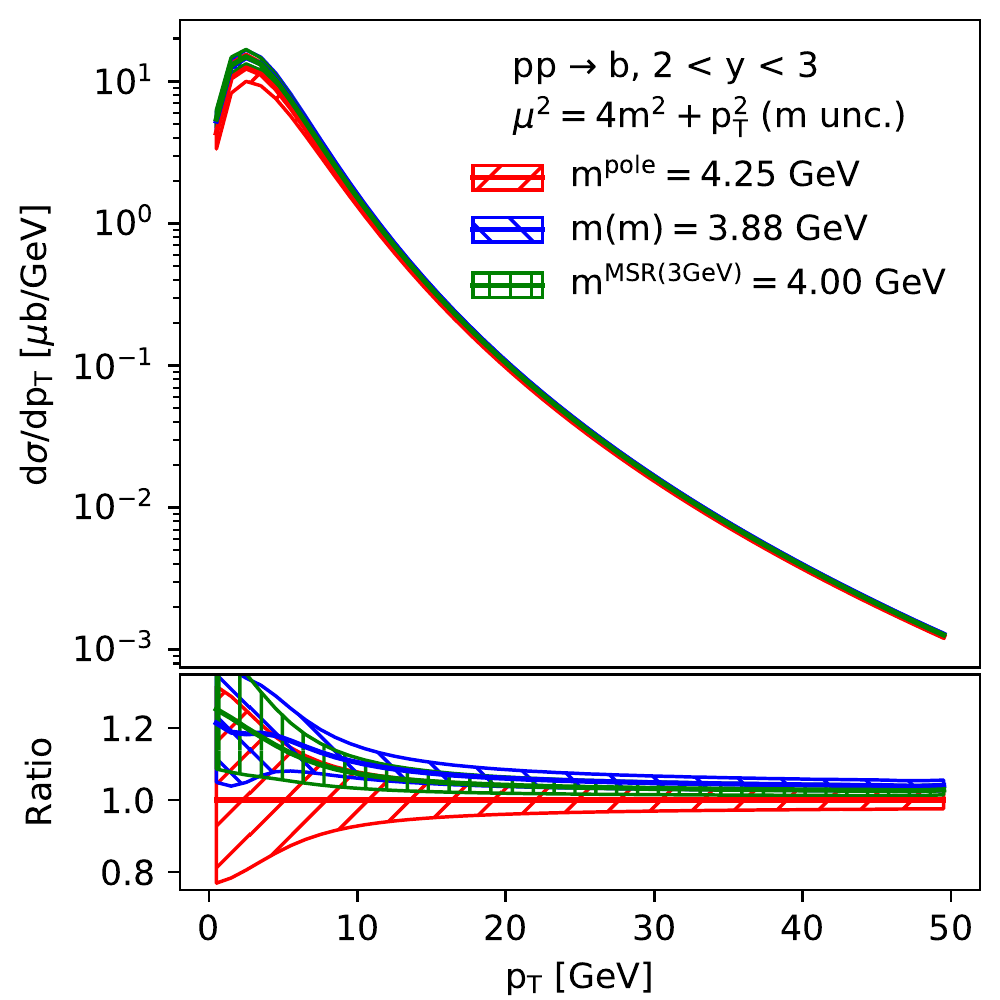}
  \includegraphics[width=0.49\textwidth]{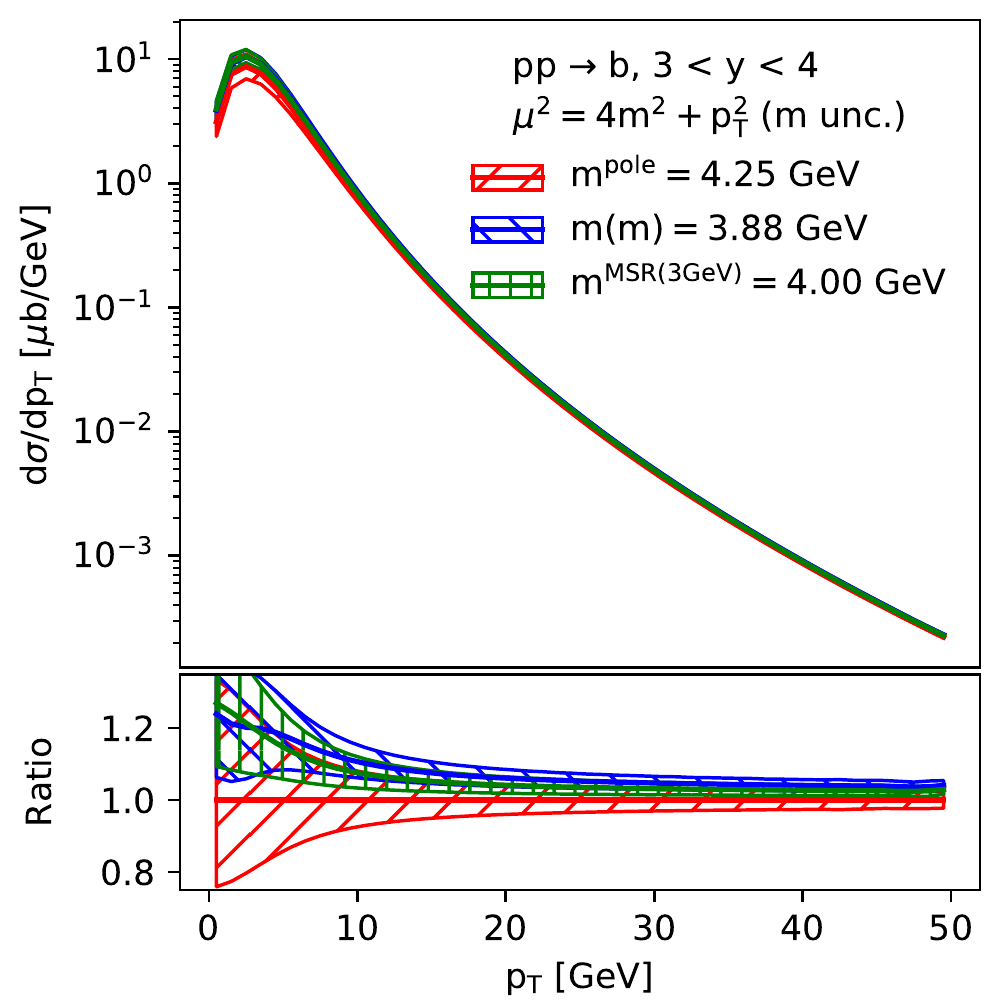}
  \caption{Same as Fig.~\ref{fig:b-pty-mass}, but for the bottom-mass value
    $\msbarm[b]$~=~3.88~$\pm$~0.13~GeV (converted to
    $\msrm[b]$(3~GeV)~=~4.00~$\pm$~0.13~GeV and
    $\msrm[b]$~=~4.25~$\pm$~0.25~GeV), as extracted in the ABMP16 NLO fit. 
   The size of the uncertainties of the predictions with the heavy-quark mass
   renormalized in the \msbar and MSR schemes are larger than in Fig.~\ref{fig:b-pty-mass} 
   because the uncertainties of the ABMP \msbar fitted masses are larger
   than the uncertainties of the \msbar masses reported by the PDG~\cite{Tanabashi:2018oca}. 
  }
  \label{fig:b-pty-mass-abm}
\end{figure}

\begin{figure}
  \centering
  \includegraphics[width=0.47\textwidth]{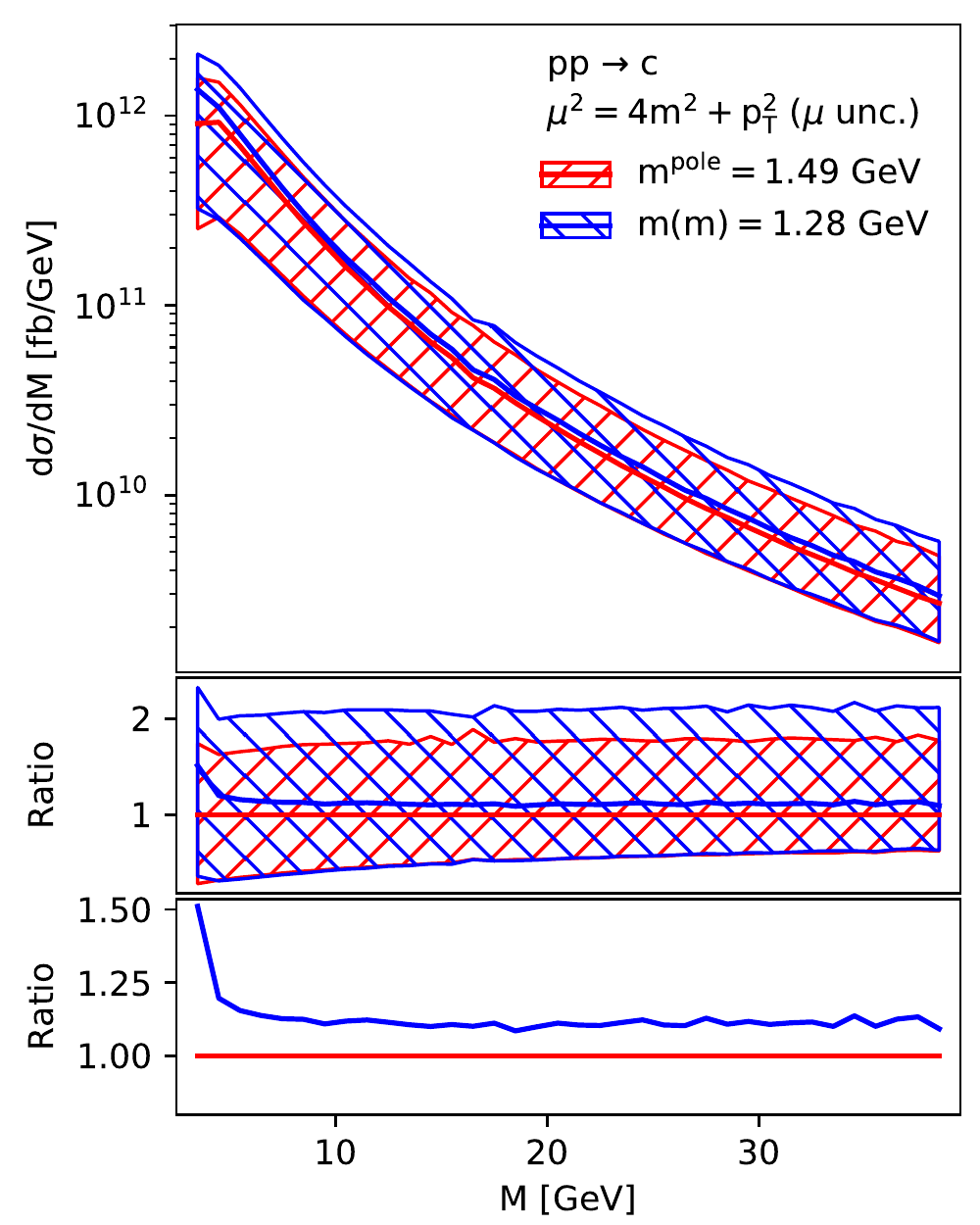}
  \includegraphics[width=0.47\textwidth]{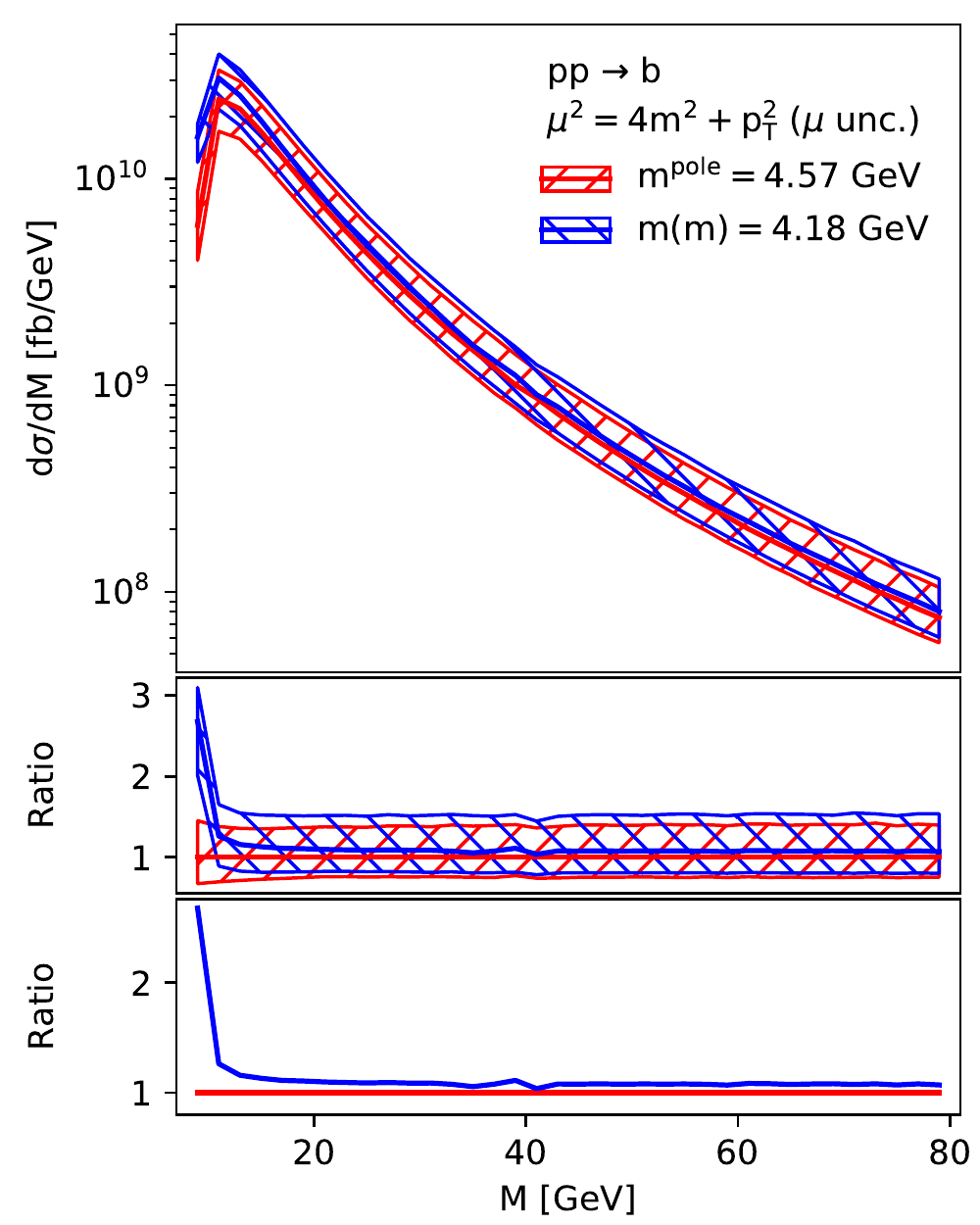}\\[5ex]
  \includegraphics[width=0.47\textwidth]{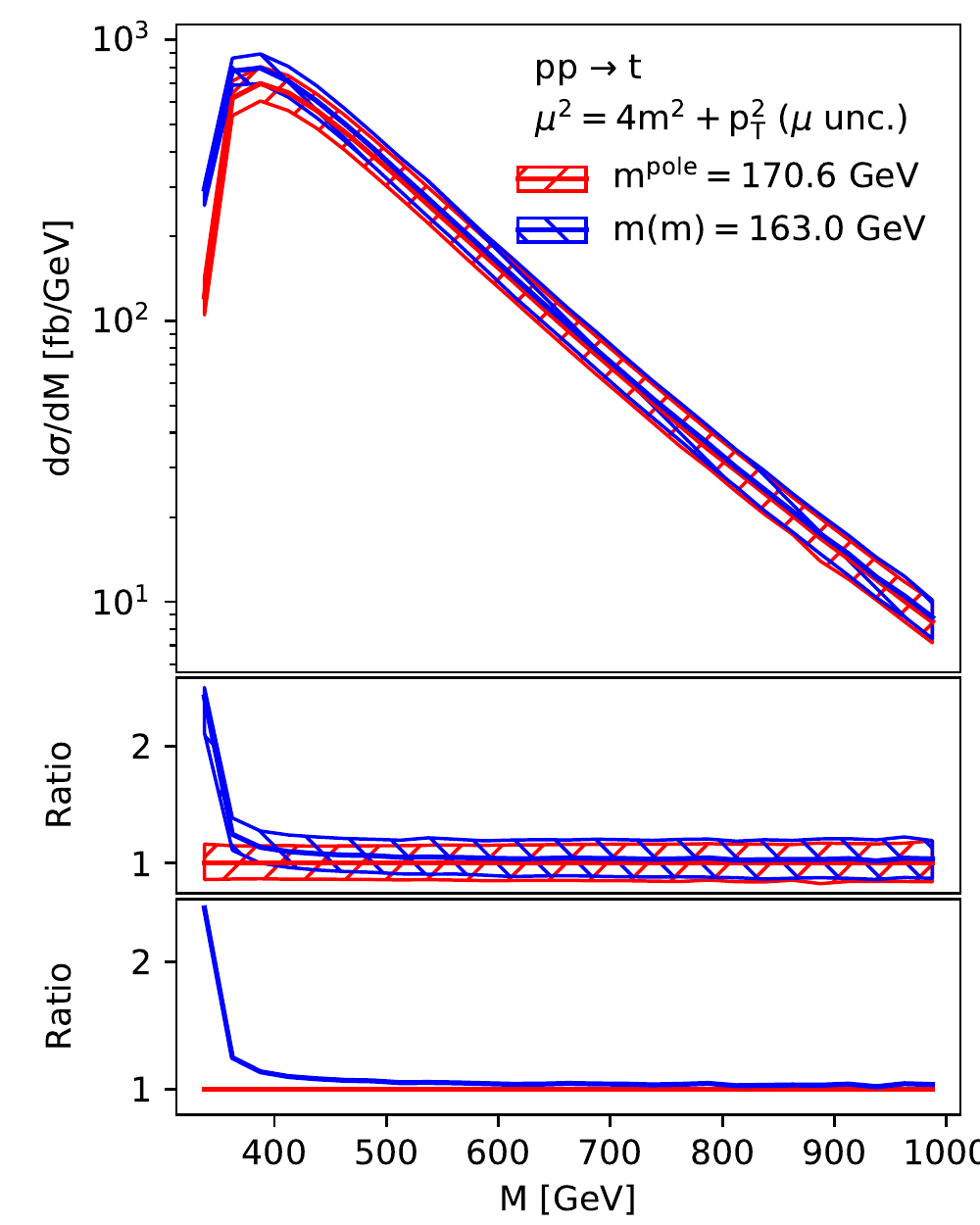}
  \caption{The NLO differential cross-sections at the LHC ($\sqrt{s} = 7$~TeV)
    for charm (upper left), bottom (upper right) and top (lower) hadro-production 
    with their scale uncertainties as a function 
    of the invariant mass $M_{Q{\bar Q}}$ of the heavy-quark pair
    in the pole and \msbar mass schemes. 
    The lower panels display the theoretical predictions normalized to the
    central values obtained in the pole mass scheme.} 
  \label{fig:cbt-m-mu}
\end{figure}

\begin{figure}
  \centering
  \includegraphics[width=0.49\textwidth]{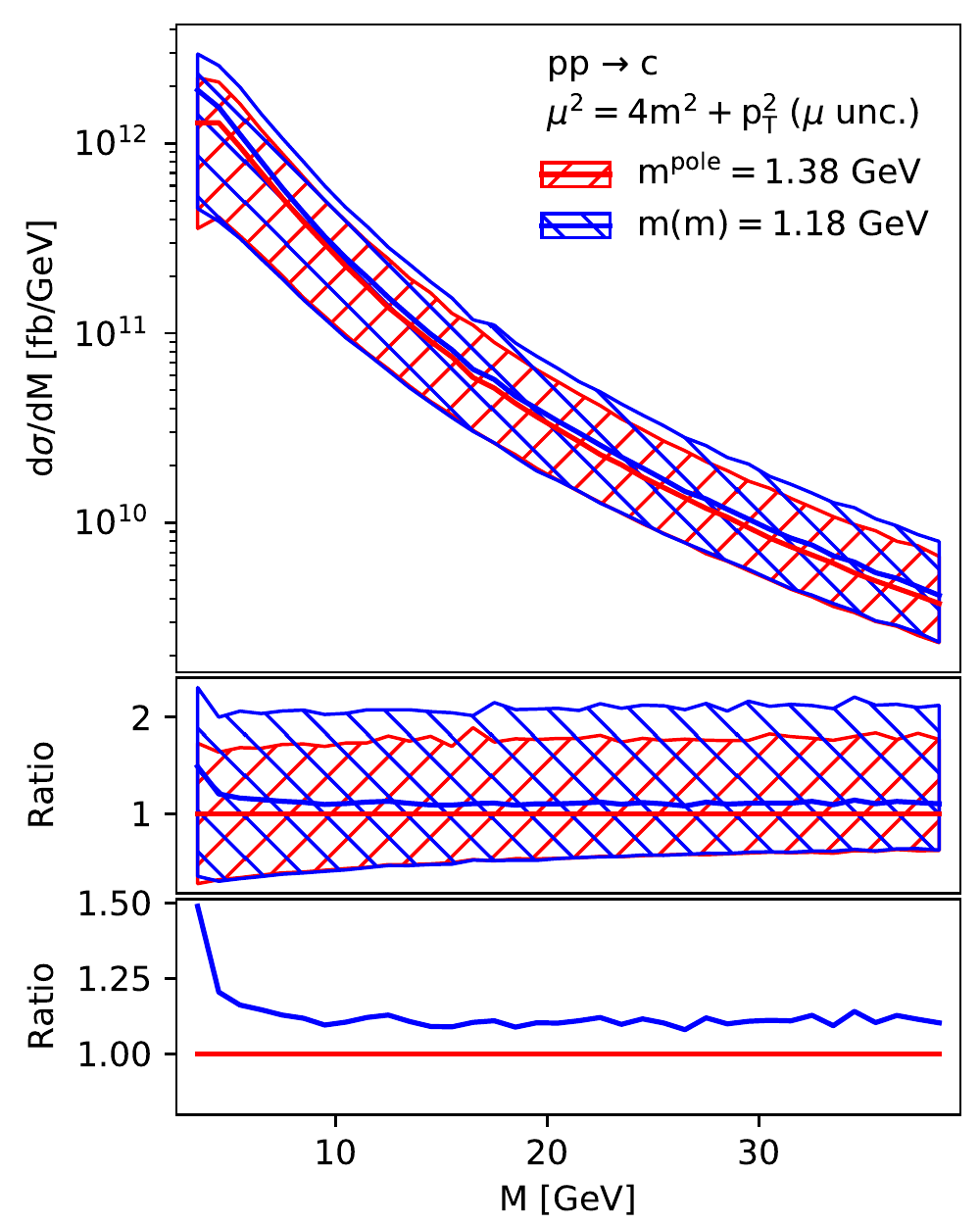}
  \includegraphics[width=0.49\textwidth]{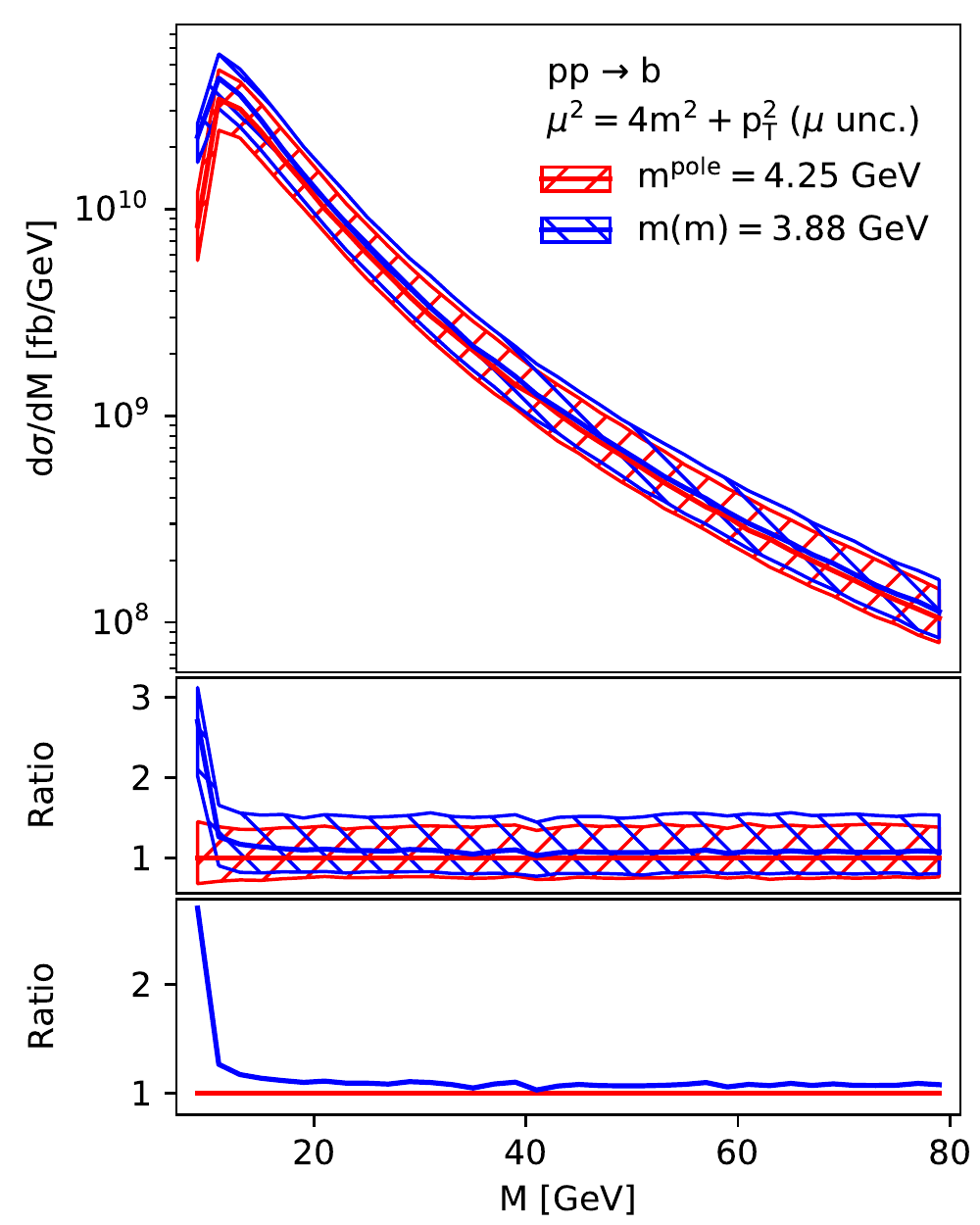}\\[5ex]
  \includegraphics[width=0.49\textwidth]{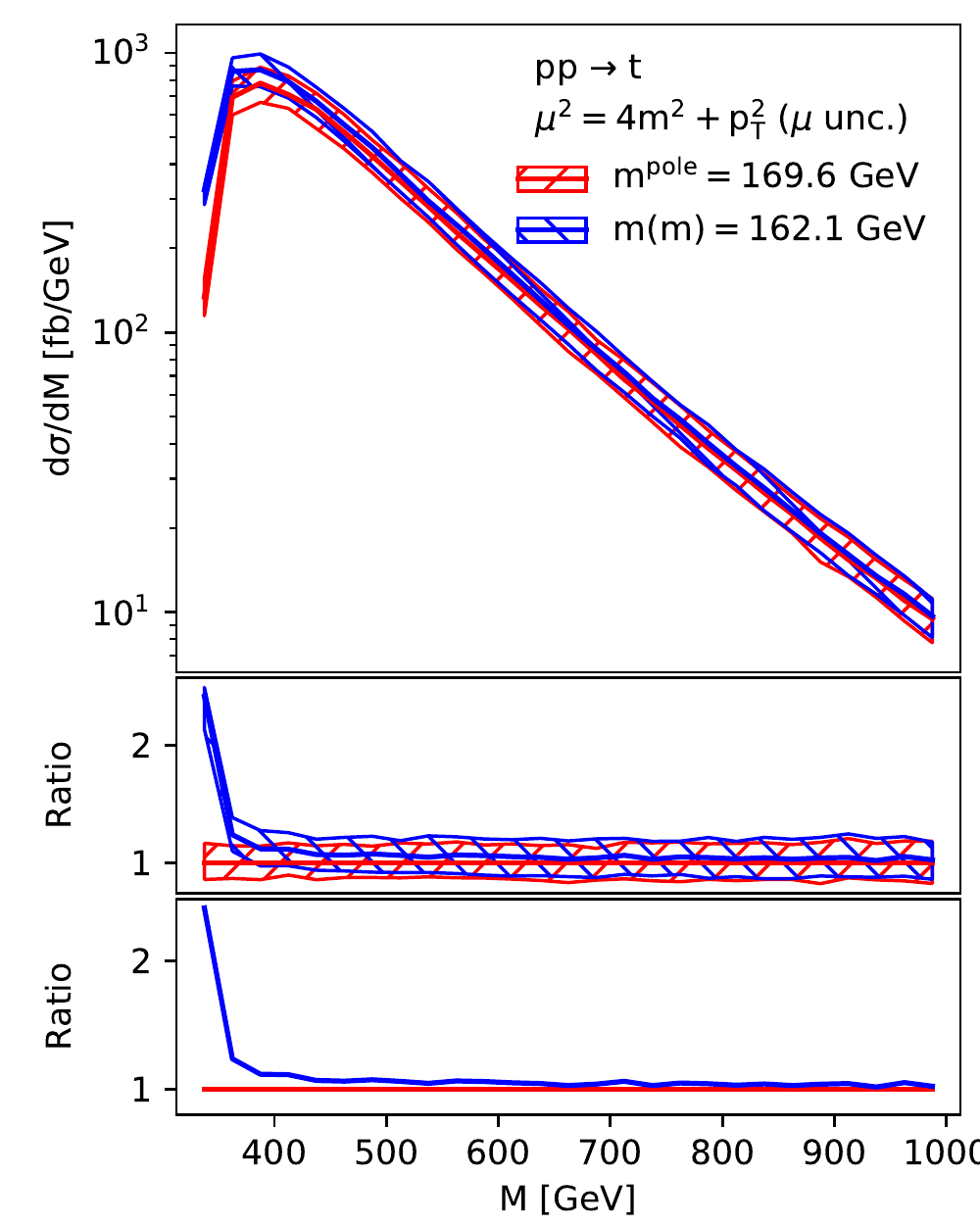}
  \caption{Same as Fig.~\ref{fig:cbt-m-mu} but for heavy-flavor \msbar mass
    values corresponding to those extracted in the ABMP16 NLO fit.} 
  \label{fig:cbt-m-mu-abm}
\end{figure}

\begin{figure}
  \centering
  \includegraphics[width=0.595\textwidth]{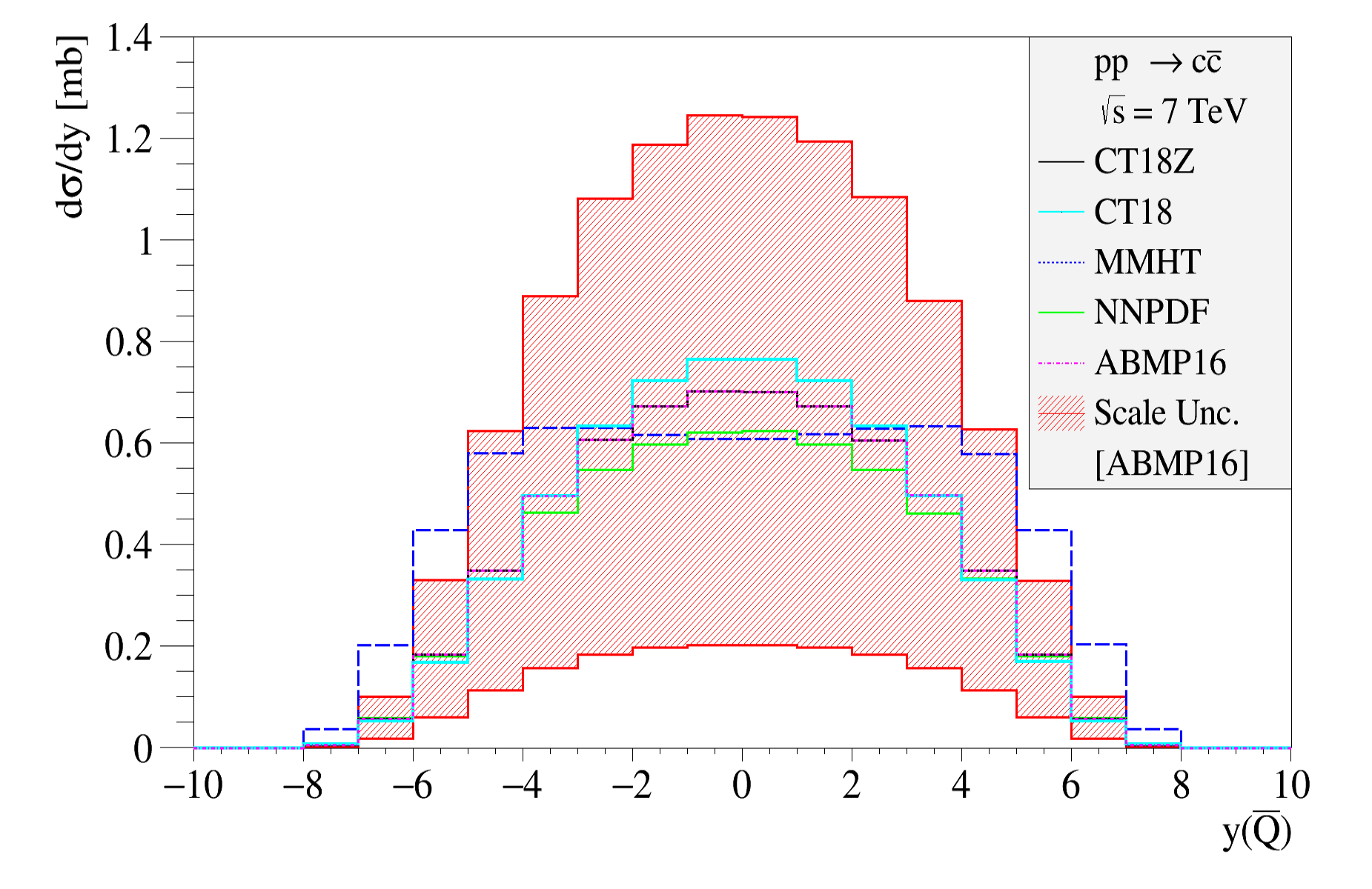}\\[2ex]
  \includegraphics[width=0.595\textwidth]{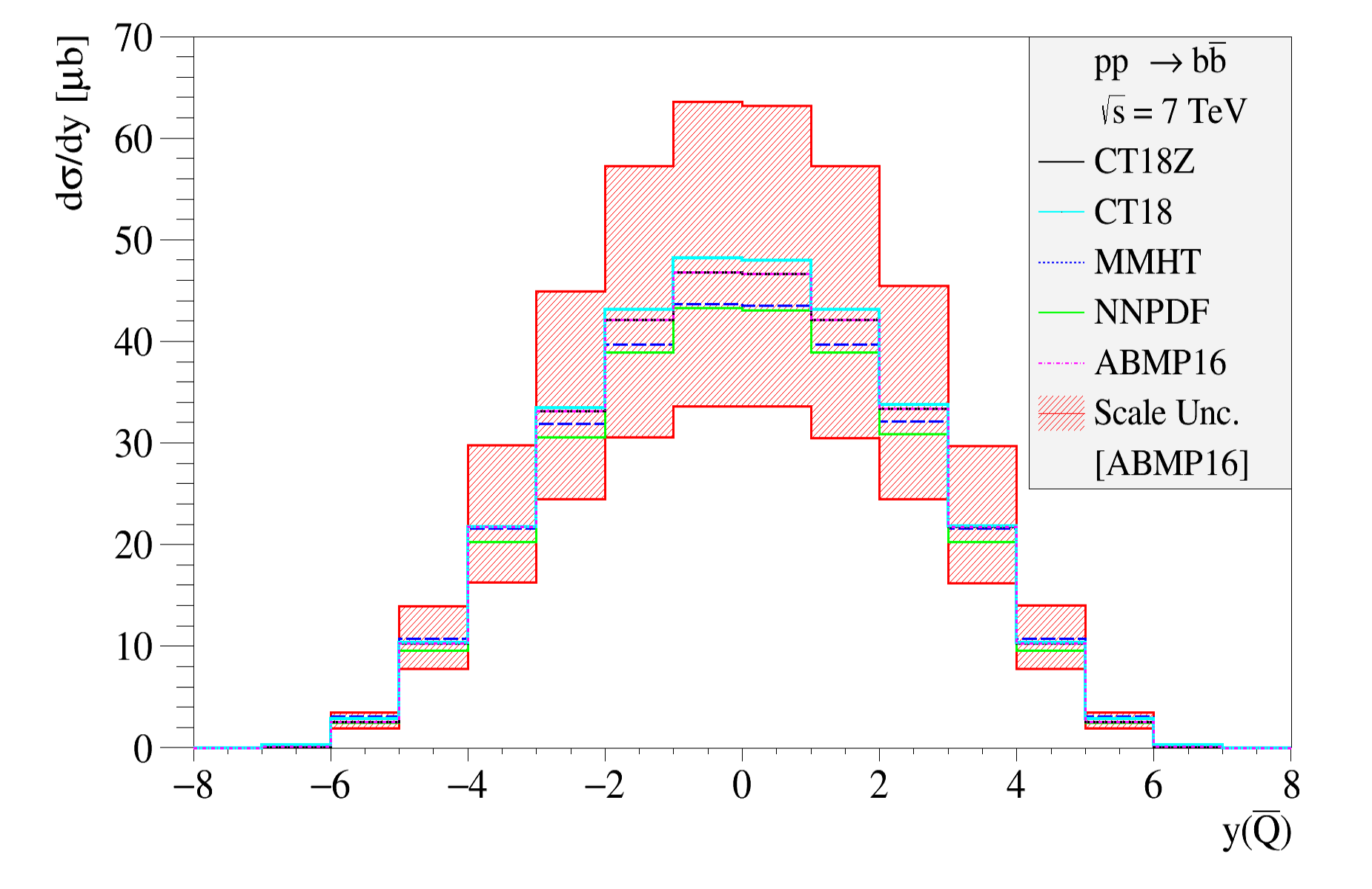}\\[2ex]
  \includegraphics[width=0.595\textwidth]{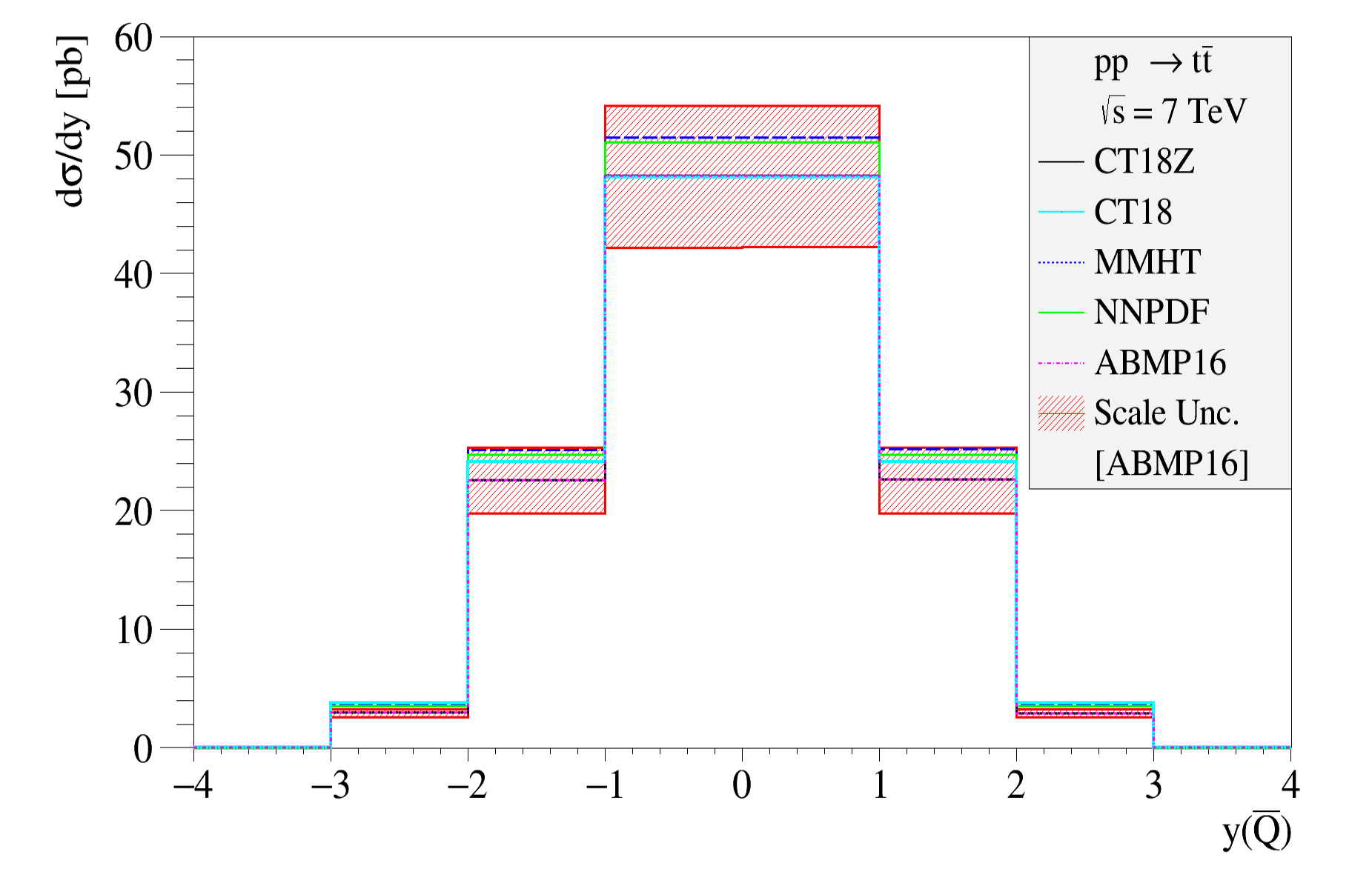}\\[1ex]
  \caption{The NLO differential cross-sections at the LHC ($\sqrt{s} = 7$~TeV)
    for charm (upper panels), bottom (intermediate panels) and top (lower panels) hadro-production
    as a function of the rapidity $y$ of the produced antiquark with mass
    renormalized in the \msbar scheme, using 
    $\mu_R = \mu_F = \sqrt{p_T^2 + 4 m_Q^2(m_Q)}$ 
    and central NLO PDF sets + $\alpha_S(M_Z)$ values from different collaborations 
    (CT18~\cite{Hou:2019efy}, CT18Z~\cite{Hou:2019efy}, MMHT14~\cite{Harland-Lang:2014zoa},
    NNPDF3.1~\cite{Ball:2017nwa}, ABMP16~\cite{Alekhin:2018pai}). Scale
    uncertainty bands computed with our nominal set (ABMP16 NLO) are also
    shown.
  }
    \label{fig:lucas-rapidity}
\end{figure}

\begin{figure}[!p]
  \centering
  \includegraphics[width=0.495\textwidth]{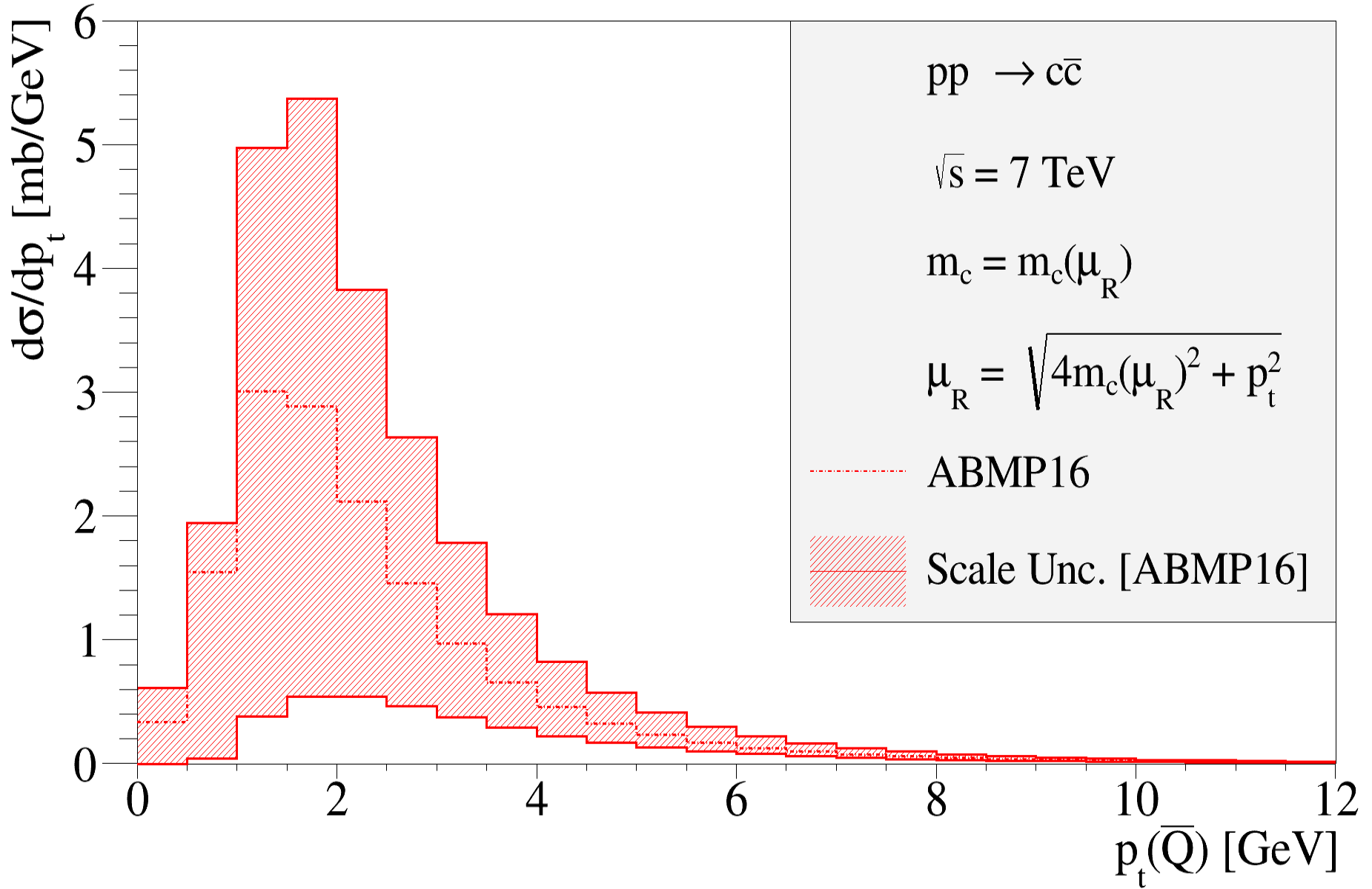}
  \includegraphics[width=0.495\textwidth]{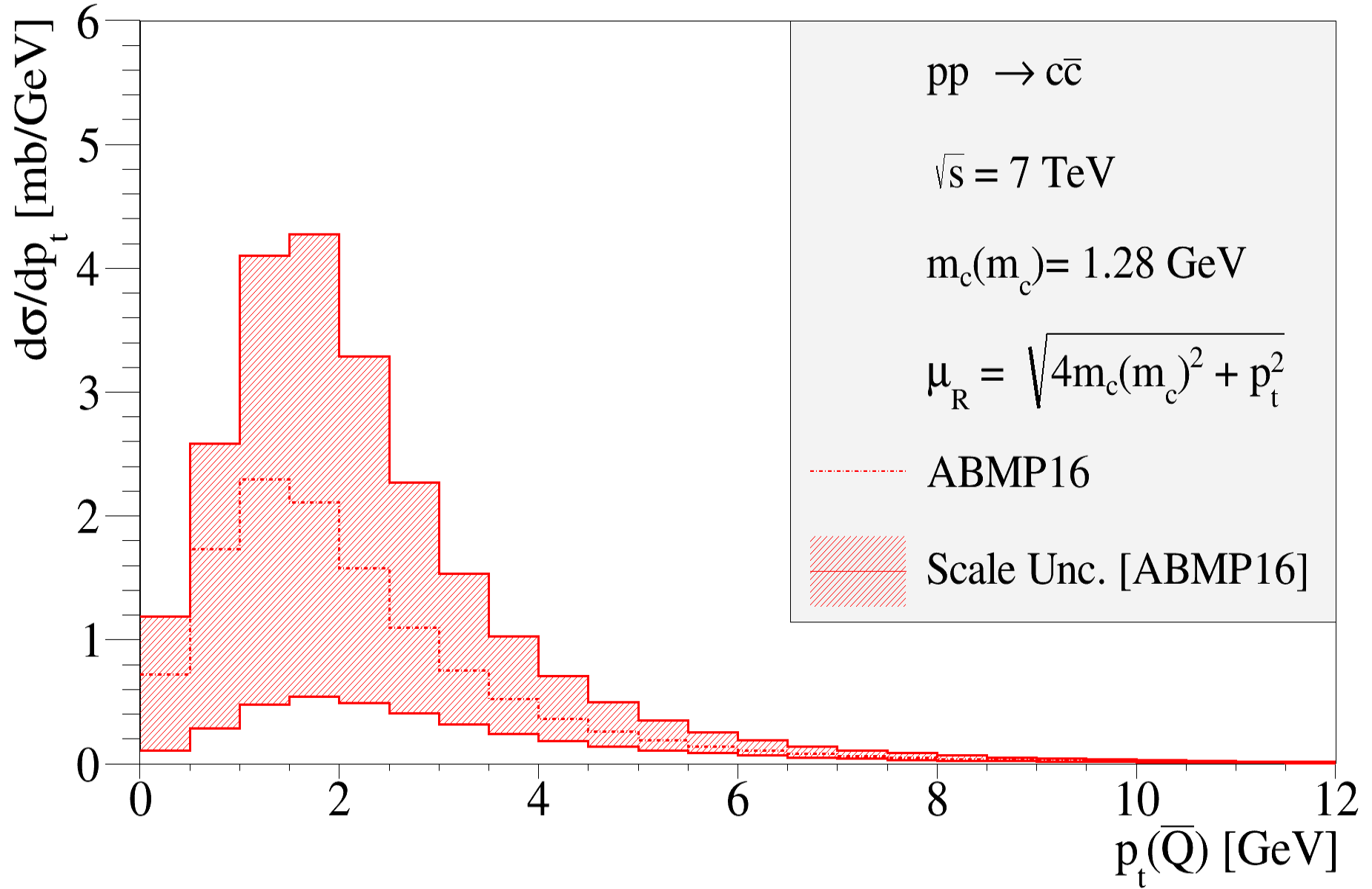}\\[5ex]
  \includegraphics[width=0.495\textwidth]{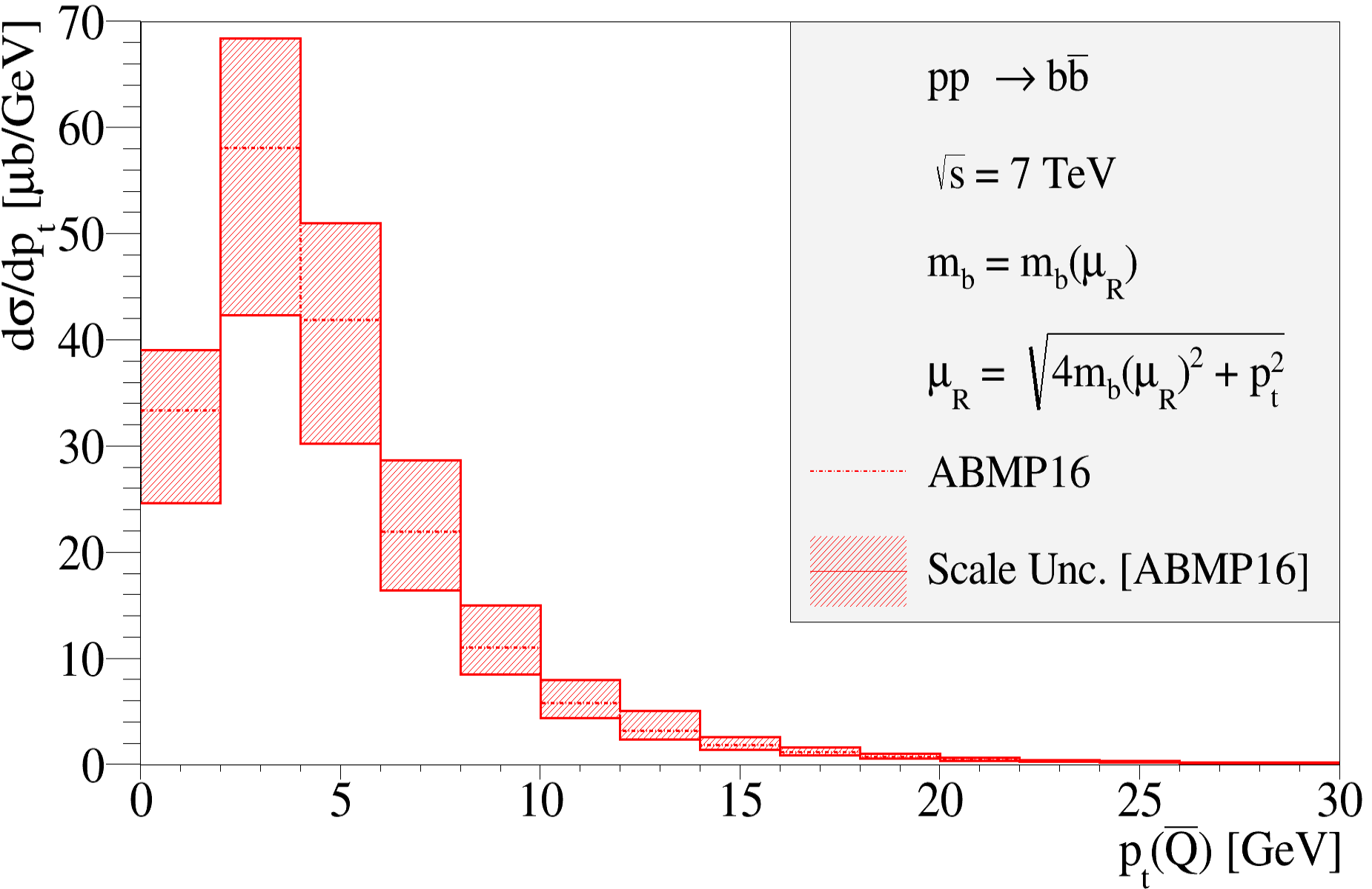}
  \includegraphics[width=0.495\textwidth]{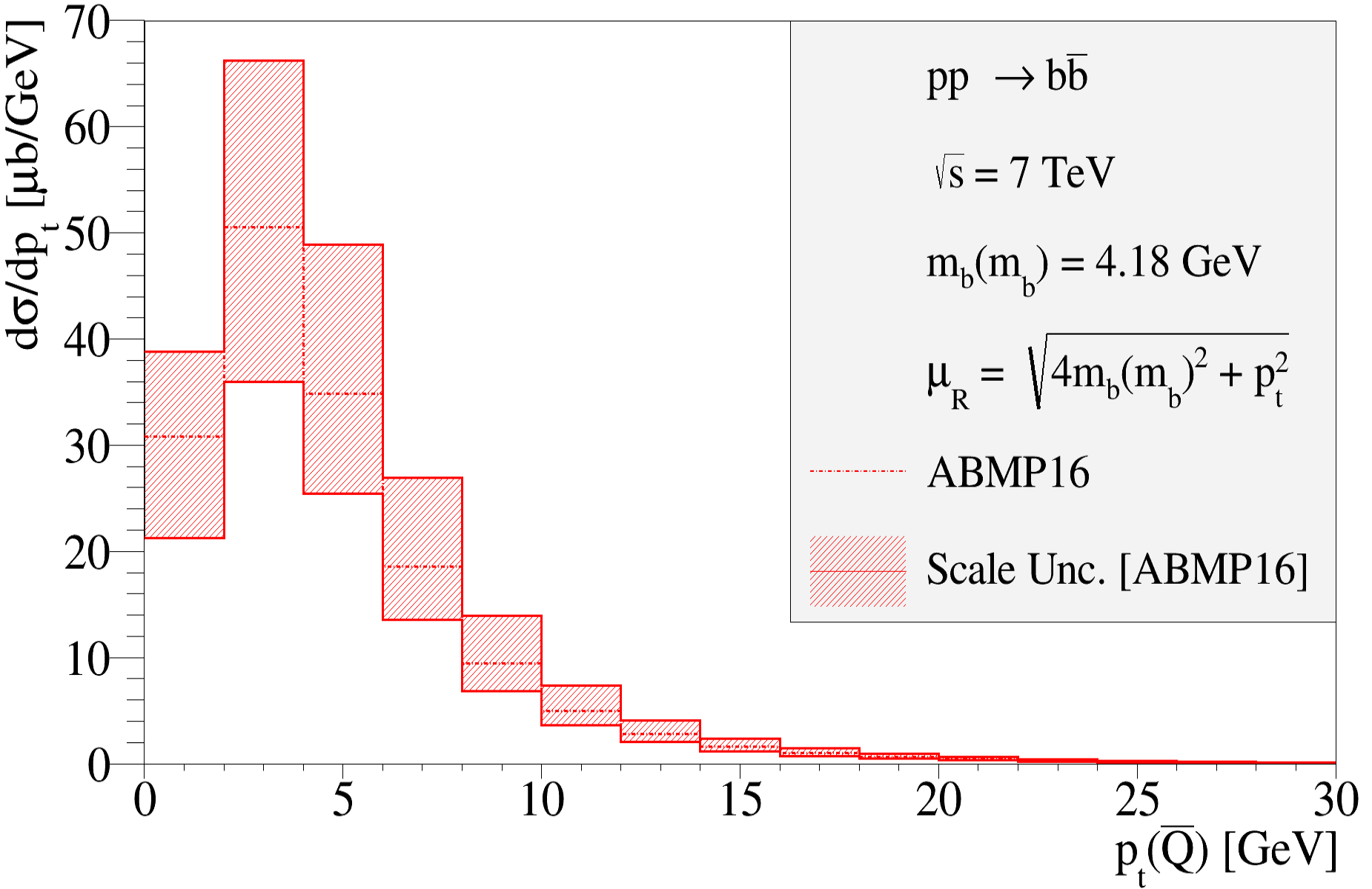}\\[5ex]
  \includegraphics[width=0.495\textwidth]{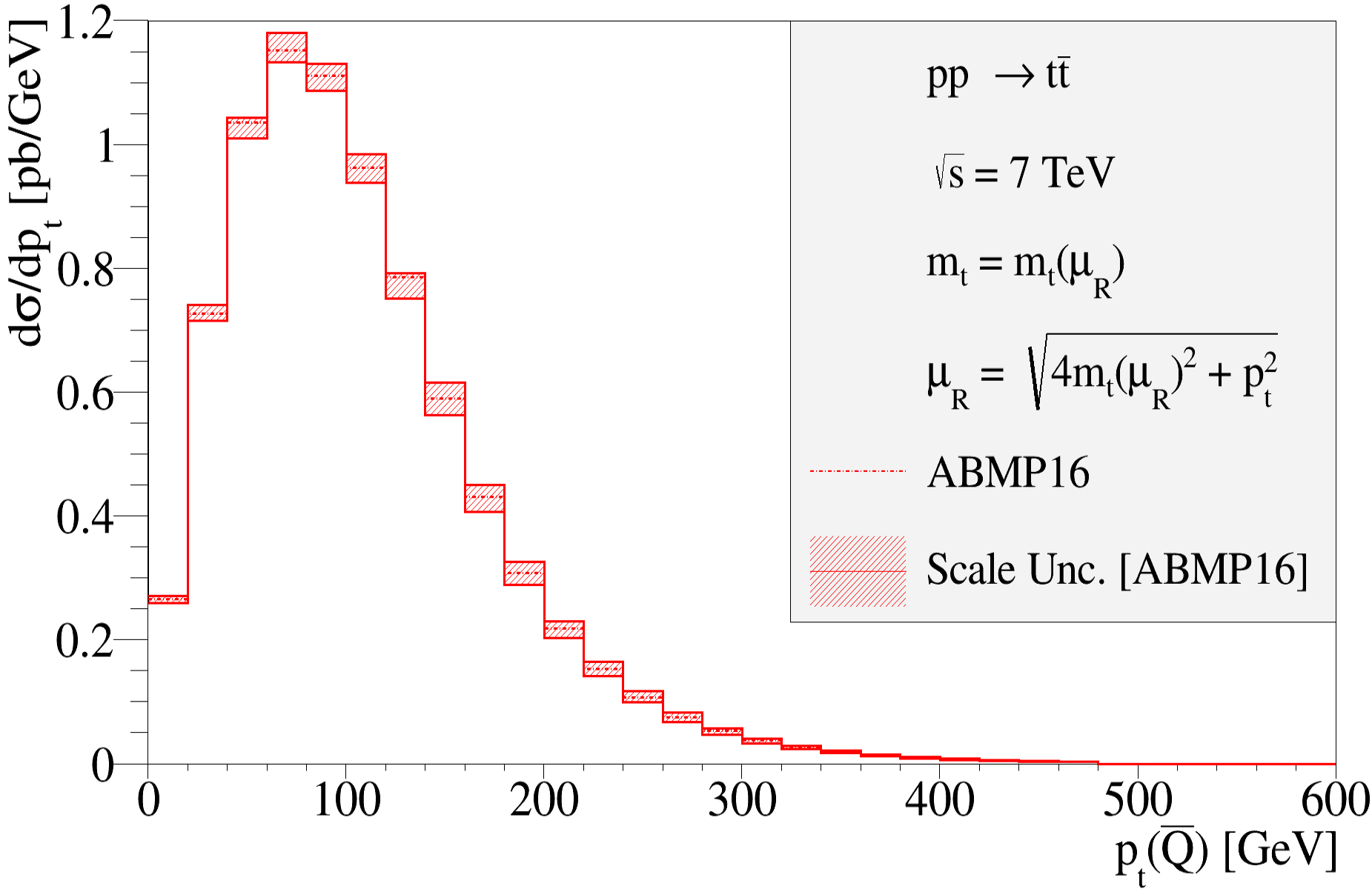}
  \includegraphics[width=0.495\textwidth]{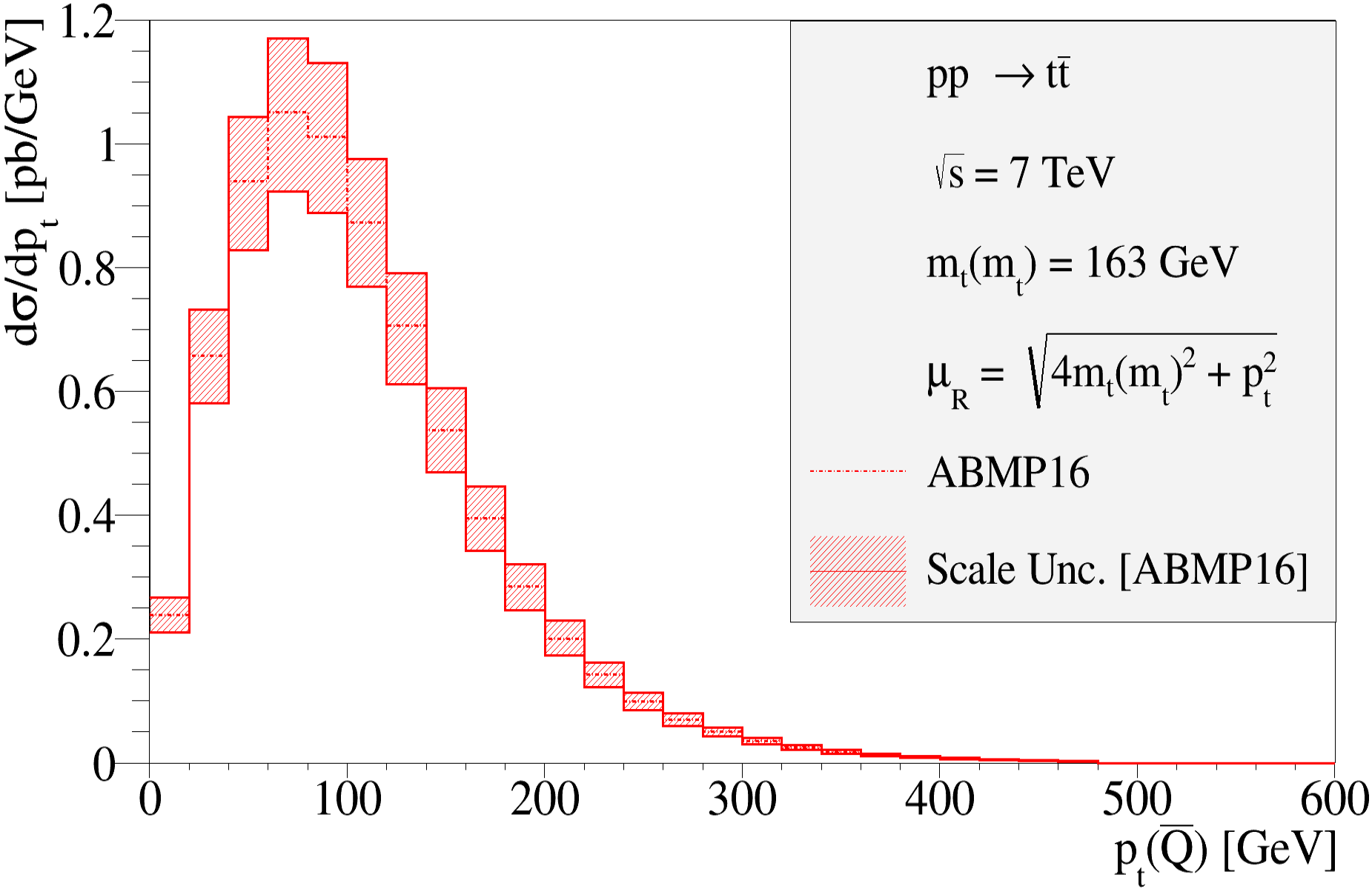}\\[5ex]
  \caption{The NLO differential cross-sections at the LHC ($\sqrt{s} = 7$~TeV)
    for charm (upper panels), bottom (intermediate panels) and top (lower
    panels) hadro-production with their
    7-point ($\mu_R$, $\mu_F$)
    scale uncertainties as a function of the $p_T$ of the produced antiquark with mass renormalized in
    the \msbar scheme, using
    for central predictions
    $\mu_R = \mu_F = \sqrt{p_T^2 + 4 m_Q^2(\mu_m)}$ and different mass renormalization scales $\mu_m$.    
    The panels on the left use a dynamical mass renormalization scale $\mu_m = \mu_R$,
    with these two scales varied simultaneously during $\mu_R$ variation (procedure (i) for scale variation discussed in the text),
    whereas the panels    on the right use the static mass renormalization scale $\mu_m = \msbarm[Q]$,    with $\msbarm[Q]$ fixed to the values of the PDG ($\msbarm[c]$~=~1.28 GeV,    $\msbarm[b]$ = 4.18 GeV, $\msbarm[t]$ = 163 GeV).    The $\alpha_S (M_Z)$ values,    the $\alpha_S$ evolution and the central PDFs extracted from the ABMP16    NLO fit are used in all parts of the computation.
    }
  \label{fig:lucas-dynmass-a}
\end{figure}

\begin{figure}[h!]
  \centering
  \includegraphics[width=0.495\textwidth]{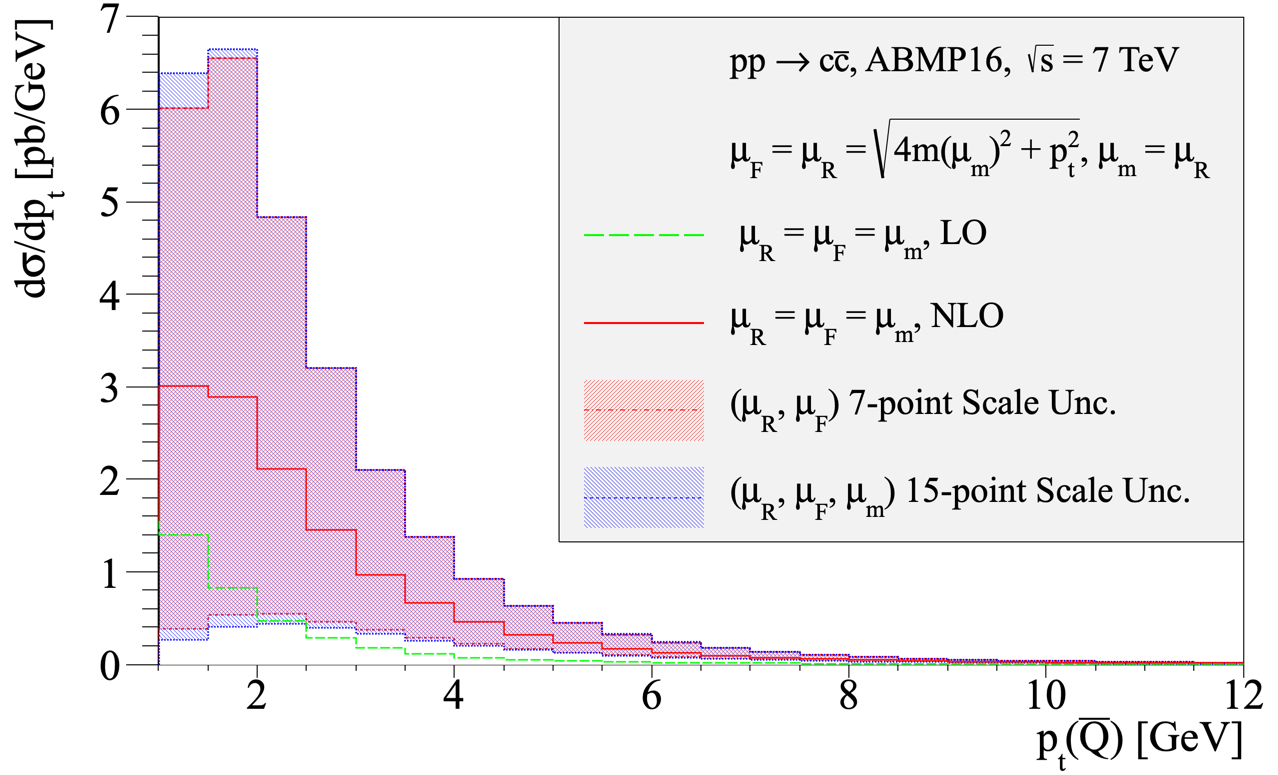}
  \includegraphics[width=0.495\textwidth]{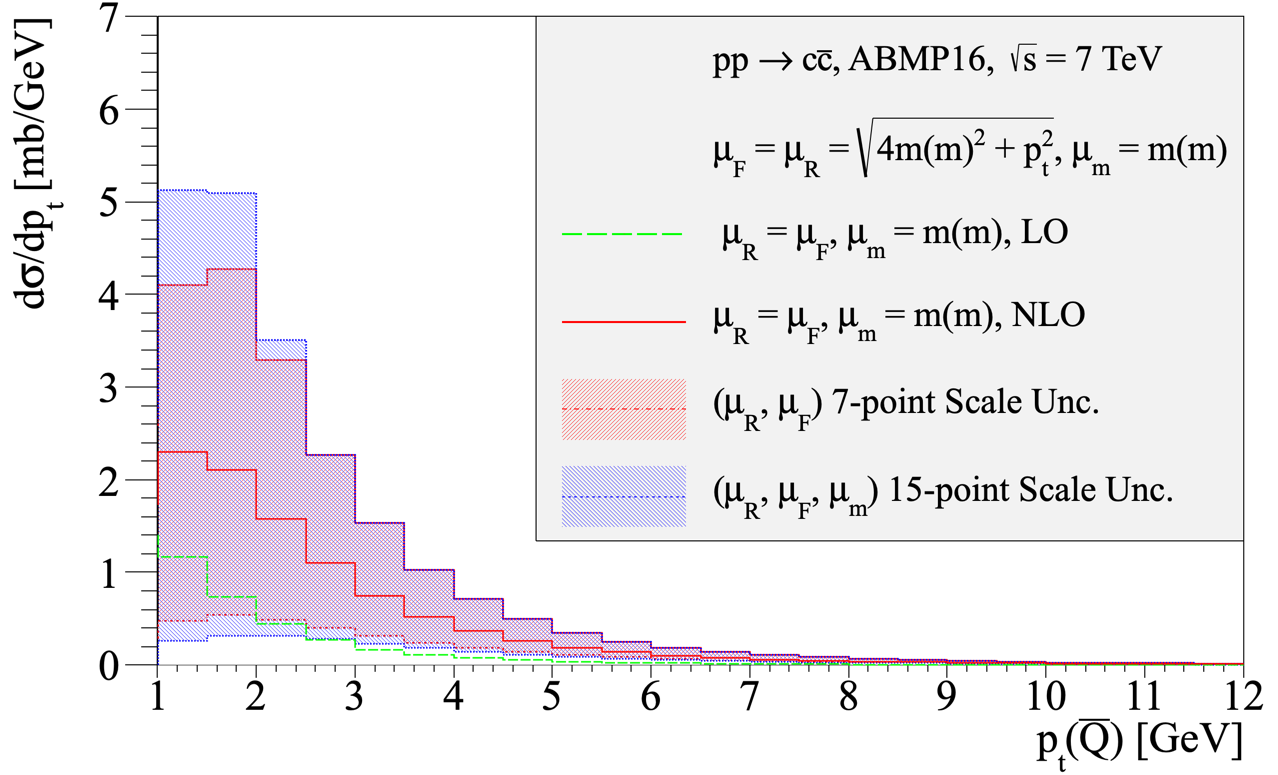}\\[5ex]
  \includegraphics[width=0.495\textwidth]{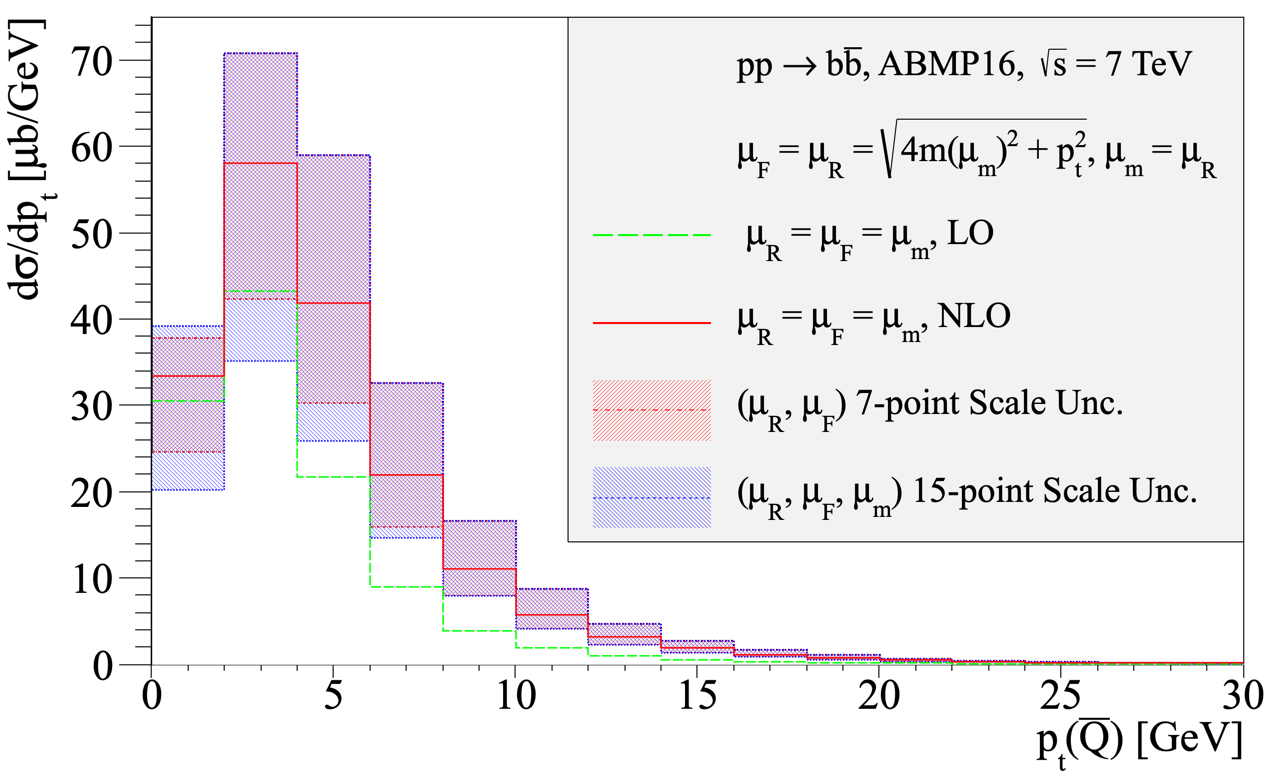}
  \includegraphics[width=0.495\textwidth]{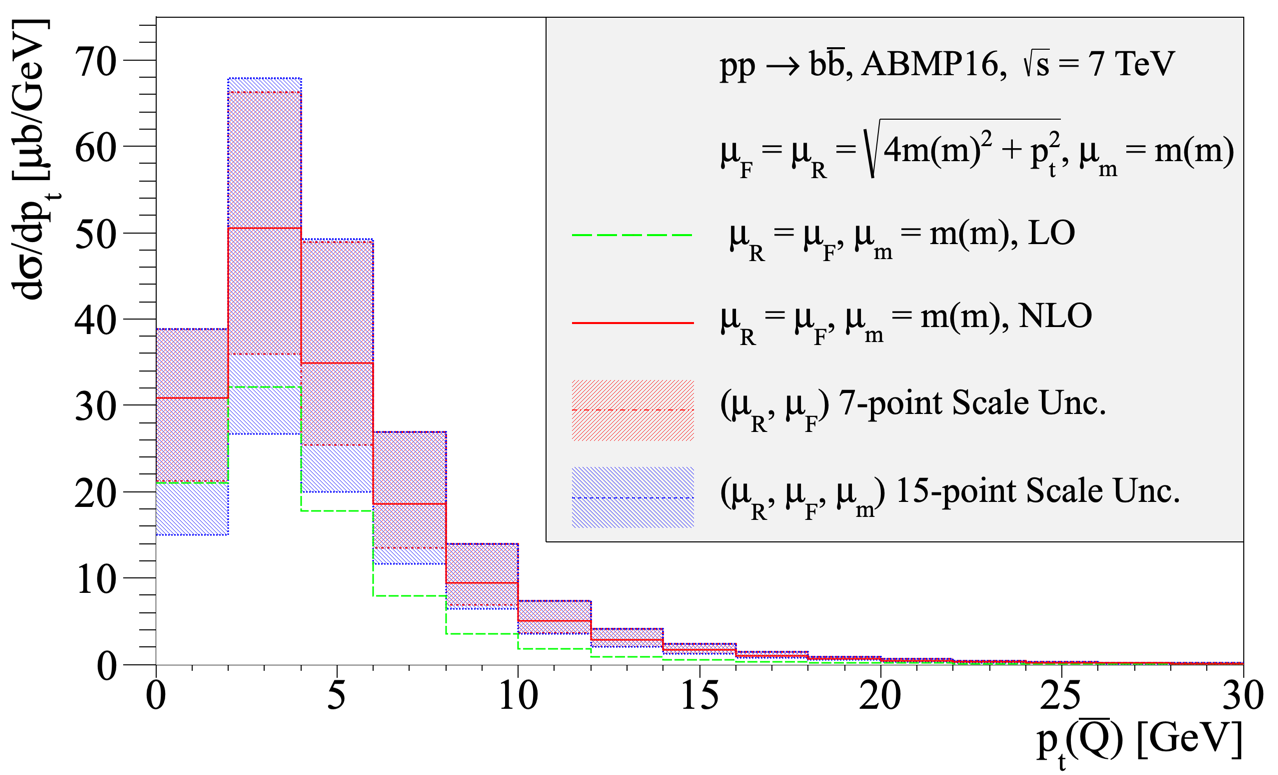}\\[5ex]
  \includegraphics[width=0.495\textwidth]{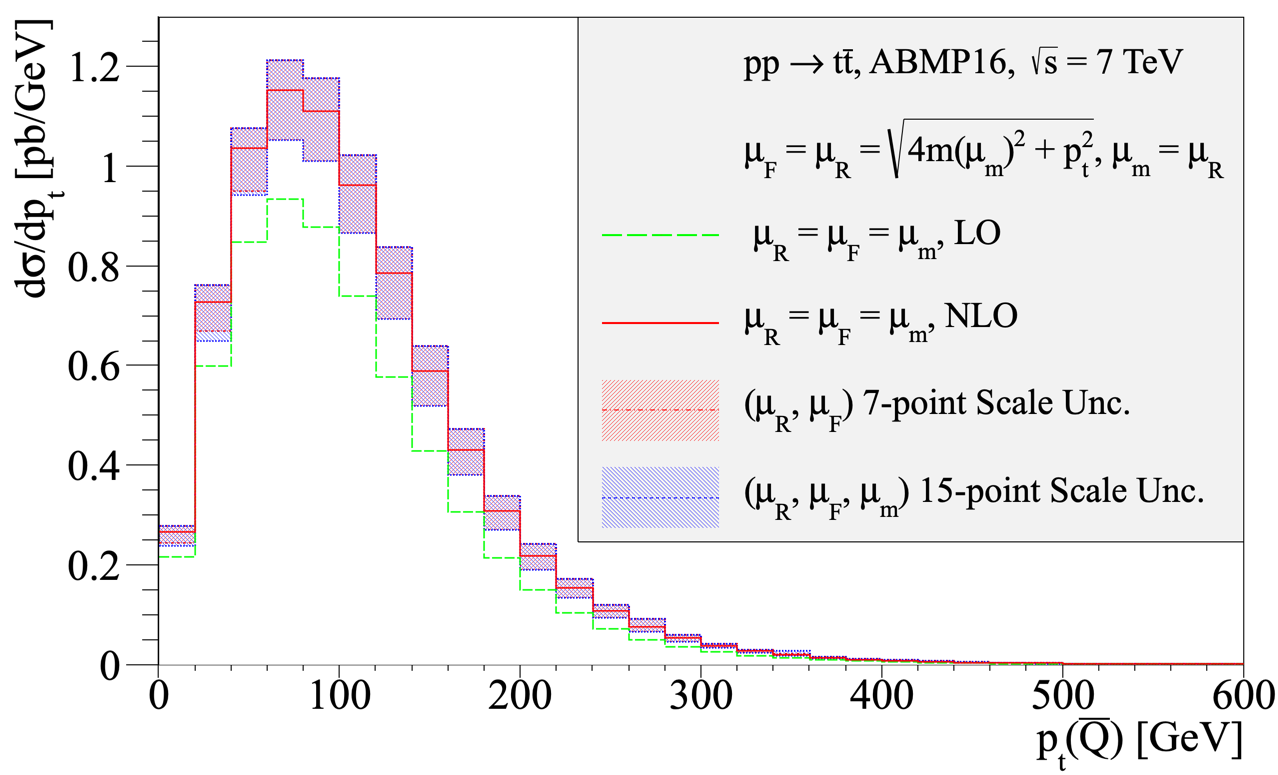}
  \includegraphics[width=0.495\textwidth]{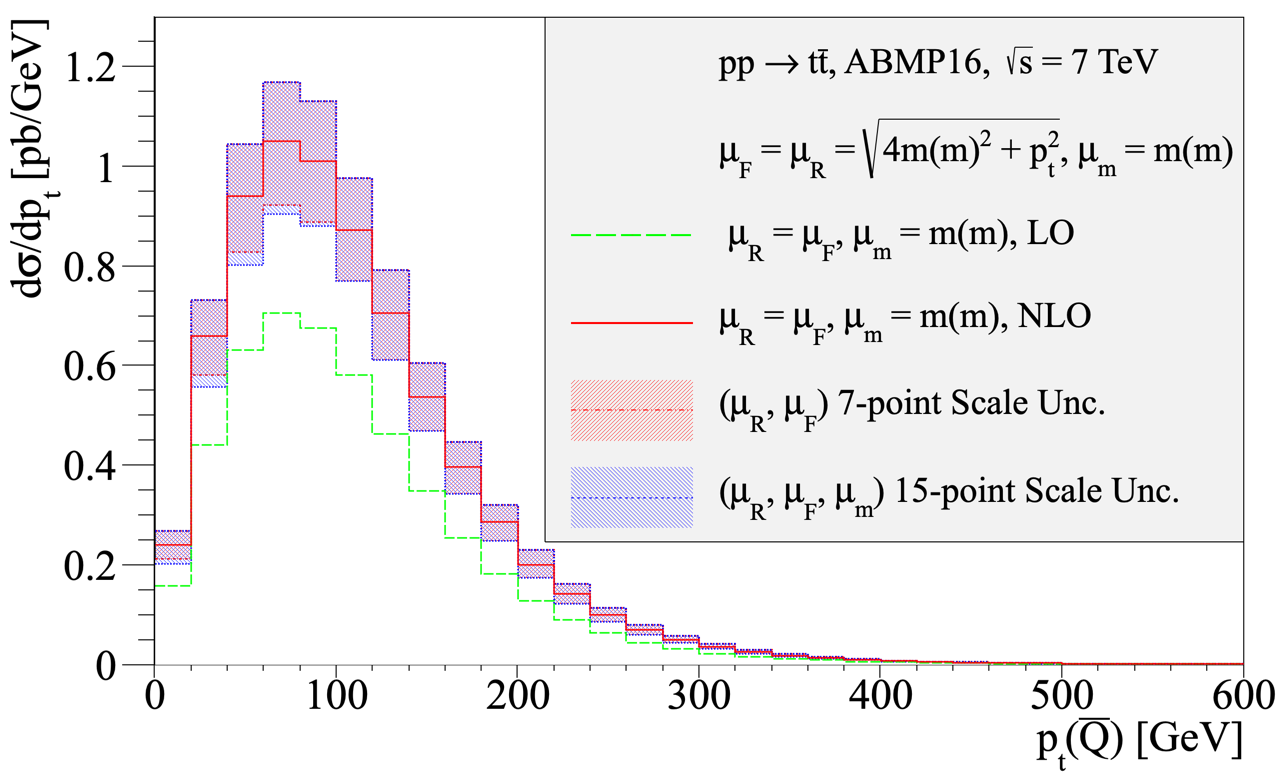}\\[5ex]
  \caption{
    Same as Figure~\ref{fig:lucas-dynmass-a}, but for a different
    scale variation procedure.
    For central predictions, computed
    with $\mu_R = \mu_F = \mu_0 = \sqrt{p_T^2 + 4 m_Q^2(\mu_m)}$
    the panels on
the left use a dynamical mass renormalization scale $\mu_m = \mu_R$, whereas
the panels on the right use the static mass renormalization scale $\mu_m =
\msbarm[Q]$, with $\msbarm[Q]$ fixed to the values of the PDG
($\msbarm[c]$~=~1.28 GeV, 
$\msbarm[b]$ = 4.18 GeV, $\msbarm[t]$ = 163 GeV).
Scale
      variations are computed according to the procedure (ii) discussed in the
      text, 
      with fully independent variations of the involved scales. The ($\mu_R$,
      $\mu_F$) 7-point scale variation bands at fixed $\mu_m = \mu_0$ are
      shown in red, whereas the ($\mu_R$, $\mu_F$, $\mu_m$) 15-point scale
      variation bands are shown in blue. $\mu_m$ is varied independently of
      $\mu_R$ and $\mu_F$ in the interval [1/2,2] around its central
      value. Central LO predictions, computed using the same ($\mu_R$, $\mu_F$,
      $\mu_m$) and $m(\mu_m)$ values as the central NLO ones, are also shown
      (green dashed line). 
    The $\alpha_S (M_Z)$ values, the $\alpha_S$ evolution and the central PDFs
    extracted from the ABMP16 NLO fit are used in all parts of the
    computation,
    both for NLO and for LO predictions.
    }
  \label{fig:lucas-dynmass-b}
\end{figure}

In Figs.~\ref{fig:c-pty-mass} and \ref{fig:b-pty-mass} we compare 
the theoretical uncertainties of the NLO calculations due to variations 
of the quark mass values in the different mass renormalization schemes.
We use $\msbarm[c] = 1.28 \pm 0.03$~GeV and $\msbarm[b] = 4.18 \pm 0.03$~GeV
in the \msbar mass scheme as quoted by the PDG~\cite{Tanabashi:2018oca} and,
correspondingly, $\msrm[c] = 1.36 \pm 0.03$~GeV and $\msrm[b] = 4.33 \pm 0.03$~GeV
in the MSR scheme~\footnote{The heavy-quark mass uncertainties in the MSR
  scheme remain the same as in the \msbar scheme, since in the conversion
  formulas between different schemes one just adds extra terms proportional to
  $\alpha_S$, for which one does not consider any uncertainty, see
  eqs.~(\ref{fromruntopol}), (\ref{eq:massrelation2}).}. 
In the pole mass scheme, we set $\polem[c] = 1.49\, \pm\, 0.25$~GeV, $\polem[b] = 4.57\, \pm \, 0.25$~GeV.
The latter variations reflect the fact that the pole mass is 
affected by an intrinsic renormalon ambiguity of the order of $\Lambda_{\text{QCD}}$, as already mentioned in Sec.~\ref{imple}.
If we made a more conservative assumption 
  for the uncertainties of charm and bottom quark pole masses, using a value
  twice as large, i.e. $\pm\, 0.5$ GeV, the size
  of the pole mass variation uncertainty bands in Figs. ~\ref{fig:c-pty-mass}
  and \ref{fig:b-pty-mass} would approximately be doubled with respect to the
  one shown. This is due to the fact that the cross-sections for charm
  (bottom) hadro-production in the on-shell scheme scale approximately linearly
  with the charm (bottom) mass (decreasing with increasing masses), at least
  for values of the charm (bottom) mass close enough to the central one
  considered in this work.
Therefore, calculations in the \msbar or MSR mass schemes afford substantially
smaller uncertainties (in particular at low $p_T$) due to precise input quark masses.
Changing the central values of the charm- and bottom-quark \msbar masses, which
are set to the PDG values in Fig.~\ref{fig:c-pty-mass} and
Fig.~\ref{fig:b-pty-mass}, to the values extracted in the ABMP16 fit, 
$\msbarm[c] = 1.18 \pm 0.03$~GeV and $\msbarm[b] = 3.88 \pm 0.13$~GeV
(corresponding to $\msrm[c] = 1.21 \pm 0.03$~GeV and $\msrm[b] = 4.00 \pm
0.13$~GeV in the MSR scheme and $\polem[c] = 1.38\,\pm\, 0.25$~GeV, 
$\polem[b] = 4.25\,\pm\, 0.25$~GeV
in the pole mass scheme) 
lead to results qualitatively similar in case of charm, shown in
Fig.~\ref{fig:c-pty-mass-abm}, whereas for the bottom the MSR and \msbar mass
uncertainty bands, shown in Fig.~\ref{fig:b-pty-mass-abm}, are enlarged with
respect to those in Fig. \ref{fig:b-pty-mass} due to the larger
uncertainty accompanying the bottom-mass extraction in the 
ABMP16 fit ($\pm$ 0.13 GeV) as compared to the PDG case ($\pm$ 0.03 GeV). 

In Fig.~\ref{fig:cbt-m-mu} the single-differential cross-sections as a 
function of the invariant mass $M_{Q{\bar Q}}$ of the heavy-quark pair 
in the pole and \msbar mass scheme are shown, as calculated using {\texttt{MCFM}} 
(no implementation of the MSR scheme is available for this distribution).
The impact of changing from the pole to the \msbar mass scheme is largest at 
the lowest values of $M_{Q{\bar Q}}$ close to the production threshold. 
At a technical level, this is due to the derivative term 
in eq.~(\ref{eq:conversion}) becoming dominant in this kinematic region.
However, this implies that the term $\delta m^{\mathrm{pole-sd}} = \polem - m^{\mathrm{sd}}$
in eq.(\ref{eq:pole-sd}) for the conversion of \polem to a short distance mass grows parametrically as 
$\delta m^{\mathrm{pole-sd}} \sim m^{\mathrm{sd}} \alpha_S $, hence is no longer small either. 
This situation is realized for the \msbar mass definition and it persists even
when changing the \msbar mass value, as follows from the comparison of
Fig.~\ref{fig:cbt-m-mu} with Fig.~\ref{fig:cbt-m-mu-abm}, where different
$m(m)$ values are employed. This excludes the \msbar scheme from being a
suitable mass renormalization scheme for observables very close to threshold, 
cf. Ref.~\cite{Dowling:2013baa} for a detailed analysis for the top-quark pair
invariant mass distribution. Alternative mass renormalization schemes for
observables dominated by the production threshold have been mentioned in Sec.~\ref{imple}. 

For comparison to current experimental data on pair-invariant mass $M_{Q{\bar Q}}$ 
distributions in hadro-production, however, this aspect is of minor relevance. 
For instance, for top production at the
LHC~\cite{Sirunyan:2019jyn,Sirunyan:2019zvx} the size of the lowest 
$M_{t{\bar t}}$ bin is large, extending to ${\mathcal{O}}(50)$~GeV above threshold,
so that sensitivity to threshold dynamics is significantly reduced and the
\msbar mass scheme is still applicable in analyses of those data. 

In Fig.~\ref{fig:lucas-rapidity} we show the impact of using different PDF
sets (together with their $\alpha_S(M_Z)$ value) in the rapidity distributions
for charm-, bottom- and top-quarks (see also Ref.~\cite{Kemmler:2019the}). 
We fix the heavy-quark \msbar masses to the PDG values. Slight changes in the
normalization of the distributions can be ascribed to the fact that different
PDF fits are accompanied by slightly different values of $\alpha_S(M_z)$. On
the other hand, larger changes in normalization and in shapes are related to
the different behaviour of different PDFs as a function of $x$ and $\mu_F$. In
particular, in case of charm production, the shape of the rapidity
distribution obtained with the central set of the MMHT14 PDF
fit~\cite{Harland-Lang:2014zoa} for $pp$ collisions at $\sqrt{s}$ = 7 TeV is
much wider with respect to that obtained with the central PDF sets from other
widely used fits. This is particularly evident when using the \msbar
heavy-quark mass, instead of the pole mass, in the computation, due to the
lower value of the first one with respect to the second one, and is related to
the peculiar and very flexible MMHT14 PDF parameterization and the particular 
behaviour of the gluon distribution at small $x$. 
At the scales relevant for the calculation, the MMHT14 NLO central gluon
distribution steeply rises for smaller $x$ and displays large uncertainties, 
in absence of data capable of constraining it for $x < 10^{-4}$ in the fit, 
see also Ref.~\cite{Accardi:2016ndt}. 
On the other hand, in case of top and bottom production, the differences among predictions making use of
different PDF sets are smaller than for the charm case, because, for fixed
rapidity values, these processes probe larger ($x$, $Q^2$) values, where more
data have been used to constrain the various PDFs. In particular, the
predictions obtained by different PDF sets,  turn out to be within the scale
uncertainty band computed using the ABMP16 NLO PDF nominal set, at least for
rapidities away from the far-forward region.  

Additionally, in this paper we explore the possibility of using a dynamical scale
in the heavy-quark \msbar mass renormalization, as an alternative to the
static value $\msbarm[Q]$ and its variations used in the previous distributions and in Ref.~\cite{Dowling:2013baa}. 
There, the $p_T$ distribution of the top-quark at NLO was computed for static
central scales $\mu_R=\mu_F=\mu_m=\msbarm[t]$, varying them simultaneously by 
factors 1/2 and 2 around their central value and finding that the scale
uncertainty band was reduced with respect to the case when $\mu_R$ and $\mu_F$
are varied and $\mu_m$ is fixed to $\msbarm[t]$. In general, we expect that
dynamical scales, catching the different kinematics of different events,
provide a more accurate description of differential distributions. Thus, 
in the following we consider the case when the
central values for
the renormalization and mass renormalization scales are chosen dynamically and coincide, i.e., $\mu_m = \mu_R = \mu_0 = \sqrt{p_T^2 + 4 m_Q^2(\mu_R)}$. We fix the central
factorization scale to the same value $\mu_F$ = $\mu_0$. For this
configuration, we compute scale uncertainties,
by two different procedures.
In the first procedure (i) we fix $\mu_R=\mu_m$ even in the scale variation but we still vary
independently 
$\mu_R$ in the interval [$\mu_{R,1}$, $\mu_{R,2}$], where
$\mu_{R,1}=0.5 \sqrt{p_T^2 + 4 m_Q^2(\mu_{R,1})}$ and 
$\mu_{R,2}=2 \sqrt{p_T^2 + 4 m_Q^2(\mu_{R,2})}$, and $\mu_F$ 
in the interval [1/2, 2] around the chosen (mass) renormalization scale,
excluding the
$(\mu_R, \mu_F)$
extreme combinations
(2, 1/2) and (1/2, 2), but keeping all the others,
as in the
conventional
7-point
scale-variation procedure. These
variations
implicitly
also encode a heavy-quark mass
variation,  with the mass value spanning the interval [$m(\mu_{R,2})$,
$m(\mu_{R,1}$)]. 
In the second variation procedure (ii), which is more general
  than (i), we fix $\mu_R=\mu_m=\mu_F=\mu_0=\sqrt{p_T^2 + 4 m_Q^2(\mu_m)}$ in
  the central predictions as before, but we vary $\mu_R$, $\mu_F$ and $\mu_m$
  independently from each other, each by factors 1/2 and 2 around $\mu_0$,
  excluding the extreme scale combinations as in the conventional
  scale-variation procedure. In other words, we release the constraint $\mu_R$
  = $\mu_m$ during the variation of these scales. This procedure leads to a
  7-point ($\mu_R$, $\mu_F$) scale variation band at fixed $\mu_m$ (not
  coinciding with the one of procedure (i), because there the $\mu_R = \mu_m$
  scales are varied simultaneously), and to a more comprehensive 15-point
  ($\mu_R$, $\mu_F$, $\mu_m$) uncertainty band.
  The $p_T$ distributions obtained
  with the scale configuration
  and variation procedures (i) and (ii)
  are shown in the upper, intermediate and
  lower left panels of Figs.~\ref{fig:lucas-dynmass-a}
  and~\ref{fig:lucas-dynmass-b}
for the charm-, bottom- and top-antiquark, respectively.

For both procedures (i) and (ii),
in case of charm, the ($\mu_R$,
$\mu_F$) uncertainty band turns out to be larger than that computed using a
fixed value of the charm-mass \msbarm[c] and making the standard 7-point
scale variation around the central choice $\mu_0 = \sqrt{p_T^2 + 4
  m_c^2(m_c)}$, shown
in red
in the right upper panel of both
Figs.~\ref{fig:lucas-dynmass-a}
and
\ref{fig:lucas-dynmass-b}.
Similar considerations on the
  size of the uncertainty bands in the comparison between charm results with
  dynamical and static scales $\mu_m$ apply also to the 15-point scale
  variation band, computed according to procedure (ii) and shown in blue in the
  left and the right upper panels of Fig.~\ref{fig:lucas-dynmass-b},
  respectively for the dynamical and static $\mu_m$  cases. In other words,
  adding $\mu_m$ variations does not modify the general conclusions inferred by comparing the
  7-point uncertainty bands.

  On the other hand, in case of top (bottom),
  close to the peak of
  the $p_T$ distribution, i.e. in the bulk of the phase-space,
  the
uncertainties accompanying the computation with dynamical $\mu_m$ are much smaller (smaller) than for
$\mu_m = \msbarm[Q]$, as can be seen by comparing the left and right lower
(intermediate) panels of Fig.~\ref{fig:lucas-dynmass-a}
for
procedure (i) and Fig.~\ref{fig:lucas-dynmass-b} for procedure (ii),
showing that a choice 
of the mass renormalization scale coinciding with relevant scales of the
hard-scattering process helps reducing uncertainties.
  Scale uncertainties computed according the most general procedure (ii) are larger than those for procedure (i) in the bulk of the phase space, with progressively reduced differences at increasing $p_T$ in the tail of the $p_T$ distributions. The largest difference of band sizes, amounting to a factor $\mathcal{O}$(2-4) depending on the bin, is visible for top production in the $p_T$ range [0, 200]~GeV when comparing the lowest left panel of Fig.~\ref{fig:lucas-dynmass-b} to the lowest left panel of Fig.~\ref{fig:lucas-dynmass-a}.
  For procedure (ii),  the
  scale-variation uncertainty band for the top $p_T$ distribution with $m(\mu)$ has a size 40~-~50\% smaller than the corresponding band obtained in the computation with $m(m)$, in the region around the peak, as follows from comparing the lowest left and right panels of Fig.~\ref{fig:lucas-dynmass-b}.
In case of top-quark pair production, including or neglecting
  the $\mu_m$ variation turns out to have a rather small impact on the size of the
  scale variation uncertainty bands, as is visible in both lowest panels of
  Fig.~\ref{fig:lucas-dynmass-b}. This confirms and extends results obtained
  previously by other authors for the case $\mu_m=m(m)$ (see
  ref.~\cite{Dowling:2013baa} and the total cross-sections in Table~1 of
  Ref.~\cite{Catani:2020tko}).
  In case of charm
  a reduction of scale uncertainties when using dynamical $\mu_m$ instead of static $\mu_m$ 
  is not visible in any of the scale variation configurations.
  Charm-quark running mass values span scale values down to  $\lesssim \mathcal{O}(1$ GeV), too close to the small scale value
$\mathcal{O}(\Lambda_{\mathrm{QCD}})$ where perturbative QCD generally stops to be valid. On the other hand, in case of bottom and top, 
the running mass values stay far from this limit (see Fig.~\ref{fig:mcmbmtrun}
left) and all scales involved are well within the domain of validity of
perturbative~QCD. 

In Fig.~\ref{fig:lucas-dynmass-b} we also add leading order (LO)
  predictions to all panels, obtained by using the same values of $m(\mu_m)$
  and $m(m)$ as in the NLO ones, and the same NLO PDFs and $\alpha_S(M_z)$ value. 
  The central LO predictions are not always included in the NLO uncertainty bands, however the LO
  uncertainty bands (not shown) are much larger than the NLO ones,
  cf. also Ref.~\cite{Dowling:2013baa} for top production.

\begin{figure}[h!]
  \begin{center}
  \vspace{0.5cm}
  \includegraphics[width=0.85\textwidth]{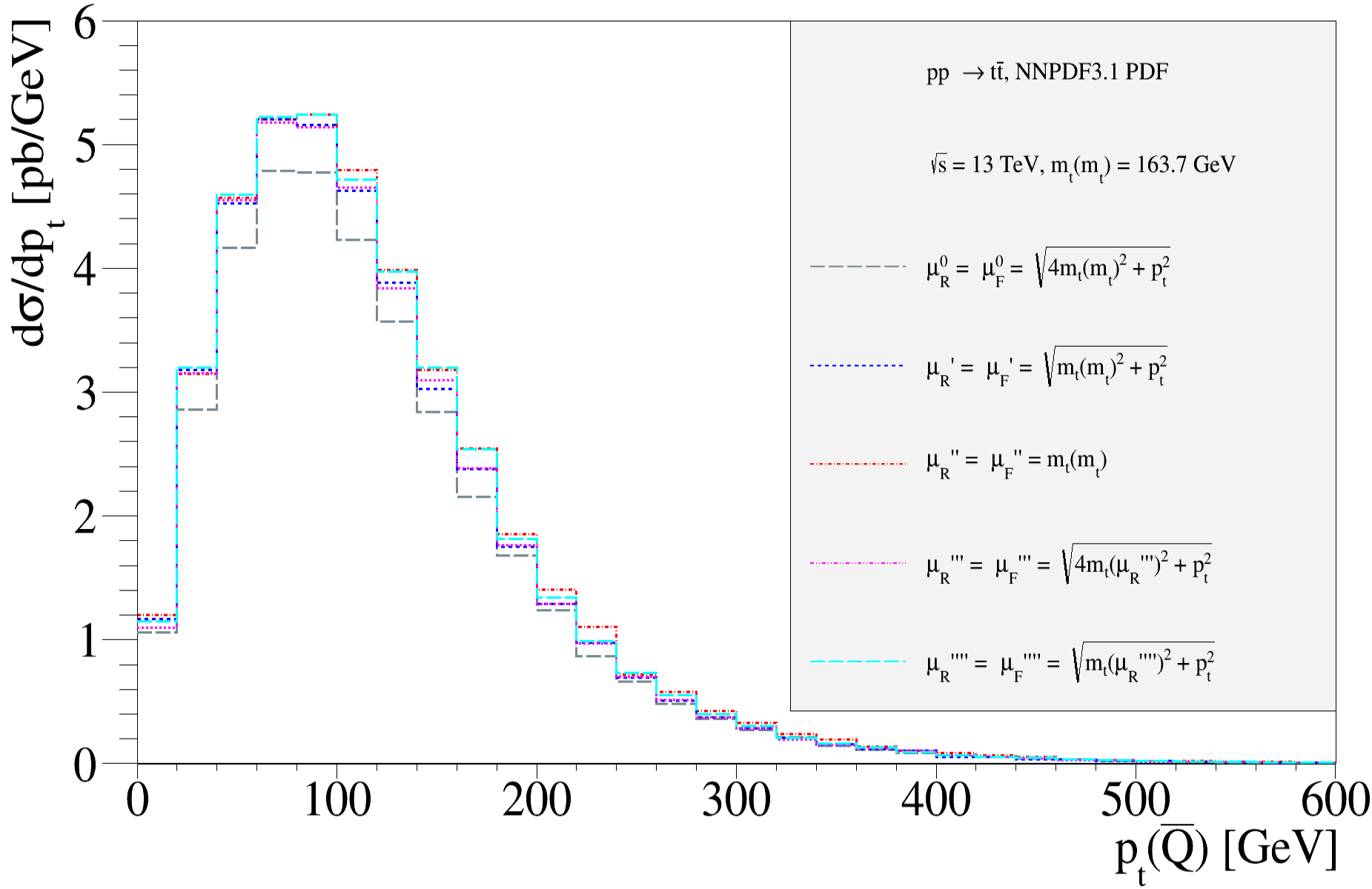}
  \vspace{1.cm}\\
    \includegraphics[width=0.85\textwidth]{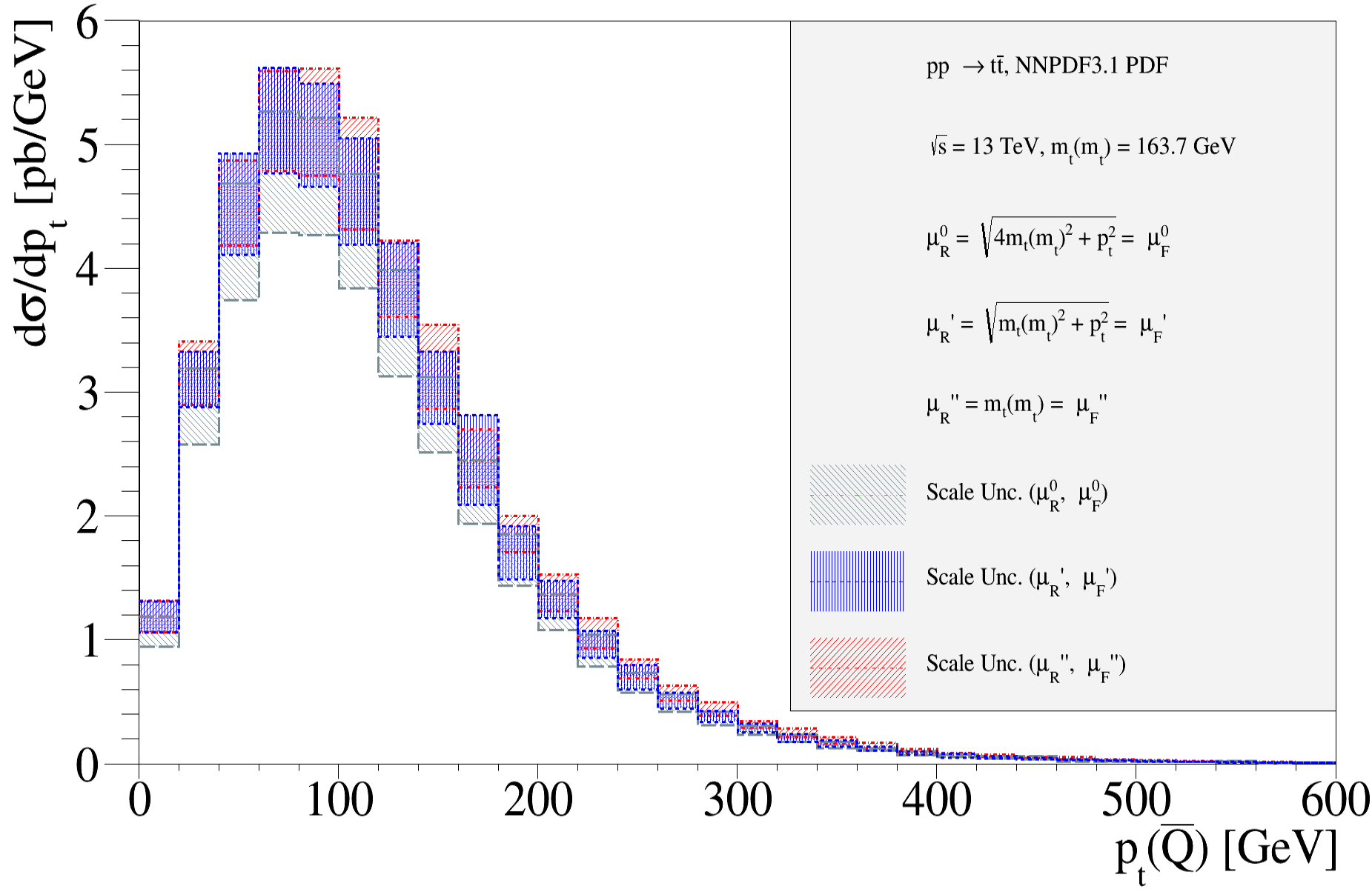}
  \vspace{0.5cm}
  \caption{The NLO differential cross-sections at the LHC ($\sqrt{s}$ = 13
    TeV) for top hadro-production as a function of the $p_T$ of the produced
    antiquark with mass renormalized in the \msbar scheme, using as input
    NNPDF3.1 NLO PDFs with their $\alpha_S(M_Z)$ value and $\alpha_S$
    evolution, and different ($\mu_R$, $\mu_F$, $\mu_m$) combinations: in the
    upper panel central predictions with static scale $\mu_R = \mu_F = \msbarm[t]$
    and $\mu_m = \msbarm[t]$ are compared to those with dynamical scales $\mu_R
    =\mu_F = \sqrt{p_T^2 + m_t^2(\mu_m)}$  and to those with dynamical scales
    $\mu_R = \mu_F = \sqrt{p_T^2 + 4 m_t^2(\mu_m)}$ for both $\mu_m = \msbarm[t]$
    and for $\mu_m = \mu_R$. Scale uncertainty bands, shown in the lower panel
    only for the cases with $\mu_m = \msbarm[t]$), refer to 7-point ($\mu_R$,
    $\mu_F$) variation of factors (1/2, 2) around the central
    values.  
    \label{fig:lucas-cat}}
  \end{center}
\end{figure}

Another example of dynamical mass renormalization scale choice was shown in
Ref.~\cite{Catani:2020tko}, where the \msbar mass renormalization scheme was
studied as an alternative to the pole mass scheme for producing predictions
for top-quark related distributions at NNLO. 
There the $t\bar{t}$-pair invariant mass distribution was studied at NNLO, using the
\msbar mass at a scale $\mu_m \sim M_{t\bar{t}}/2$, setting
$\mu_R=\mu_F=M_{t\bar{t}}/2$ and making a 15-point scale variation of factors
(1/2, 2) around the ($\mu_R$, $\mu_F$, $\mu_m$) central value. Predictions
were compared to the case when $\mu_m = \msbarm[t]$, seeing small
differences. On the other hand, $p_T$ and rapidity distributions were computed
using static $\mu_m$ values. To the best of our knowledge, our paper is the
first work where the use of a dynamical $\mu_m$ scale in computing $p_T$
distributions for heavy-quark hadro-production is investigated. 

We checked that our NLO predictions are consistent with those reported in
Ref.~\cite{Catani:2020tko}, when using their configuration. In
Fig.~\ref{fig:lucas-cat} we present the $p_T$ distribution of the 
antitop-quark for $t\bar{t}$ production in $pp$ collisions at $\sqrt{s}$ = 13 TeV,
using as input the NNPDF3.1 NLO PDF set with its $\alpha_S(M_z)$ default value
and $\alpha_S$ evolution, i.e. one of the configurations already considered in
Ref.~\cite{Catani:2020tko}, and multiple choices for the ($\mu_R$, $\mu_F$,
$\mu_m$) scales. For fixed $\mu_m = \msbarm[t]$ = 163.7 GeV, we observe that
central predictions using $\mu_R^0 = \mu_F^0 = \sqrt{p_T^2 + 4 m_t^2(m_t)}$
have larger ($\mu_R$, $\mu_F$) uncertainty bands (especially in the peak
region) and have smaller absolute values than those using central scales
$\mu_R^\prime = \mu_F^\prime = \sqrt {p_T^2 + m_t^2(m_t)}$ or
$\mu_R^{\prime\prime} = \mu_F^{\prime\prime} = \msbarm[t]$, with differences
between central values at the peak amounting to $\sim$ 10\%, as shown in the
lower panel.  
The latter two scale choices can be considered better scale choices
(i.e. scale choices leading to a faster perturbative convergence) for
$t\bar{t}$ production  than the first one, as also proven by the fact that
NNLO corrections, reported in Ref.~\cite{Catani:2020tko} for the
($\mu_R^{\prime\prime},\mu_F^{\prime\prime}$) case in comparison to the NLO
ones, are quite small. On the other hand, the central predictions we obtained
using $\mu_R^{\prime\prime\prime} = \mu_F^{\prime\prime\prime} = \sqrt{p_T^2 + 4 m_t^2(\mu_R^{\prime\prime\prime})}$
are in much better agreement with the
previous ones than those with ($\mu_R^0$, $\mu_F^0$), as shown in the upper
panel, and have smaller scale uncertainty bands (not reported in the plot),
which shows that the use of $m_t(\mu_R^{\prime\prime\prime})$ instead of $\msbarm[t]$ in the
dynamical scale definition improves the perturbative convergence
of the calculation, corresponding to smaller NNLO corrections. The predictions
with ($\mu_R^{\prime\prime\prime}$, $\mu_F^{\prime\prime\prime}$) are larger
than those with ($\mu_R^0$, $\mu_F^0)$ because 
$m_t(\mu_R^{\prime\prime\prime}) < \msbarm[t]$ and $\mu_R^{\prime\prime\prime} < \mu_R^0$. 
The differences at the peak of the $p_T$ distribution amount to 
$\Delta\mu_R \sim -14.5$ GeV and $\Delta m \sim - 7.4$ GeV. 
A similar behaviour emerges when comparing the lowest left and right panel of
Figs.~\ref{fig:lucas-dynmass-a}
or \ref{fig:lucas-dynmass-b},
for which analogous considerations and conclusions apply. 
On the other hand, if one uses a scale ($\mu_R^{\prime\prime\prime\prime}$,
$\mu_F^{\prime\prime\prime\prime}$) = $\sqrt{p_T^2 +  m_t^2(\mu_R^{\prime\prime\prime\prime})}$, 
one finds central predictions only slightly larger than for the case ($\mu_R^\prime$, $\mu_F^\prime$), as 
also shown in the upper panel of Fig.~\ref{fig:lucas-cat}, considering that both
$(m_t(\mu_R^{\prime\prime\prime\prime})-\msbarm[t])$ and
$(\mu_R^{\prime\prime\prime\prime} - \mu_R^{\prime}) \sim$ -0.9 GeV 
at the peak of the $p_T$ distribution. 

In summary, the heavy-quark $p_T$-distributions in Figs.~\ref{fig:lucas-dynmass-a}, \ref{fig:lucas-dynmass-b} and \ref{fig:lucas-cat} 
with dynamical renormalization and factorization scales of the type 
($\mu_R$, $\mu_F$) = $\sqrt{p_T^2 +  \kappa\, m_Q^2(\mu_m)}$ for some number $\kappa = 1 \dots 4$ 
and the quark masses in the \msbar scheme 
$m_Q(\mu_m)$ evaluated at the dynamical scale $\mu_m = \mu_R$
directly incorporate the running effects of the mass parameter. 
If compared to sufficiently precise experimental data, this offers new and
complementary ways to test the running, e.g., of the top-quark mass, cf. \cite{Sirunyan:2019jyn}.

\newpage
\cleardoublepage

\section{Phenomenological applications}
\label{appli}

The use of the theory results can be illustrated with a number of applications in phenomenology, 
determining the strong coupling constant $\alpha_S(M_Z)$ and values of the top-quark mass
in the different renormalization schemes as well as constraints on PDFs by using available LHC data.

\subsection{Extraction of \msbarm[t] and \msrm[t] + $\alpha_S(M_Z)$ from differential $t\bar{t}$ cross-sections at NLO}
\label{alphas-mt-fit}

\begin{table}
  \setlength{\tabcolsep}{3pt}
  \renewcommand{\arraystretch}{1.35}
  \begin{tabular}{|l|l|}
    \hline
    Settings & Fit results \\
    \hline
    
    pole mass & $\chi^2/\mathrm{dof} = 1364/1151$, $\chi^2_{t\bar{t}}/\mathrm{dof} = 20/23$\\
    $\mu_R = \mu_F = H'$ & $\polem[t] = 170.5 \pm 0.7 (\text{fit}) \pm 0.1 (\text{mod}) {}^{+0.0}_{-0.1} (\text{par}) \pm 0.3 (\mu)$ GeV \\
    Ref.~\cite{Sirunyan:2019zvx} & $\alpha_S(M_Z) = 0.1135 \pm 0.0016 (\text{fit}) {}^{+0.0002}_{-0.0004} (\text{mod}) {}^{+0.0008}_{-0.0001} (\text{par}) {}^{+0.0011}_{-0.0005} (\mu)$ \\
    \hline
        
    pole mass & $\chi^2/\mathrm{dof} = 1363/1151$, $\chi^2_{t\bar{t}}/\mathrm{dof} = 19/23$\\
    $\mu_R = \mu_F = \polem[t]$ & $\polem[t] = 169.9 \pm 0.7 (\text{fit}) \pm 0.1 (\text{mod}) {}^{+0.0}_{-0.0} (\text{par}) {}^{+0.3}_{-0.9} (\mu)$ GeV \\
    this work & $\alpha_S(M_Z) = 0.1132 \pm 0.0016 (\text{fit}) {}^{+0.0003}_{-0.0004} (\text{mod}) {}^{+0.0003}_{-0.0000} (\text{par}) {}^{+0.0016}_{-0.0008} (\mu)$ \\
    \hline
    
    \msbar mass & $\chi^2/\mathrm{dof} = 1363/1151$, $\chi^2_{t\bar{t}}/\mathrm{dof} = 19/23$\\
    $\mu_R = \mu_F = \msbarm[t]$ & $\msbarm[t] = 161.0 \pm 0.6 (\text{fit}) \pm 0.1 (\text{mod}) {}^{+0.0}_{-0.0} (\text{par}) {}^{+0.4}_{-0.8} (\mu)$ GeV \\
    this work & $\alpha_S(M_Z) = 0.1136 \pm 0.0016 (\text{fit}) {}^{+0.0002}_{-0.0005} (\text{mod}) {}^{+0.0002}_{-0.0001} (\text{par}) {}^{+0.0015}_{-0.0009} (\mu)$ \\
    \hline
    
    MSR mass, $R=3$~GeV & $\chi^2/\mathrm{dof} = 1363/1151$, $\chi^2_{t\bar{t}}/\mathrm{dof} = 19/23$\\
    $\mu_R = \mu_F = \msrm[t]$ & $\msrm[t] = 169.6 \pm 0.7 (\text{fit}) \pm 0.1 (\text{mod}) {}^{+0.0}_{-0.0} (\text{par}) {}^{+0.3}_{-0.9} (\mu)$ GeV \\
    this work & $\alpha_S(M_Z) = 0.1132 \pm 0.0016 (\text{fit}) {}^{+0.0003}_{-0.0004} (\text{mod}) {}^{+0.0002}_{-0.0000} (\text{par}) {}^{+0.0016}_{-0.0008} (\mu)$ \\
    
    \hline
  \end{tabular}
  \caption{The values for $\alpha_S(M_Z)$ and the top-quark mass in different
    mass schemes obtained in Ref.~\cite{Sirunyan:2019zvx} and in this work
    by fitting the CMS data on $t\bar{t}$ production and the HERA DIS data~\cite{Abramowicz:2015mha}
    to theoretical predictions. 
    The fit, model (mod), parametrisation (par) and scale variation ($\mu$) uncertainties are reported. 
    Also the values of $\chi^2$ are reported, as well as the partial $\chi^2$
    values per number of degrees of freedom (dof) for
    the $t\bar{t}$ data ($\chi^2_{t\bar{t}}$) for $23$ $t\bar{t}$ cross-section bins in the fit. 
    The scale $H'$ is defined in the text.
  }
  \label{tab:mtfit}
\end{table}

\begin{figure}[t!]
  \centering
  \includegraphics[width=0.55\textwidth]{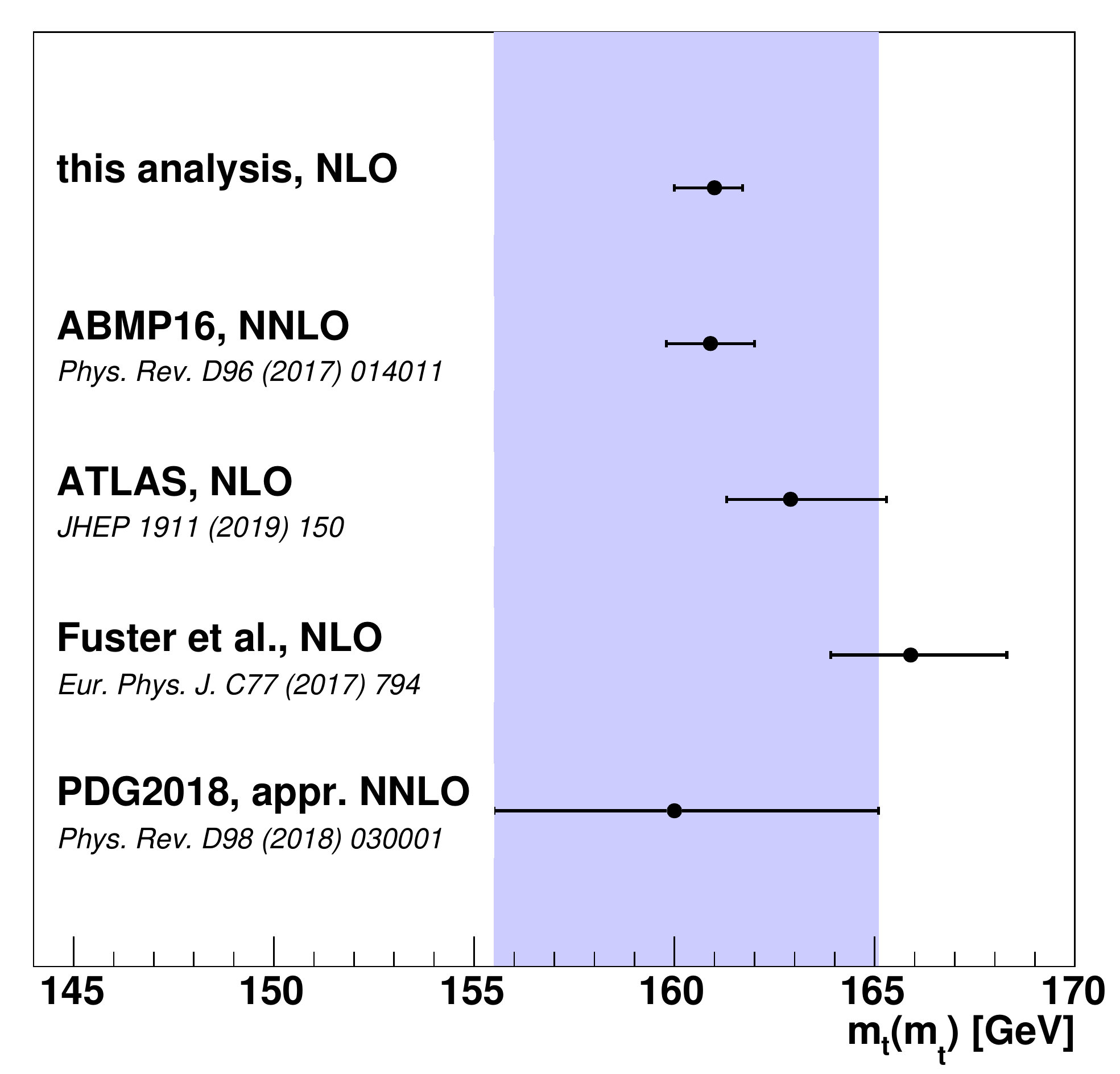}
  \caption{The extracted value of \msbarm[t] compared to other
    determinations~\cite{Tanabashi:2018oca,Alekhin:2017kpj,Fuster:2017rev,Aad:2019mkw}. 
    The world average labelled as `PDG2018, appr. NNLO' is based on a single determination of the D0
    collaboration~\cite{Abazov:2011pta}.} 
  \label{fig:mtmt}
\end{figure}

The top-quark mass can be extracted using measurements of the total or
differential $t\bar{t}$ production cross-sections. 
As an example, we use the recent CMS measurement~\cite{Sirunyan:2019zvx} of normalized
triple-differential $t\bar{t}$ cross-sections as a function of invariant mass
and rapidity of the $t\bar{t}$ pair, and the number of additional jets. 
These observables provide decent sensitivity to the values of \msbarm[t] and \msrm[t] 
in a simultaneous fit with $\alpha_s(M_Z)$ and the PDFs, i.e. the complete set of input
theoretical parameters of fixed-order calculations for stable top-quark pair
production. We compare the results with the ones obtained in the CMS
analysis~\cite{Sirunyan:2019zvx}. 
In particular, the distributions of the $t\bar{t}$ invariant mass and
the additional jet multiplicity are sensitive to the top-quark mass through
threshold and cone effects~\cite{Alioli:2013mxa}. 

The QCD analysis setup follows the original CMS
analysis~\cite{Sirunyan:2019zvx}, 
and the main settings are summarized in the following paragraph~\footnote{The detailed description 
of the fitting procedure can be found in Ref.~\cite{Sirunyan:2019zvx}, see Section~10 in particular.}.
The QCD analysis is done using the {\texttt{xFitter}} framework~\cite{Alekhin:2014irh}. 
Theoretical predictions for the $t\bar{t}$ data are obtained at NLO in the pole mass scheme using the
{\texttt{MadGraph5\_aMC@NLO}} program~\cite{Alwall:2014hca}, 
interfaced with {\texttt{aMCfast}}~\cite{Bertone:2014zva} and
{\texttt{ApplGrid}}~\cite{Carli:2010rw} to store the calculated cross-sections
bin-by-bin in the format which is suitable for PDF fits with {\texttt{xFitter}}. 
The dependence of the theoretical predictions on the top-quark mass 
is taken into account by generating several sets of predictions
with different values of this parameter and smoothly interpolating them in the fit. 
The HERA combined inclusive DIS data~\cite{Abramowicz:2015mha} 
are included in the fit to provide constraints on the valence and sea quark
distributions and to probe the gluon distribution and $\alpha_s$ through scaling
violations, while the CMS $t\bar{t}$ data provide direct constraints on the
gluon PDF and $\alpha_s$, as well as on the top-quark mass as discussed in Ref.~\cite{Sirunyan:2019zvx}. 

In our analysis, we convert the NLO calculations for the $t\bar{t}$ production
cross-sections from the pole mass scheme into the \msbar or MSR mass schemes
according to eq.~(\ref{eq:conversion}). 
Due to the fact that the calculated cross-sections are stored in {\texttt{ApplGrid}} 
tables as bin integrated cross-sections, 
it is not possible to use a dynamic scale $\mu_R = \mu_F = H' = (\sum_i\, m_{T,i})/2$, 
defined as one half of the sum of transverse masses $m_{T,i} = \sqrt{m_i^2+p_{T,i}^2}$
of the final-state partons $i$, since $H'$ is not constant within the
bin~\footnote{At the best of our understanding, the publicly available version
  of \texttt{MadGraph5\_aMC@NLO} is not yet capable of computing integrals
  over bins using a running mass, but only using a pole mass.}.  
Instead, we use a static scale $\mu_R = \mu_F = \polem[t]$, and we 
perform the extraction of the pole mass with this scale choice.

As the analysis of triple-differential $t\bar{t}$ cross-sections requires NLO
predictions not only for inclusive $t\bar{t}$ production ($N_{jet} \ge 0$), 
but even for inclusive $t\bar{t}+1$ jet production ($N_{jet} \ge 1$), 
{\texttt{MadGraph5\_aMC@NLO}} is the only public code, 
among those providing such calculations, 
that is already interfaced to {\texttt{ApplGrid}}. 
In general, also other frameworks implementing NLO QCD corrections 
could be adopted, even beyond the fixed-order studies considered here, 
but they are not yet interfaced with {\texttt{ApplGrid}}.  

The fit results obtained using different mass schemes are given in Table~\ref{tab:mtfit}. 
The values of $\chi^2$ characterize the fit quality. 
These values are very similar in all fit variants and illustrate a general good
description of the $t\bar{t}$ data. 
To estimate uncertainties, we follow the procedure from Ref.~\cite{Sirunyan:2019zvx} and determine fit, model, parametrization and scale variation uncertainties. 
As in the CMS analysis, the scales are varied coherently in all bins of the
measured cross-sections. 
As shown in Table~\ref{tab:mtfit}, in the pole mass scheme, switching from the dynamic scale
$H^\prime$ to the static scale $\polem[t]$ modifies the extracted pole mass by about
$0.6$~GeV, a value still smaller than the fit uncertainties amounting to $0.7$~GeV, 
but enlarges the scale uncertainties substantially. 
Therefore, the larger scale uncertainties obtained in this analysis using the \msbar or MSR mass schemes,
as compared to Ref.~\cite{Sirunyan:2019zvx}, are explained by the usage of the
static scale in the calculations. 
Switching from the pole mass $\polem[t]$ to the \msbar mass $\msbarm[t]$ or the MSR mass $\msrm[t]$(3 GeV) 
does not affect the scale uncertainties significantly.
On the other hand, if in the future one would know the value of the \msbar
masses very precisely (from some other measurement), one could use them to get
accurate predictions for differential cross-sections with smaller heavy-quark
mass uncertainties, while the pole mass would be affected by
$\mathcal{O}$($\Lambda_{QCD}$) uncertainties.  

In the light of these observations, it will be worth to implement the
transition to the other mass schemes directly in the
{\texttt{MadGraph5\_aMC@NLO}} program~\footnote{At the moment the program 
{\texttt{MadGraph5\_aMC@NLO}} does not compute directly cross-section
integrals using running masses and we have developed the {\texttt{xFitter}}
interface to it to convert the predictions in the pole mass scheme to the \msbar and MSR schemes.}
and in further Monte Carlo integrators/event generators, such that predictions for differential
$t\bar{t}$ cross-sections in association with jets can be 
obtained in the format which is suitable for PDF fits in different mass 
schemes and with dynamical scales. 
The advantages of the latter for the running masses have been illustrated in
the previous Sec.~\ref{diff}.

Furthermore,
in Table 3
we do not observe any noticeably larger theoretical uncertainty when
fitting
the \msbar running mass instead of
the pole mass,
contrary to what
was reported in Refs.~\cite{Fuster:2017rev,Aad:2019mkw}.
A direct comparison of our analysis to those ones is not
  possible because of the different data sets used. 
  Our analysis is based on triple differential
  distributions in the invariant mass and rapidity of the $t\bar{t}$-pair and
  in the number of light jets, using recent precise data obtained by the CMS
  collaboration during the LHC Run 2 at $\sqrt{s}=$~13~TeV, whereas the studies in
  Refs.~\cite{Fuster:2017rev,Aad:2019mkw} make use of a different differential
  distribution ($d\sigma/d\rho_s$, where $\rho_s$ is an observable related to the
  inverse of the $M_{t\bar{t}j}$ invariant mass), as measured by the ATLAS
  collaboration, at different center-of-mass energies ($\sqrt{s}=$ 7 and 8
  TeV, respectively), during Run 1.
Switching from the pole mass scheme to the MSR mass scheme with $R=3$~GeV changes the extracted mass value by $0.3$~GeV only, which is
well within the current experimental and theoretical uncertainties. On the
other hand, the value of $\alpha_s(M_Z)$ extracted from the fit does not
change significantly when using different schemes, as also shown in Table~\ref{tab:mtfit}. 
The obtained values are compatible with $\alpha_s(M_Z)=0.1191 \pm 0.0011$ 
extracted in the ABMP16 fit at NLO~\cite{Alekhin:2018pai} within two standard deviations~\footnote{
The PDG value of $\alpha_s(M_Z)=0.1179 \pm 0.0010$ is based on comparisons to theory at NNLO and on lattice data.}.

The extracted value of $\msbarm[t]$ is compared with several other determinations in Fig.~\ref{fig:mtmt}.
In the ABMP16 analysis, the running top-quark mass was determined from
measurements of total top-quark pair and single-top production cross-sections
in a global QCD fit at NNLO~\cite{Alekhin:2017kpj}. 
In Ref.~\cite{Aad:2019mkw} ATLAS extracted a \msbarm[t] value at NLO from their measurement of $t\bar{t}+1$ jet
production cross-sections, while Ref.~\cite{Fuster:2017rev} has obtained \msbarm[t] 
at NLO using the ATLAS measurement of $t\bar{t}+1$ jet production~\cite{Aad:2015waa} on the basis of LHC Run-1 data. 
Currently, the world average value of \msbarm[t] by the PDG~\cite{Tanabashi:2018oca} is based on a single
determination of this parameter by the D0 collaboration at approximate NNLO~\cite{Abazov:2011pta}. 
When comparing to the other determinations of $\msbarm[t]$ displayed in Fig.~\ref{fig:mtmt}, 
it is worth to note that only the results of this work and of the ABMP16
analysis are obtained in a simultaneous fit of $\msbarm[t]$, $\alpha_s(M_Z)$ and PDFs, preserving 
correlations among these quantities, while the other determinations were done
using a value of $\alpha_s(M_Z)$ and a PDF set fixed a-priori.

In line of principle, the applied methodology can be extended to the extraction 
of the \msbarm[c] and \msbarm[b] mass values from measurements of charm and bottom production 
in association with jets at colliders. 
However, this is a great challenge from the experimental point of view, because measuring 
jets at low $p_T$, where the sensitivity to the charm- and bottom-quark mass
would be particularly large, is hard.

\subsection{NLO PDF fits with differential charm hadro-production cross-sections}
\label{nlo-charm-fits}

The application of differential distributions for charm hadro-production 
with the \msbar mass definition allows for an update of PDF fits 
which use heavy-flavor measurements from the LHC, 
to constrain the gluon distribution at low values of the longitudinal momentum
fraction $x$~\cite{Zenaiev:2015rfa,Gauld:2015yia,Gauld:2016kpd}. 
In particular, constraints at the lowest $x$ values explored nowadays 
($x \gtrsim 10^{-6}$) can be obtained by considering the charm hadro-production
process at high rapidities ($|y| \lesssim 4.5$) at the LHC, whereas the bottom
hadro-production process at similar rapidities at the LHC is sensitive to
slightly larger $x$ values ($x \gtrsim 10^{-5}$), with a region of sensitivity
that partially overlaps with the one of charm data. 
Because of the large scale dependence of the NLO calculations for charm hadro-production, it is customary  
to include in such fits only ratios of cross-sections, which are constructed
using measurements at different values of rapidity and/or transverse momentum,
or at different center-of-mass energies.  

As an example, in the PROSA analysis~\cite{Zenaiev:2015rfa} charm
and bottom hadro-production cross-sections~\cite{Aaij:2013mga,Aaij:2013noa}
as a function of rapidity were used in ratios 
to the respective cross-section in the rapidity interval $3 < y < 3.5$  
for each $p_T$ bin, 
together with the inclusive DIS data~\cite{Aaron:2009aa} and the heavy-flavor
production DIS data~\cite{Abramowicz:1900rp,Abramowicz:2014zub} from HERA. 
These ratios feature a reduced scale dependence, but, at the same time, they have reduced 
sensitivity to the heavy-quark mass. 
We repeat this PROSA analysis using the \msbar heavy-quark masses in the
calculations of both the DIS structure functions~\cite{Alekhin:2010sv} 
and the charm and bottom hadro-production cross-sections, instead of pole masses, 
while all other settings are as in Ref.~\cite{Zenaiev:2015rfa}. 
As a result, we observe only a small impact on the $\chi^2$ value and the fitted PDFs, 
with a new central PDF that is well within the previously found PDF uncertainties. 
These small differences are driven mainly by the change in the predictions for the HERA
data, because the LHCb data used in the format of normalised cross-sections do
not provide any notable sensitivity neither to the heavy-quark mass scheme, nor to
the value of the heavy-quark mass. 
As a result, the fitted \msbar heavy-quark masses are determined as 
\begin{eqnarray}
\msbarm[c] &=& 1.17 \pm 0.05~\textrm{GeV}\, ,
\label{eq:dztocc0}
\\
\msbarm[b] &=& 3.98 \pm 0.14 ~\textrm{GeV}\, ,
\label{eq:dztocc1}
\end{eqnarray}
which can be compared with the fitted values of heavy-quark masses 
that arise when using the pole masses in the theory predictions in the fit,
\begin{eqnarray}
\polem[c] &=& 1.26 \pm 0.06 ~\textrm{GeV}\, ,
\label{eq:dztocc2}
\\
\polem[b] &=& 4.19 \pm 0.14 ~\textrm{GeV}\, .
\label{eq:dztocc3}
\end{eqnarray}
The quoted uncertainties are fit uncertainties only.
In the
  evaluation of $m_b(m_b)$ we neglect the uncertainties related to the charm
  mass value used in the fermionic loops appearing in virtual corrections for
  $b\bar{b}$ hadro- and DIS-induced production, all evaluated with $m_c=0$.
The \msbar masses in eqs.~(\ref{eq:dztocc0}), (\ref{eq:dztocc1}) are compatible with the results obtained in
Refs.~\cite{Abramowicz:1900rp,Abramowicz:2014zub,Bertone:2016ywq,H1:2018flt}, 
where the HERA data alone were analyzed to determine the heavy-quark \msbar masses. 
The \msbar masses are also in better agreement with the world average values~\cite{Tanabashi:2018oca}, 
than the pole masses of eqs.~(\ref{eq:dztocc2}), (\ref{eq:dztocc3}), indicating that the latter carry a significant
intrinsic theoretical uncertainty. 
Therefore in our most recent PDF analysis~\cite{Zenaiev:2019ktw} 
we have solely adopted heavy-quark running masses.  

\subsection{NNLO PDF fits with total charm hadro-production cross-section}
\label{nnlo-charm-fits}

\begin{table}
  \setlength{\tabcolsep}{8pt}
  \renewcommand{\arraystretch}{1.35}
  \begin{tabular}{lll}
    \hline
    Measurement & Final state & Kinematic region \\
    \hline
    ALICE 5 TeV~\cite{Acharya:2019mgn} & \dz, \dch, \dstar, \ds & $0 < p_T < 36$ GeV, $|y|<0.5$ \\
    ALICE 7 TeV~\cite{Acharya:2019mgn} & \dz, \dch, \dstar, \ds & $0 < p_T < 36$ GeV, $|y|<0.5$ \\
    ATLAS 7 TeV~\cite{Aad:2015zix} & \dch, \dstar, \ds & $3.5 < p_T < 100$ GeV, $|\eta|<2.1$ \\
    LHCb 5 TeV~\cite{Aaij:2016jht} & \dz, \dch, \dstar, \ds & $0 < p_T < 10$ GeV, $2<y<4.5$ \\
    LHCb 7 TeV~\cite{Aaij:2013mga} & \dz, \dch, \dstar, \ds, \lamc & $0 < p_T < 8$ GeV, $2<y<4.5$ \\
    LHCb 13 TeV~\cite{Aaij:2015bpa} & \dz, \dch, \dstar, \ds & $0 < p_T < 15$ GeV, $2<y<4.5$ \\
    \hline
  \end{tabular}
  \caption{
    Summary of the most precise measurements of open charm production at the LHC.
  }
  \label{tab:charmdata}
\end{table}

The NLO predictions for differential charm hadro-production at the LHC 
have very large scale uncertainties ($>100\%$ in some phase space regions), 
as illustrated in Sec.~\ref{diff}.
The lack of theory predictions for differential cross-sections on charm and
bottom hadro-production at NNLO prevents including the corresponding existing data 
in the state-of-the-art PDF fits, which nowadays are mostly provided at NNLO accuracy. 
In this context measurements of the total charm hadro-production cross-section 
would be beneficial, because they can be confronted in the PDF fits with 
the already available inclusive NNLO predictions~\cite{Baernreuther:2012ws,Czakon:2012zr,Czakon:2012pz,Czakon:2013goa}
which have significantly reduced scale uncertainties. 
However, no such measurements have been performed to date. 

On the other hand, the ALICE~\cite{Acharya:2017jgo,Acharya:2019mgn}, ATLAS~\cite{Aad:2015zix} and
LHCb~\cite{Aaij:2013mga,Aaij:2015bpa,Aaij:2016jht} experiments have provided
measurements of charm production in different kinematic regions which cover
more than one half of the phase space. One can reliably determine the total
cross-section by extrapolating these measurements to the full phase space. 
The extrapolation procedure is analogous to that adopted for extracting reduced
cross-sections for charm production in $ep$ collisions at
HERA~\cite{H1:2018flt} from measurements in a fiducial phase space. 
These reduced cross-sections are then routinely used in global PDF fits. 
In the following, we perform such extrapolations and provide the inferred values of
the total $c\bar{c}$ production cross-section at different center-of-mass energies and their
ratios, together with experimental and theoretical uncertainties arising from
the extrapolation procedure.
We then compare the results to theoretical predictions at NNLO in QCD which are
obtained using different PDF sets, and demonstrate how these data can help
to reduce PDF uncertainties. 

The existing most precise LHC measurements of open charm production are summarized
in Table~\ref{tab:charmdata}. 
The ALICE measurements at $\sqrt{s}$ = 5 and 7~TeV 
cover the central region $|y| < 0.5$, the LHCb measurements at 5, 7 and 13~TeV 
provide coverage of the forward region $2<y<4.5$, and the ATLAS
measurement at 7~TeV essentially bridges the gap by providing data at $|\eta| < 2.1$. 
However, while both ALICE and LHCb provide measurements nearly 
in the full $p_T$ range starting from 0~GeV, ATLAS reports the cross-sections
only for $p_T > 3.5$~GeV, thus leaving the bulk of the corresponding $p_T$
kinematic range unmeasured. Furthermore, it turns out that the most precise
data of ALICE and LHCb among all open $D$-meson data are those for \dz
production, while this final state was not measured by ATLAS. 

Given these arguments, we extrapolate ALICE and LHCb measurements of \dz
production at 5 and 7~TeV to the full phase space. 
In order to maintain the least dependence on the theoretical predictions, 
both ALICE and LHCb measurements are extrapolated to nearby regions of $y$,
namely to $0<|y|<1.5$ and $|y|>1.5$, respectively: 
\begin{equation}
\sigma_{\textrm{total}} \,=\, 
\sigma_{\textrm{ALICE}} \times K_{\textrm{ALICE}} + \sigma_{\textrm{LHCb}} \times K_{\textrm{LHCb}} \times 2
\, ,
\label{eq:extrapolation1}
\end{equation}
where
\begin{equation}
K_{\textrm{ALICE}} \,=\, \frac{\sigma^{\textrm{NLO}}_{|y|<1.5}}{\sigma^{\textrm{NLO}}_{|y|<0.5}}
\, ,
\qquad\qquad
K_{\textrm{LHCb}} \,=\, \frac{\sigma^{\textrm{NLO}}_{|y|>1.5}}{\sigma^{\textrm{NLO}}_{2<|y|<4.5}}
\, .
\label{eq:extrapolation2}
\end{equation}
Here $\sigma_{\textrm{ALICE}}$ and $\sigma_{\textrm{LHCb}}$ denote the ALICE and
LHCb data on fiducial cross-sections, respectively, and
$\sigma^{\mathrm{NLO}}$ in different rapidity ranges are the theoretical
predictions. The factor $2$ in the second term takes into account that the
LHCb data are provided for only $2<y<4.5$ and need to be extrapolated to
$2<|y|<4.5$. 
We exploit the symmetry around $y=0$ and assume that the 
cross-sections for $2<y<4.5$ and $-4.5<y<-2$ are equal, as reasonably expected in
case of $pp$ collisions. 
Also the measurements are extrapolated into the full range of $p_T$ (not
shown in eqs.~(\ref{eq:extrapolation1}), (\ref{eq:extrapolation2}) for brevity), 
which implies only a $1\%$ correction for the LHCb data at 7~TeV provided for $0 < p_T < 8$~GeV, 
and even smaller corrections for the ALICE data sets. 
This procedure is used to obtain the total cross-section for \dz production at
collision energies $\sqrt{s}$ = 5 and 7~TeV, while at 13~TeV we extrapolate solely the LHCb
measurement since no other data are available at this
energy~\footnote{\label{footnotealice} Preliminary predictions on $D^0$
  production in $pp$ collisions at $\sqrt{s}$ = 13~TeV were reported by the
  ALICE collaboration in a conference proceeding~\cite{Hamon:2018zqs} in
  2018, but they have neither been confirmed yet nor further refined in a
  regular article. 
  In addition, the data are presented in plots, but no 
  numerical tables are provided in Ref.~\cite{Hamon:2018zqs}.}.  

We calculate the total charm production cross-section from the \dz production
cross-section dividing the latter by the fragmentation fraction from
Ref.~\cite{Lisovyi:2015uqa}: 
\begin{equation}
  \sigma({c\bar{c}}) \,=\, \sigma({\dz} + {\bar{D}^0}) / (2f(c \to \dz))
  \, ,\qquad\qquad
  f(c \to \dz) \,=\, 0.6141 \pm 0.0073
  \, .
  \label{eq:dztocc}
\end{equation}
The factor $2$ in eq.~(\ref{eq:dztocc}) accounts for the fact that both $c$ and $\bar{c}$
fragment into charmed hadrons. 
We assume $f(c\rightarrow D^0) = f(\bar{c} \rightarrow \bar{D}^0)$, and
$f(c\rightarrow \bar{D}^0) = f(\bar{c} \rightarrow D^0) = 0$. 
The uncertainty on $f(c \to \dz)$, which amounts to $1\%$, is neglected. 
We also compute ratios of cross-sections at different center-of-mass energies
$R_{7/5} = \sigma_{7~\textrm{TeV}} / \sigma_{5~\textrm{TeV}}$ and 
$R_{13/7} = \sigma_{13~\textrm{TeV}} / \sigma_{7~\textrm{TeV}}$, which benefit from
a partial cancellation of theoretical uncertainties~\cite{Cacciari:2015fta}. 

The theoretical predictions $\sigma^{\mathrm{NLO}}$ in eqs.~(\ref{eq:extrapolation1}), (\ref{eq:extrapolation2})
are computed using the \msbar masses as described in the previous
sections. 
The hard-scattering cross-sections for heavy-quark hadro-production are
supplemented with phenomenological non-perturbative FFs to describe the $c \to \dz$ transition. 
The factorization and renormalization scales are chosen to be
$\mu_R = \mu_F = \mu_0 = \sqrt{4m_c^2(m_c)+p_T^2}$ and varied by a factor of
two up and down (both simultaneously and independently for $\mu_R$ and $\mu_F$) 
to estimate the scale uncertainties with the conventional 7-point scale variation,
leaving out the combinations ($\mu_R$, $\mu_F$) = (0.5, 2)$\mu_0$ and (2, 0.5)$\mu_0$.
The \msbar charm-quark mass is set to $\msbarm[c] = 1.275 \pm 0.030$~GeV~\cite{Tanabashi:2018oca}. 

The proton is described by the PROSA PDF set~\cite{Zenaiev:2015rfa}, which is expected to have a reliable gluon
distribution at low $x$ thanks to the heavy-quark data used for its determination. 
Furthermore, to estimate the PDF uncertainties, the extrapolation is performed
using the ABMP16~\cite{Alekhin:2018pai}, CT14~\cite{Dulat:2015mca}, MMHT2014~\cite{Harland-Lang:2014zoa}, 
JR14~\cite{Jimenez-Delgado:2014twa}, NNPDF3.1~\cite{Ball:2017nwa} and HERAPDF2.0
 FF3A~\cite{Abramowicz:2015mha} NLO PDF sets. 
Then, the envelope covering the PROSA PDF uncertainties and the difference obtained using 
any of the additional PDF sets is constructed. 
This conservative procedure is essential, because the theoretical calculations for
the highest $y$ values involve the gluon PDF at the lowest $x$ values (up to
$4\times 10^{-8}$), which are not directly covered by data in any of the PDF
fits (not even in the PROSA fits which include the charm data up to $y = 4.5$
as measured by LHCb, for which PDFs at $x < 10^{-6}$ and their uncertainties
are extrapolated from the results obtained up to $x \sim 10^{-6}$, using
built-in procedures in the LHAPDF library~\cite{Buckley:2014ana}). 

The fragmentation of charm-quarks into \dz mesons is described by the
Kartvelishvili function with $\alpha_K = 4.4 \pm 1.7$~\cite{Zenaiev:2015rfa},
while the fragmentation fraction $f(c \to \dz)$ cancels for the extrapolation
factors in eqs.~(\ref{eq:extrapolation1}), (\ref{eq:extrapolation2}). 

All theoretical uncertainties are
assumed to be fully correlated for cross-sections in different kinematic
regions and at different center-of-mass energies. 
The robustness of the extrapolation procedure is checked by varying the boundary $y$ between the
kinematic regions into which the ALICE and LHCb measurements are extrapolated
by $\pm 0.5$ (at the same time, this variation tests consistency of the ALICE and LHCb data). 
As a further check of the method, we have computed predictions for the ALICE and LHCb data using
NLO matrix elements matched, according to the Powheg
method~\cite{Nason:2004rx, Frixione:2007vw}, to parton shower and
hadronization implemented in \texttt{PYTHIA8}~\cite{Sjostrand:2019zhc},  
and found these predictions to be consistent with our NLO~+~FF predictions within theoretical uncertainties. 

\begin{table}
  \renewcommand{\arraystretch}{1.5}
  \begin{tabular}{|c||c|c|c|c|c|c||c||c|}
    \hline
    \multirow{2}{*}{Observable \textbackslash Unc. [\%]} & \multirow{2}{4.8em}{($\mu_R$,~$\mu_F$) var.~at~NLO} & \multirow{2}{2em}{\msbar mass} & \multirow{2}{*}{$\alpha_{K}$} & \multirow{2}{*}{PDF} & \multirow{2}{*}{$y$} & \multirow{2}{2.3em}{Total th.} &  \multirow{2}{*}{Exp.} &  \multirow{2}{*}{Total} \\[3ex] \hline 
    $\sigma(c\bar{c})_{\rm 5TeV}/{\rm mb} = 5.254$ & ${}^{+0.8}_{-0.6}$ & ${}^{-0.1}_{+0.1}$ & ${}^{-2.0}_{+1.1}$ & ${}^{+4.8}_{-1.5}$ & ${}^{-2.0}_{+2.2}$ & ${}^{+5.0}_{-2.5}$ & $\pm{4.3}$ & ${}^{+6.6}_{-5.0}$ \\ \hline 
    $\sigma(c\bar{c})_{\rm 7TeV}/{\rm mb} = 6.311$ & ${}^{+0.7}_{-0.6}$ & ${}^{-0.1}_{+0.1}$ & ${}^{-2.0}_{+1.1}$ & ${}^{+7.8}_{-1.9}$ & ${}^{-2.2}_{+2.4}$ & ${}^{+7.9}_{-2.8}$ & $\pm{6.5}$ & ${}^{+10.2}_{-7.1}$ \\ \hline 
    $\sigma(c\bar{c})_{\rm 13TeV}/{\rm mb} = 11.298$ & ${}^{+0.7}_{-2.9}$ & ${}^{+0.2}_{-0.2}$ & ${}^{+1.5}_{-0.6}$ & ${}^{+0.0}_{-2.9}$ & n/a & ${}^{+1.6}_{-4.1}$ & $\pm{6.1}$ & ${}^{+6.3}_{-7.3}$ \\ \hline 
    $R_{7/5} = 1.201$ & ${}^{+0.1}_{-0.0}$ & ${}^{+0.0}_{-0.0}$ & ${}^{-0.0}_{+0.0}$ & ${}^{+2.9}_{-0.4}$ & n/a & ${}^{+2.9}_{-0.4}$ & $\pm{7.8}$ & ${}^{+8.3}_{-7.8}$ \\ \hline 
    $R_{13/7} = 1.790$ & ${}^{+1.3}_{-3.5}$ & ${}^{+0.2}_{-0.2}$ & ${}^{+3.6}_{-1.7}$ & ${}^{+1.0}_{-8.5}$ & n/a & ${}^{+3.9}_{-9.3}$ & $\pm{8.9}$ & ${}^{+9.7}_{-12.9}$ \\ \hline 
  \end{tabular}
  \caption{Extrapolated total charm production cross-sections and their
    ratios at different center-of-mass energies together with uncertainties
    from parametric variations of the scales at NLO, the mass $\msbarm[c] \pm
    0.03$~GeV, $\alpha_{K} \pm 1.7$, PDFs and the rapidity $y_{\rm ALICE,LHCb} \pm 0.5$.
    The correlation factor between $R_{7/5}$ and $R_{13/7}$ is $-0.61$.
    $\alpha_S$ uncertainties are negligible compared to the PDF ones, computed
    using as a baseline the CT14 PDF set of eigenvectors at NLO.
  }
  \label{tab:extrap}
\end{table}

\begin{figure}
  \centering
  \includegraphics[width=1.0\textwidth]{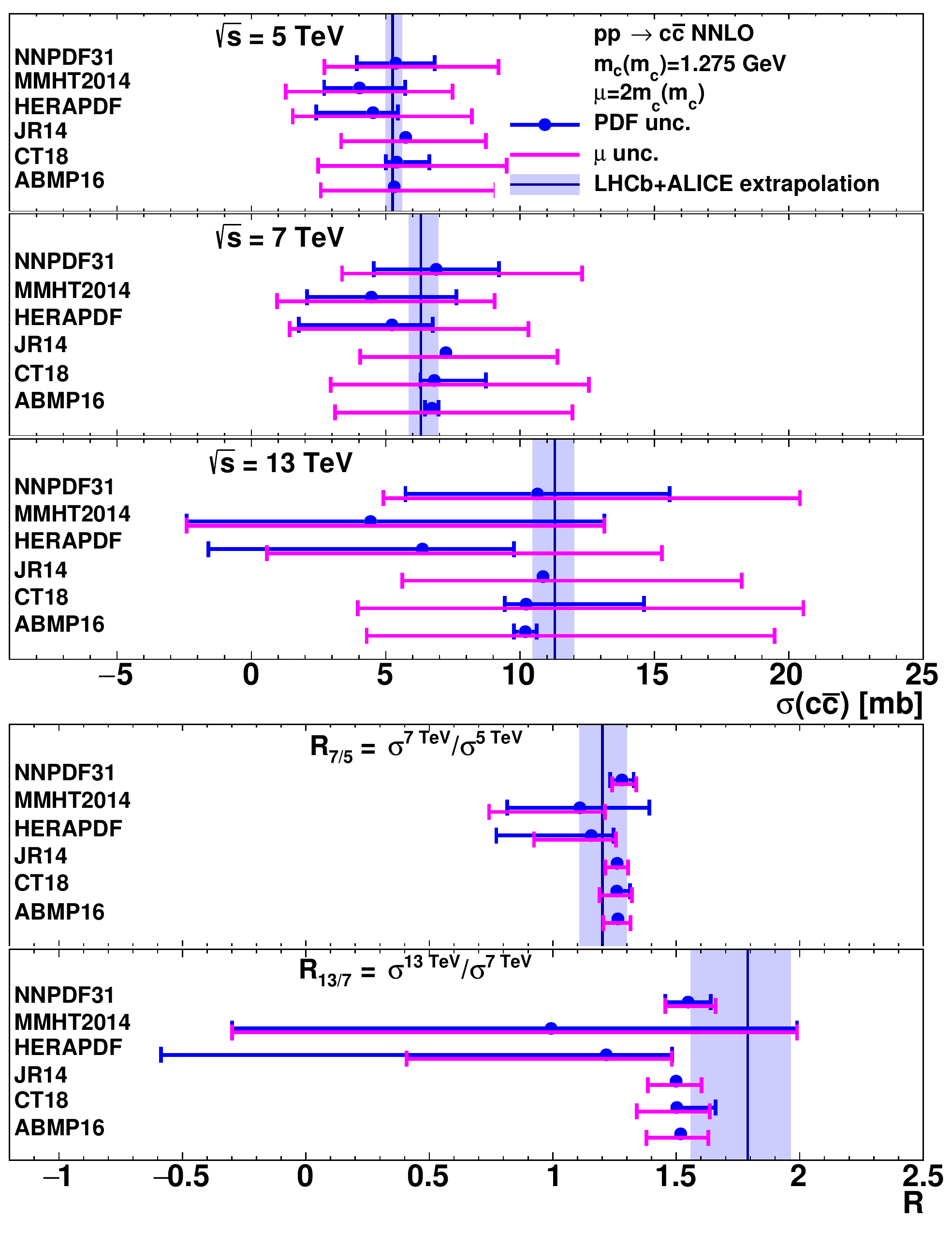}
  \caption{Comparison of the extrapolated total charm production 
    cross-sections and their ratios with the NNLO theoretical predictions using
    different PDF sets.
    Uncertainties from scale variations at NNLO ($\mu$) and PDFs are shown separately.
  }
  \label{fig:extrap}
\end{figure}

The results of the extrapolation are reported in Table~\ref{tab:extrap}. 
The scale, mass, PDF and fragmentation uncertainties are added in quadrature to
obtain the total theoretical uncertainty assigned to the extrapolated results. 
The experimental uncertainties of the input data are propagated to the extrapolated cross-sections and reported separately. 
The experimental uncertainties of the input data sets are assumed to be fully
uncorrelated~\footnote{We are confident this is quite a reasonable assumption,
  already also adopted in e.g. Ref.~\cite{Zenaiev:2019ktw, Zenaiev:2015rfa},
  in absence of more detailed information on correlation matrices in the
  experimental papers.}.  
The experimental and theoretical extrapolation uncertainties are approximately of the same size. 
The total uncertainties are obtained by adding the experimental and
theoretical uncertainties in quadrature, and amount to $\approx 10\%$. 
Our value for the total charm cross-section at 7~TeV is in agreement with the
extrapolated cross-sections reported in 
Refs.~\cite{Aad:2015zix,Acharya:2019mgn,Bhattacharya:2015jpa} within uncertainties. 

While the central values for the extrapolation factors in
eqs.~(\ref{eq:extrapolation1}), (\ref{eq:extrapolation2}) 
were obtained at NLO, their uncertainties are
calculated such that they should cover potential deviations from the unknown true QCD result. 
Therefore the resulting total cross-sections, with these
uncertainties included, can be compared to calculations in any QCD scheme to
any order. 
Furthermore, for determining these extrapolation factors, only the
shape of the predictions for the $p_T$ and $y$ differential cross-sections is
relevant, while a large part of theory uncertainties related to normalization cancels. 

The extrapolated cross-sections and their ratios are compared to NNLO 
predictions obtained using the NNLO PDF sets ABMP16~\cite{Alekhin:2017kpj}, CT18~\cite{Hou:2019efy}, 
MMHT2014~\cite{Harland-Lang:2014zoa}, JR14~\cite{Jimenez-Delgado:2014twa},
NNPDF3.1~\cite{Ball:2017nwa} and HERAPDF2.0~\cite{Abramowicz:2015mha}. 
The cross-sections are computed using the {\texttt{Hathor}} program~\cite{Aliev:2010zk} interfaced in
{\texttt{xFitter}}~\cite{Alekhin:2014irh}.  
The factorization and renormalization scales are chosen to be $\mu_R = \mu_F = 2\msbarm[c]$ and $\mu_R$ and $\mu_F$ are varied
by a factor of two up and down
according to the 7-point scale variation procedure to estimate scale uncertainties. 
The \msbar charm-quark mass is set to $\msbarm[c] = 1.275$~GeV~\cite{Tanabashi:2018oca}.

\begin{figure}
    \centering
    \includegraphics[width=0.49\textwidth]{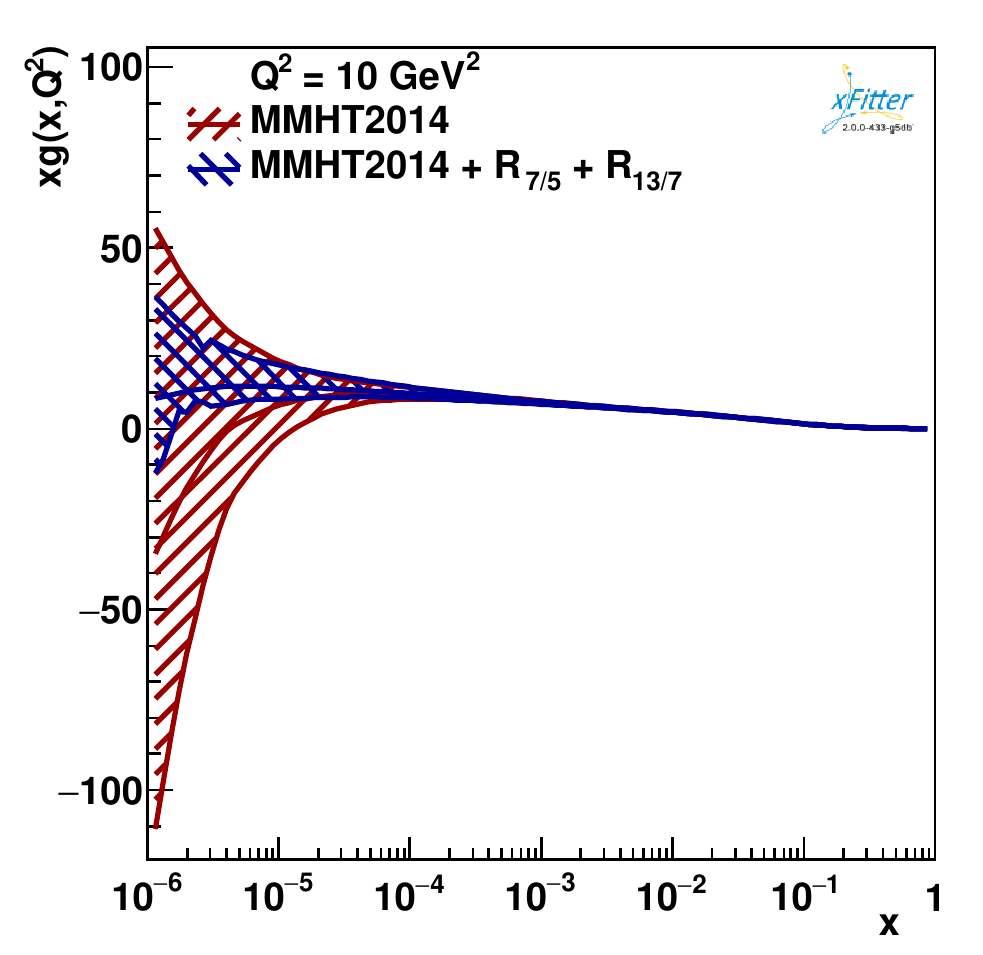}
    \includegraphics[width=0.49\textwidth]{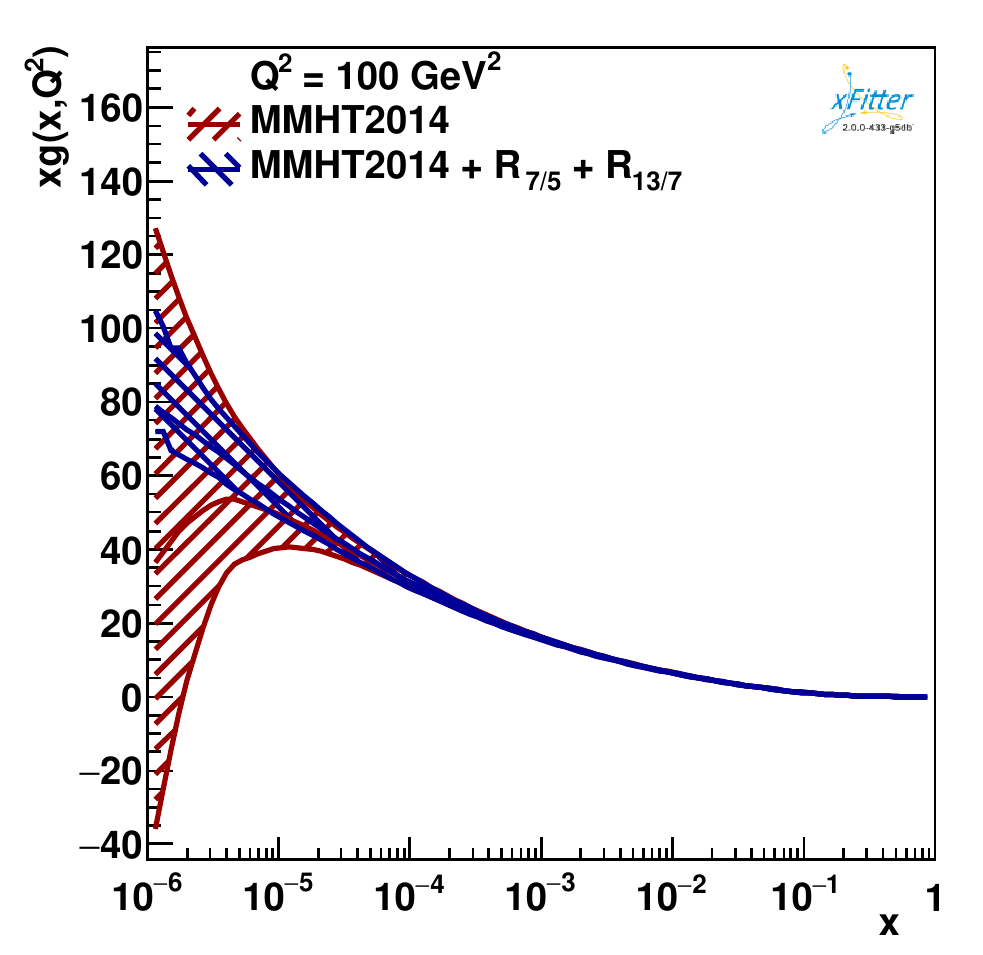}
    \caption{The gluon distribution of the original and profiled MMHT2014 NNLO PDF set at $Q^2 = 10$~GeV$^2$ (left) and $Q^2 = 100$~GeV$^2$ (right).
      }
    \label{fig:extrap-profile}
\end{figure}

\begin{figure}
\centering
\includegraphics[width=0.49\textwidth]{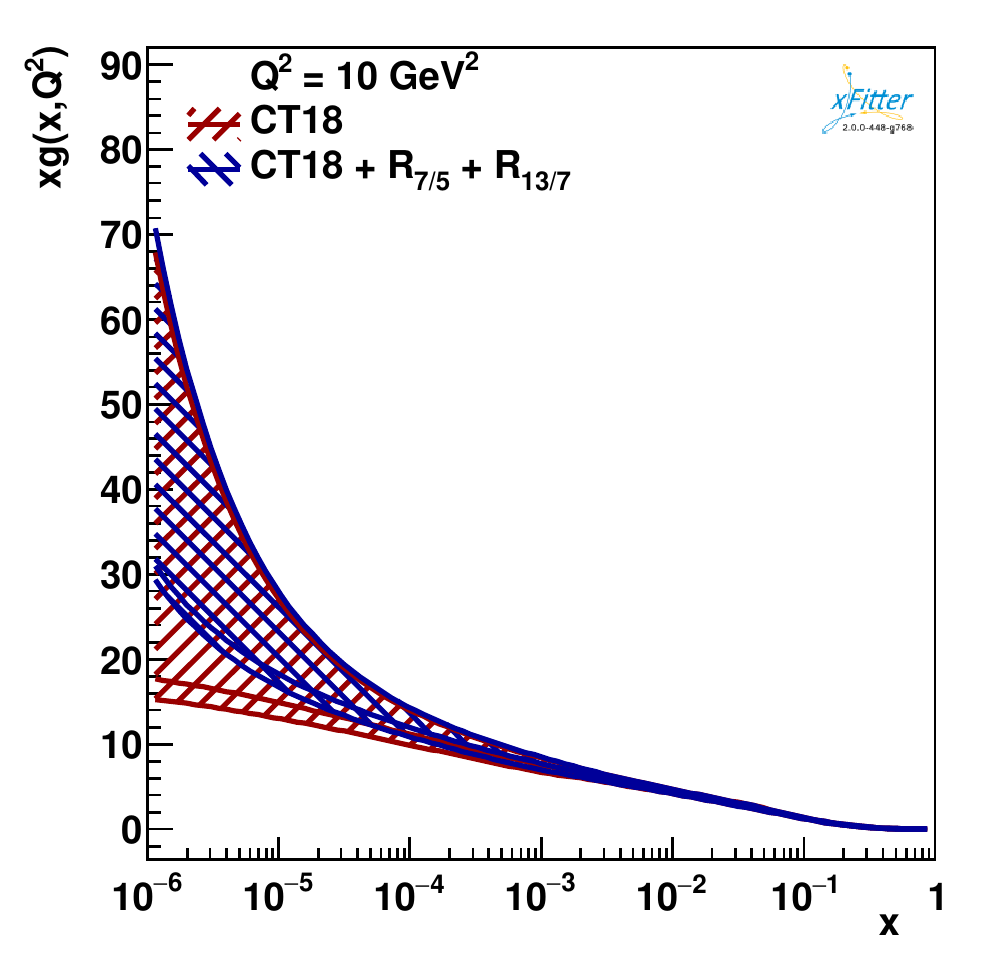}
\includegraphics[width=0.49\textwidth]{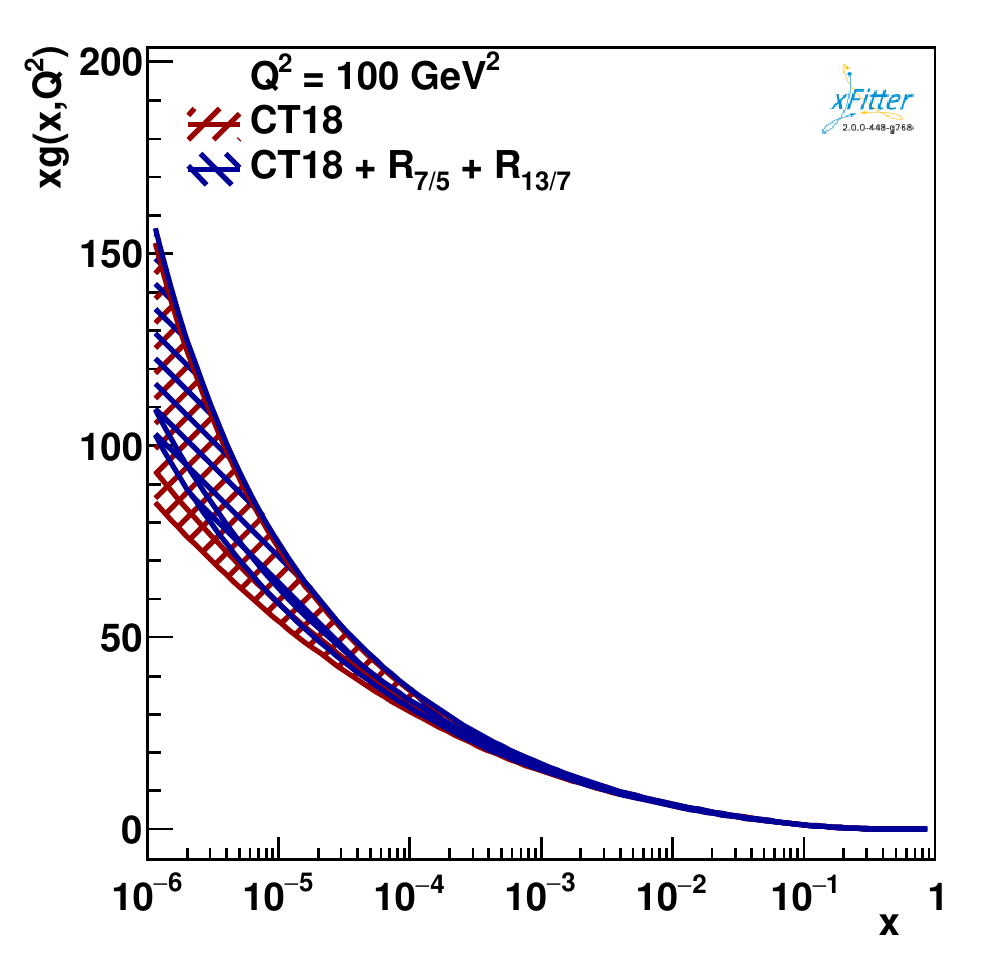}\\
\includegraphics[width=0.49\textwidth]{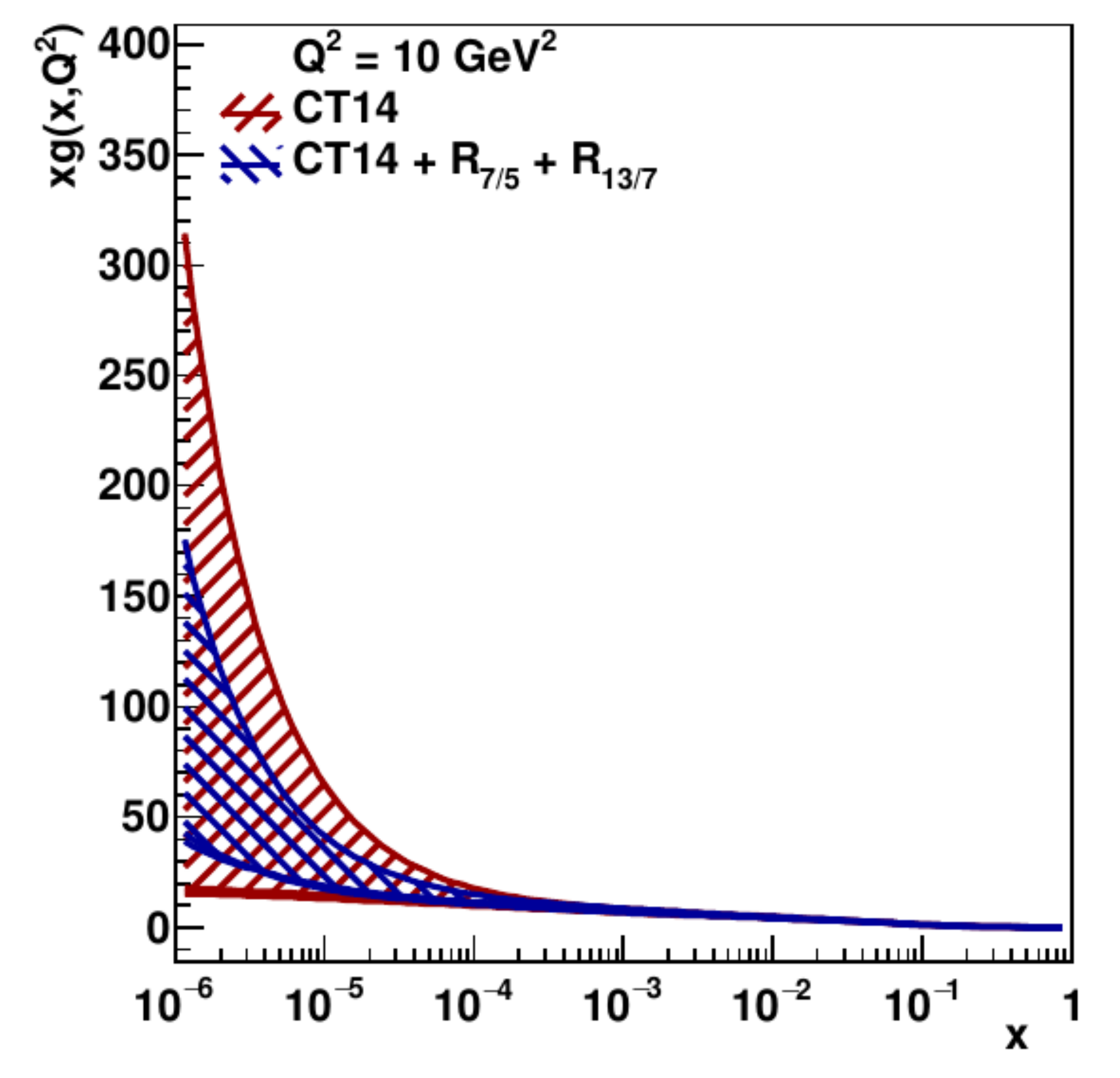}
\includegraphics[width=0.49\textwidth]{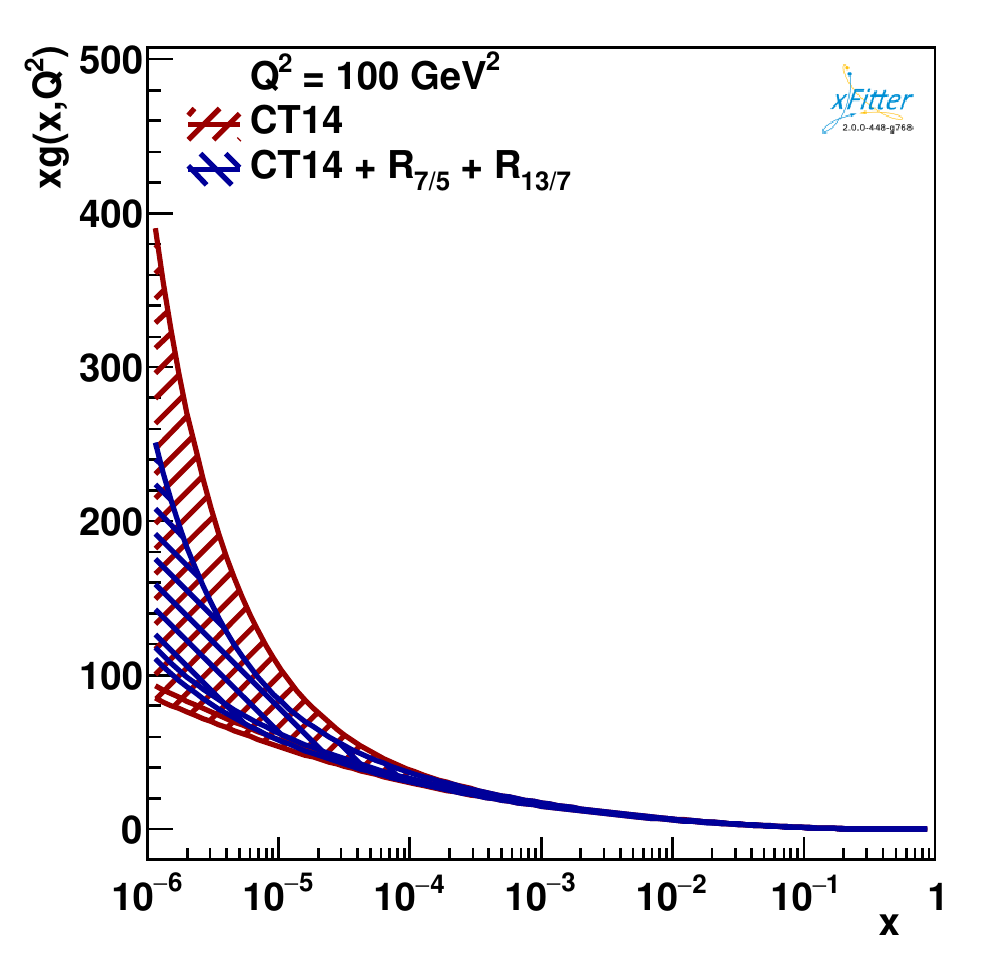}
\caption{Same as Fig.~\ref{fig:extrap-profile}, but for the CT18 PDF at 
  $Q^2$ = 10~GeV$^2$ (up left) and $Q^2$ = 100~GeV$^2$ (up right)
  and for the CT14 PDF at $Q^2$ = 10~GeV$^2$ (down left) and $Q^2$ =
  100~GeV$^2$ (down right), all at NNLO.
  }
    \label{fig:extrap-profile2}
\end{figure}

In Fig.~\ref{fig:extrap} we show the extrapolated cross-sections and their ratios compared to NNLO predictions. 
For the NNLO predictions, the theoretical uncertainty arising from scale
variations and the PDF uncertainty are shown separately. All theoretical
predictions agree with the data within uncertainties, but noticeably the
MMHT2014, HERAPDF2.0 (and CT14, not plotted in the figure)
PDF sets have uncertainties which are larger
than both scale and data extrapolation uncertainties for some of the
observables. In particular, the MMHT2014 and HERAPDF2.0 predictions for the
cross-sections at $\sqrt{s}$ = 13~TeV are consistent with negative values
within uncertainties (see also Ref.~\cite{Accardi:2016ndt}). 
Predictions based on the new CT18 PDFs (and unlike those using the previous PDF set CT14)
do not show anymore a large positive uncertainty which greatly exceeds the extrapolated cross-section.
These PDF sets could benefit from including in their fits data on charm
production cross-sections or on their ratios.

Remarkably, also the scale uncertainties appear to be different when using different PDF sets. 
Among the considered observables, the most conclusive one is $R_{7/5}$ for which both
data extrapolation and theoretical scale uncertainties are moderate ($\approx 10\%$). 
Our extrapolated value for this observable can be used in future NNLO
PDF fits to constrain the gluon PDF at low $x$. 
The other ratio $R_{13/7}$ has a larger extrapolation uncertainty suffering from a lack of experimental
measurements of charm production in the central rapidity region at 13~TeV. 
We are confident that this lack will be solved by the data which will appear
in forthcoming experimental studies at the LHC (see footnote~\ref{footnotealice}). 

As a demonstration that the provided observables can indeed constrain the PDFs, 
we employ a profiling technique~\cite{Paukkunen:2014zia} based on
minimizing the $\chi^2$ function built from data and theoretical predictions,
taking into account both data and theoretical uncertainties arising from PDF variations. 
The analysis is performed using the {\texttt{xFitter}} program~\cite{Alekhin:2014irh}. 
We consider the MMHT2014 PDF set at NNLO and the ratios $R_{7/5}$, which
exhibits the least scale uncertainties, and $R_{13/7}$. 
The correlation of $R_{7/5}$ and $R_{13/7}$ due to the common input of 7~TeV
data sets is taken into account. 
The PDF uncertainties are included in the $\chi^2$ functional 
through nuisance parameters, and the values of these nuisance parameters at
the minimum are interpreted as optimised (or profiled) PDFs, while their
uncertainties determined using the tolerance criterion of $\Delta\chi^2 = 1$
correspond to the new PDF uncertainties. 

The original and the profiled MMHT2014 gluon PDF are shown in
Fig.~\ref{fig:extrap-profile} at the scales $Q^2 = 10$ and $100$~GeV$^2$. 
The profiled distribution exhibits greatly reduced uncertainties at low $x$,
and in this region the distribution is shifted towards larger values of the gluon density.  
In case of the MMHT2014 set, the original gluon PDF is negative at low $x$ values, 
while the profiled one remains positive down to at least $x \sim 5 \cdot 10^{-6}$, 
thanks to the constraint realized by adding the ratios of charm data in the PDF fit. 
We emphasize that the strong impact at low $x$ is obtained as well when
working with other PDF sets. 
As an example, in Fig.~\ref{fig:extrap-profile2} the CT14 and CT18 gluon distributions are shown before and after profiling. 
For these sets the gluon PDF is always positive in the entire $x$ range 
for all eigenvectors by construction.
In case of CT14, adding the aforementioned data strongly reduces PDF
uncertainties at low $x$, whereas the effect is milder for CT18, but still
sizable at low $Q^2$.

\section{Conclusions}
\label{conclu}

The hadro-production of heavy-quarks is an important class of processes at LHC. 
Not only for top, but also for bottom and charm, a wealth of very precise
high-statistics data has been collected by the LHC collaborations, 
differential in the relevant kinematic variables, 
such as the transverse momentum $p_T$, the rapidity $y$ or 
the pair-invariant mass $M_{Q{\bar Q}}$ of the heavy-quarks (or of the respective heavy hadrons).
In comparison to theory predictions in perturbative QCD, these data can be
directly  used for the extraction of heavy-quark masses, which are typically
correlated with the value of the strong coupling constant $\alpha_S(M_Z)$. 
The data also have an impact on fits of fundamental non-perturbative QCD
parameters such as PDFs, where they give unique kinematic constraints.

In order to provide meaningful determinations of heavy-quark masses, the value
of $\alpha_S(M_Z)$ and PDFs, QCD predictions with good accuracy are needed.
  Theory predictions are available at NNLO for top-quark hadro-production since
  some time, also for differential distributions. Very recently differential
  predictions have also appeared for bottom pair hadro-production, but not for
  charm pair.
In the latter case, the predictions at NLO accuracy are generally not
sufficiently precise enough considering the large theoretical uncertainties,
which stem predominantly from scale variations. In view of the much smaller
experimental uncertainties reached in modern analyses improvements in the
theoretical descriptions are clearly required. 

One such aspect, which has been studied in this paper, 
is the choice of a suitable renormalization scheme for the heavy quark masses.
We have investigated different heavy-quark mass renormalization schemes with
emphasis on the \msbar and MSR masses as representative short-distance mass definitions.
The choice of a particular mass scheme as well as the values for the scales $\mu_R$ and $\mu_F$
have an impact on the rate of apparent convergence of the perturbative expansion of the cross-sections. 
We have investigated a range of dynamical scale choices for the cross-sections 
and, in case of the \msbar mass,
also for the mass renormalization scale $\mu_m$.
In particular, we have found that dynamical renormalization and factorization scales of the type ($\mu_R$, $\mu_F$) $\simeq$ $\sqrt{p_{T}^2 + \kappa\, m^2_Q(\mu_m)}$ for heavy-quark $p_T$-distributions with the running of mass $m_Q(\mu_m)$ included and $\mu_m$ set equal to $\mu_R$, when $m_Q(\mu_m)$ does not reduce to a value below about 1 GeV, can
lead to
reduced
residual scale uncertainties, with respect to the use of the analogous functional form with $m_Q(\mu_m)$ replaced by $m_Q(m)$.
    The amount of reduction depends on the input parameters of the computation and on the convention adopted for scale variation.
        The most general scale variation procedure that we have considered corresponds to
  independent variations of $\mu_R$, $\mu_F$ and
$\mu_m$ in the range [1/2, 2] around their respective central values.
The resulting 7-point envelope in ($\mu_R$, $\mu_F$) keeping
$\mu_m=\mu_m^0$
dominates the total uncertainty band, i.e. adding an independent 
$\mu_m$ variation for a 15-point envelope in ($\mu_R$, $\mu_F$, $\mu_m$)
increases the size of the band only by a moderate amount. 
This confirms previous findings by other authors in case of top quark pairs.
    The maximum amount of scale uncertainty reduction that we have observed when adopting this scale variation procedure, when comparing $p_T$ distributions with $m_Q(\mu_m)$ to those with $m_Q(m)$ with the input configuration of Fig.~\ref{fig:lucas-dynmass-b} ($\kappa=4$), occurs in case of top-antitop production and amounts to a few tens percent in the region around the peak of the top-quark $p_T$ distribution.
At NLO accuracy in QCD
scale uncertainties
are, however, in general still large for all mass schemes, but theory
predictions using \msbar or MSR masses carry smaller parametric uncertainties
in the mass values, being theoretically well-defined and free of renormalon
ambiguities. 

We have demonstrated these features in extractions of the top-quark 
\msbar and MSR masses at NLO from
recent
differential distributions measured by CMS, finding
good consistency with other determinations.
These extractions
  were performed simultaneously to the one of $\alpha_S$ and PDFs, preserving
  correlations between these quantities.
Using differential charm hadro-production cross-sections we have also been
able to improve available constraints on PDFs and, using the \msbar mass
scheme, to decrease extrapolation uncertainties when determining total 
cross-section from open charm data measured in fiducial regions of phase space 
by the LHC collaborations.
In the latter case, ratios of cross-sections are particularly useful
observables to cancel residual theoretical uncertainties.
In order to carry out theses studies, we have developed software frameworks 
using the {\texttt{MCFM}} and {\texttt{xFitter}} programs to determine 
differential distributions at NLO in QCD efficiently.

Avenues for theoretical improvements include the obviously needed QCD predictions 
for charm hadro-production at NNLO accuracy, possibly combined with
the resummation of large logarithms in specific kinematics, but also an
improved description of charm- and bottom-quark fragmentation to mesons, an
issue which has been side-stepped in the present analysis.
In addition, further systematic studies of different ($\mu_R$,
$\mu_F$, $\mu_m$) dynamical scale choices for different differential
distributions of heavy-quark hadro-production are de\-si\-ra\-ble.

The extended {\texttt{xFitter}} program, implementing the MSR and \msbar mass
renormalization schemes, as an alternative to the on-shell scheme in
heavy-quark hadro-production, is publicly available on the web, and further
extensions of the {\texttt{MCFM}} and {\texttt{Hathor}} programs used to
perform calculations in this paper are available from the authors upon request.

\subsection*{Acknowledgments}
This work has been supported in part by Bundesministerium f\"ur Bildung und Forschung (contract 05H18GUCC1).
S.M. acknowledges support by the {\it MTA Distinguished Guest Scientists Fellowship Programme in Hungary}.

M.V.G, S.M. and O.Z. would also like to thank the 
Mainz Institute for Theoretical Physics (MITP) 
of the DFG Cluster of Excellence PRISMA+ (Project ID 39083149), 
for its hospitality and support during the 
Program {\it Heavy-Quark Hadro Production from Collider to Astroparticle Physics}.


\providecommand{\href}[2]{#2}\begingroup\raggedright\endgroup

\end{document}